# OPTICS OF SUBWAVELENGTH GRADIENT NANOFILMS.

*A. Shvartsburg,[1] V. Kuzmiak [2] and G. Petite [3]*

[1] Science and Technology Center for Unique Instrumentation, Russian Acad. of Sciences, p/o 117342, Butlerov Str. 15, Moscow, Russia.
[2] Institute of Photonics and Electronics, v.v.i., Czech Acad. of Sciences, Chaberska 57, 182 51 Praha 8, Czech Republic.
[3] Laboratoire des Solides Irradies, Ecole Polytechnique, CEA – DSM, CNRS, 91128, Palaiseau, France.[1]

**Abstract :** Propagation and tunneling of light through subwavelength photonic barriers, formed by dielectric layers with continuous spatial variations of dielectric susceptibility across the film are considered. Effects of giant heterogeneity-induced non-local dispersion, both normal and anomalous, are examined by means of a series of exact analytical solutions of Maxwell equations for gradient media. Generalized Fresnel formulae, visualizing a profound influence of gradient and curvature of dielectric susceptibility profiles on reflectance/transmittance of periodical photonic heterostructures are presented. Depending on the cutoff frequency of the barrier, governed by technologically managed spatial profile of its refractive index, propagation or tunneling of light through these barriers are examined. Non-attenuative transfer of EM energy by evanescent waves, tunneling through dielectric gradient barriers, characterized by real values of refractive index, decreasing in the depth of medium, is shown. Scaling of the obtained results for different spectral ranges of visible, IR and THz waves is illustrated. Potential of gradient optical structures for design of miniaturized filters, polarizers and frequency–selective interfaces of subwavelength thickness is considered.

---

[1] corresponding author : *guillaume.petite@polytechnique.edu*, tel :+33 1 69 33 44 96, fax : +33 1 69 33 30 22)

**Contents.**

**I. Introduction.**

Nanotechnology has facilitated the manufacturing of materials and structures with unique optical properties. During the past two decades these materials have attracted a growing interest in science, technology and defense, due to their ability to control the propagation of electromagnetic waves at and below the wavelength scale. Some of these goals can be achieved by means of subwavelength dielectric films with continuous spatial variations of refractive index across the film, so called gradient photonic barriers. Understanding of the processes associated with the propagation of electromagnetic (EM) waves in heterogeneous nanostructures belongs to the central challenges of this newly shaping field of gradient optics. In our Review we consider these hot topics and their potential in design of miniaturized wave filters, polarizers, reflectors, photonic heterostructures and frequency-selective interfaces.

Physical fundamentals and mathematical basis of wave theory for heterogeneous media have a long history. Due to coordinate-dependent velocity of waves propagation in these media the leading and trailing parts of both electric (E) and magnetic (H) waveforms are traveling with different velocities which imply differences in distortions of waveforms. In particular, when a wave with harmonic envelope of E and H is incidenting on a surface of stratified medium, the spatial shapes of these waves inside the medium become non-sinusoidal; moreover, the envelopes of E and H can differ significantly. Therefore, a new trend in the wave theory – Electromagnetics of non-sinusoidal waves – has emerged in the studies of electromagnetic fields in media with continuously varying parameters. In this case exact analytical solutions of Maxwell equations for stratified media, that visualize the peculiarities of EM fields structure in such media, take a principal importance.

While discussing one-dimensional (1D) stationary propagation of EM wave, several models of coordinate-dependent dielectric susceptibility $\varepsilon(z)$, that allow exact analytical solutions of Maxwell equations, can be outlined. One of the first such profiles was found by Rayleigh in solution of wave equation for acoustic problem of sound propagation with a coordinate-dependent velocity v(z) [1]. Application of this result to wave equation, governing the EM wave propagation, brought later the widely used model of $\varepsilon(z)$ [2], known as the Rayleigh profile:

$$\frac{\varepsilon(z)}{\varepsilon(z=0)} = U_R^2(z) = (1 + z/L)^{-2} \qquad (1.1)$$

Almost half a century ago the development of optics and microwave physics stimulated the using of new exactly solvable models for $U_1^2$ [3], $U_2^2$ [4] and $U_3^2$ [5]:

$$U_1^2 = 1 + z/L \ ; \ U_2^2 = (1 + z/L)^{-1} \ ; \ U_3^2 = \exp(2z/L) \qquad (1.2)$$

where the characteristic length L is a free parameter. Later on these and more complicated models, containing, e.g., the combinations of several exponents [6], attracted much attention in fields as different as optics of liquid crystals [7], plasma radiophysics [8], magnetic hydrodynamics [9], inhomogeneous superconductivity [10]. Treatment of a series of such problems in the framework of WKB approximation was summarized in [11].

Advent of lasers stimulated a burst of interest into thin transparent layers and multilayer systems, used as optical filters, polarizers and antireflection coatings. Modeling of continuous distributions of refractive index across the film by means of step-like piece-wise profiles, developed in [12], was complemented in [13] by an analysis of reflectance of films with some continuously varying profiles of refractive index n(z); the difference between bulk values of optical constants and those of thin films was marked in [14]. The new interest in this area was sparked with the progress in fabrication of artificial materials with controlled electromagnetic parameters. The artificial dielectric [15], designed as long ago as in 1948, was treated as a large-scale model of an actual dielectric, obtained by arranging of identical conducting obstacles in a regular 3D lattice. The combined effect of all these obstacles in the lattice, excited by external EM waves, was to produce a dipole polarization and an effective electric displacement in the microwave lenses; an artificial dielectric with $\varepsilon < 1$ was reported soon [16]. Later on, especially during the past two decades, the engineered dielectric properties of thin films became a well developed field of microelectronics and nanotechnology [17].

However, addressing this review to the physical problems of gradient media, we will refer the reader to the original papers, describing the technique of manufacturing of technologically managed profiles $\varepsilon(z)$, such as, e.g., etching and photolitography [18], controlled regimes of doping [19], molecular beam epitaxy [20], ion implantation [21], nanoscale porosity variations [22], fabrication of graded metallo-dielectric composites [23], physical vapor deposition of multiple materials [24], production of sculptured thin films [25]. A special attention was given to thin periodical graded-index filters and coatings, characterized by continuous sinusoidal index variation [26]–[28]. Methods of non-destroying optical control of thin dielectric films by means of inverse scattering [29], ellipsometry [30] and Brewster angle monitoring [31] were developed. The traditional ways for mathematical treatment of these problems are exemplified by using of some exactly solvable models [32], transfer matrix approach [33] and numerical simulation [34].

During the last decade the interest into optics of thin films is strengthened by the overall attention towards the tunneling phenomena [35] and, especially, by the tunneling of photons through nanostructured metal films [36]. The enhanced optical transmittance of these periodical structures, supported by surface plasmon polariton modes in metal nanofilms [37], was shown to be an effective mechanism for resonant transfer of EM energy in optoelectronic devices [38]. The problems of surface plasmon optics will not be discussed further here, because they are covered by other reviews [39] and monographs [40].

Unlike the aforesaid structures, this review is centered on gradient dielectric photonic barriers, whose optical parameters are not determined by free carriers and, thus, are not connected with any solid plasma effects. On the contrary to polariton modes, propagating near by resonances of macroscopically homogeneous spatially dispersive media [41], we will consider the effects of non-local dispersion, arising in a macroscopically heterogeneous media. This artificial dispersion of gradient layer, arising due to technological treatment [18-28] of the host dielectric, is determined by the shape and spatial scales of refractive index profile across the layer. We restrict ourselves to layers, whose thickness d is much greater than the spatial dimension in which the refractive index is formed (d > few nm). The salient features of gradient photonic barriers under discussion are the following:

(1) Flexibility of the reflectance-transmittance spectra of heterogeneous nanofilms, controlled by the gradient and curvature of refractive index profiles of photonic barriers, and arising of cutoff frequency in barriers, fabricated from dispersionless host materials (generalized Fresnel formulae).

(2) Strong heterogeneity-induced non-local dispersion of gradient layers, both normal and anomalous, which can be formed in an arbitrary spectral range by means of appropriate geometry of refractive index profile, in the host material being given.

(3) Energy transfer by evanescent EM modes, tunneling without attenuation through dielectric photonic barrier with concave profile of dielectric susceptibility $\varepsilon$ due to interference of forward- and backward-propagating modes inside the barrier, where $\varepsilon > 0$.

(4) Subwavelength spatial scales and heterogeneity-induced birefringence of gradient photonic barriers.

(5) Scaling of results, obtained by means of exactly solvable models of gradient photonic barriers, for different spectral ranges and different thicknesses of barriers.

The mathematical fundamentals of gradient optics of thin layers are based on exact analytical solutions of Maxwell equations for media with continuous spatial variations of dielectric parameters. These solutions, obtained beyond of the scope of any truncations, perturbations or other WKB – like suppositions about smallness or slowness of variations of wave fields or media, are based on some special transforms of spatial variables in Maxwell equations. Allowing frequently to visualize the structure of EM fields in gradient media by means of elementary functions, this approach opens the way to new physical results.

Harnessing materials with strong artificial non-local dispersion opens up new avenues for synthesis of optoelectronic and microwave systems. To visualize these peculiarities of gradient optics, this review is organized as follows. Section 2 is devoted to the exactly solvable models of heterogeneous wave barriers, revealing the sensitivity of their heterogeneity-induced dispersion to the parameters of spatial distribution of dielectric susceptibility across the barrier. The spatial structure of EM fields inside convex and concave wave barriers, described by exact analytical solutions of Maxwell equations for spatially varying media, obtained above, is used in Section 3 for the analysis of the reflectance-transmittance spectra of photonic heterostructures. The intriguing effects of reflectionless tunneling of EM wave through the gradient photonic barrier with a positive decreasing dielectric susceptibility (normal incidence) as well as polarization-dependent tunneling in the case of oblique incidence are discussed in Section 4. Some concepts of gradient optics are generalized for non-planar and nonlinear barriers in Sections 5 and 6. The microwave analogies of gradient optics (gradient radio-optics) are discussed in Conclusion.

## 2. Heterogeneity–induced dispersion of gradient dielectric films (exactly solvable models).

This Chapter is devoted to the analytical methods in optics of thin gradient layers. These methods can be illustrated in the framework of the simple problem of propagation of a plane EM wave in a heterogeneous, non-magnetic dielectric medium, whose dielectric susceptibility $\varepsilon(z)$ is varying continuously along the z-coordinate. To point out the effects caused by this heterogeneity let us at assume that the wave absorption of medium and its material dispersion are insignificant in the spectral range considered. In this case the spatial dependence of dielectric susceptibility can be presented as

$$\varepsilon(z) = n_0^2 U^2(z) \ ; \ U\big|_{z=0} = 1 \qquad (2.1)$$

Here $n_0$ is the medium's refractive index of at the interface $z = 0$ and the dimensionless function U(z) describes the spatial distribution of dielectric susceptibility. Maxwell equations for a linearly polarized wave with components $E_x$ and $H_y$, traveling along the z-axis through the medium (2.1), read as

$$\frac{\partial E_x}{\partial z} = -\frac{1}{c}\frac{\partial H_e}{\partial t} \ ; \ -\frac{\partial H_y}{\partial z} = \frac{n_0^2 U^2(z)}{c}\frac{\partial E_x}{\partial t} \qquad (2.2)$$

Several models U(z), which provide exact analytical solutions for the system (2.2), were mentioned above (1.1) – (1.2). It is remarkable, that the models (1.20) can be viewed as the particular cases of one generalized distribution, containing two free parameters ( L and m ) :

$$U^2(z) = \left(1 - \frac{2}{m-2}\frac{z}{L}\right)^{-m}; \quad m \neq 2; \tag{2.3}$$

The values m = 1 and m = -1 relate to the profiles $U_1^2(z)$ and $U_2^2(z)$ (1.2). Moreover, by using the classical formula

$$\lim\left(1 + \frac{1}{x}\right)^x\bigg|_{x \to \infty} = e \tag{2.4}$$

one can show that, in the limit $m \to \infty$, distribution (2.3) tends to the exponential profile $U_3^2(z)$ in (1.2).

Models (1.1) – (1.2) describe only monotonic variations of the refractive index. A model of non-monotonic barrier was built from broken straight lines (the "trapezoidal barrier" [43]), but contains unphysical sharp points, formed by the crossing of these lines. The restricted flexibility of models (1.1) – (1.2), containing only one free parameter, hampers the optimization of the regimes of wave propagation through these layers, treated as photonic barriers. To visualize the physically meaningful parameters, important for such optimization, one has to use more flexible models. Moreover, a simple mathematical approach, providing an algorithm for generation of different profiles U(z), is in need. This approach, based on the mapping of wave equation from physical space into a phase space, is shown to reveal the decisive role of heterogeneity-induced dispersion in wave processes inside the gradient photonic barriers (chapter 2.1). Chapter 2.2 is focused on the presentation of EM fields for the series of such non-local dispersive barriers by means of standardized analytical procedures. Interaction of local and non-local dispersive effects is discussed in Chapter 2.3.

*2.1. Fourier optics in the phase space.*

Unlike the exactly solvable models (2.3) – (2.4) another series of analytical solutions of Maxwell equations (2.2) can be derived by means of a systematic use of special transforms of these equations into a space of phase trajectories (phase space). This approach permits to visualize strong non-local dispersive effects, determined by the geometry of the photonic barriers. Presentation of $E_x$ and $H_y$ components of the wave field via some generating function $\Psi$ provides the reduction of the pair of first-order equations (2.2) to one second-order equation, governing the function $\Psi$. This transform can be accomplished by two different ways:

(1) the function $\Psi$ is chosen in such a way that the first equation in (2.2) becomes an identity, while the function $\Psi$ is defined by the second equation in (2.2);

(2) in contrast, the function which transforms the second equation in the pair (2.2) to an identity, is determined by the first equation of this pair.

Let us study separately the EM fields generated in both ways.

(1). By choosing the function $\Psi$ in the form that resembles the vector potential $\vec{A}$ with components $A_x = \Psi$, $A_y = A_z = 0$

$$E_x = -\frac{1}{c}\frac{\partial \Psi}{\partial t}; \quad H_y = \frac{\partial \Psi}{\partial z} \tag{2.5}$$

one obtains the equation, governing the function $\Psi$

$$\frac{\partial^2 \Psi}{\partial z^2} - \frac{n_0^2 U^2(z)}{c^2}\frac{\partial^2 \Psi}{\partial t^2} = 0 \tag{2.6}$$

Eq. (2.6) can be viewed as a wave equation with a coordinate-dependent velocity v(z)=c/(U(z)). To solve it, let us introduce a new function F and new variable $\eta$ [44]

$$F = \Psi\sqrt{U(z)} \; ; \; \eta = \int_0^z U(z_1)dz_1 \qquad (2.7)$$

One can immediately see that the space-like variable η defined this way is proportional to the optical path traversed by light at point z. By assuming a harmonic temporal dependence of $\Psi \propto \exp(-i\omega t)$, one can rewrite Eq. (2.6) in the form

$$\frac{d^2F}{d\eta^2} + F\left(\frac{\omega n_0}{c}\right)^2 = F\left[\frac{1}{2U^3}\frac{d^2U}{dz^2} - \frac{3}{4U^4}\left(\frac{dU}{dz}\right)^2\right] \qquad (2.8)$$

Owing to transform (2.7) the unknown distribution U(z) is eliminated from the left side of the wave Eq. (2.6). Using (2.8), one can consider a multitude of different models U(z) that would provide analytical solutions of this propagation equation; however the most important features of wave field in heterogeneous media can be revealed in the framework of a simple model, when the expression in the brackets in (2.8) is equal to some constant $p^2$. This condition is for instance fulfilled for the profiles

$$U(z) = \left(1 + \frac{s_1 z}{L_1} + \frac{s_2 z^2}{L_2^2}\right)^{-1} \; ; \; s_1 = 0, \pm 1 \; ; \; s_2 = 0, \pm 1 \qquad (2.9)$$

containing two arbitrary spatial scales $L_1$ and $L_2$, which are two free parameters. Herein the case $s_1 = -1$, $s_2 = 1$ corresponds to the convex profile, while the case $s_1 = 1$, $s_2 = -1$ describes the concave one. In the case of opposite signs of $s_1$ and $s_2$ profile (2.9) has either a maximum ($s_1 = -1$, $s_2 = 1$) or a minimum ($s_1 = 1$, $s_2 = -1$) with a value $U_m$ The scales $L_1$ and $L_2$ are linked in these cases with the layer thickness d and the limits $U_m$

$$U_m = \left(1 + s_1 y^2\right)^{-1} \qquad y = L_2/2L_1 ; \quad L_2 = d(2y)^{-1}; \quad L_1 = d(4y^2)^{-1} \qquad (2.10)$$

So that $y$ (or equivalently $U_m$) and d can also be chosen as the two free parameters of the model. Substitution of U(z) (2.9) to (2.8) yields the value

$$p^2 = \frac{s_1^2}{4L_1^2} - \frac{s_2}{L_2^2} \qquad (2.11)$$

Replacing the right side of (2.8) by this value $p^2$, one obtains the simple equation for the function F:

$$\frac{d^2 F}{d\eta^2} + F\left[\left(\frac{\omega n_0}{c}\right)^2 - p^2\right] = 0 \qquad (2.12)$$

Unlike the wave equation (2.6), written for the heterogeneous medium, Eq. (2.12) describes the traveling wave in an homogeneous medium. With definition (2.7) of the variable $\eta$, it is natural that phase accumulation is linear in $\eta$, thus corresponding to standard harmonic solutions in this phase space. By combining the function F from (2.7) and transform (2.12), one obtains the solution of wave equation (2.6) [44]

$$\Psi = [U(z)]^{-\frac{1}{2}} \exp[i(q\eta - \omega t)] \qquad (2.13)$$

$$q = \frac{\omega n_0 N}{c} \; ; \; N = \sqrt{1 - \frac{(pc)^2}{\omega^2}} \qquad (2.14)$$

where q is the wave vector corresponding to propagation in the $\eta$-space. Expressions (2.13)-(2.14) present the exact solution of the wave equation for the gradient dielectric layer (2.9), the wavelengths, the thickness of layer d and depth of modulation of its refractive index $U_m$ being arbitrary. It is worth noticing that the Rayleigh model (1.1) is a limiting case of the more general model (2.9), related to the value $s_2 = 0$.

Depending on the characteristic lengths $L_1$ and $L_2$ and signs $s_1$, $s_2$ the value $p^2$ can assume positive, negative or zero values, providing the different regimes of wave propagation through the barrier. Substitution of (2.13) into (2.5) brings the components of EM field inside the photonic barrier with curvilinear profile of U(z) (2.9).

(2) Alternatively to presentation (2.7), the system of Maxwell equations (2.2) can be reduced to one equation by introducing another generating function $\theta$, such as in [42]:

$$H_y = \frac{1}{c} \frac{\partial \theta}{\partial t} \; ; \; E_x = -\frac{1}{n_0^2 U^2} \frac{\partial \theta}{\partial z} \qquad (2.15)$$

Here the function U(z) describes again an unknown barrier, which will be determined below. By means of the representation (2.15) the second equation in the pair (2.2) becomes an identity, while the function $\theta$ is governed by the equation, following from the first equation in (2.2):

$$\frac{d^2\theta}{dz^2} + \left(\frac{\omega n_0}{c}\right)^2 U^2\theta = \frac{2}{U}\frac{dU}{dz}\frac{d\theta}{dz} \qquad (2.16)$$

To find the exactly solvable model U(z) for Eq. (2.16) let us introduce the new variable $\eta_1$, distinguished from $\eta$ (2.7); the dependence of the dielectric function on variable $\eta_1$ will be described by a new function $V(\eta_1)$. Moreover, we impose some special link between the functions U(z) and $V(\eta_1)$:

$$\eta_1 = \int_0^z U^2(z_1)dz_1 \; ; \; U^2(z)V^2(\eta_1) = 1 \qquad (2.17)$$

Herein, the dependences $\eta_1 = \eta_1(z)$, obtained from both equalities (2.17), have to coincide. Let us also note that, contrary to that considered above, the variable $\eta_1$ has nothing to do with the optical path. However, it defines a new phase space, in which Eq. (2.16) is reduced to a form that coincides with (2.6):

$$\frac{d^2\theta}{d\eta_1^2} + \left(\frac{\omega n_0}{c}\right)^2 V^2(\eta_1)\theta = 0 \qquad (2.18)$$

To continue this analogy, using the solution (2.13)-(2.14), one can proceed with any exactly solvable model U(z), replacing there $z \to \eta_1$. Let us use, e.g., the simplified model (2.9), where $s_1 = 0, s_2 = -1$

$$V(\eta_1) = \left(1 - \frac{\eta_1^2}{L_2^2}\right)^{-1} \; ; \; dV(\eta_1) = \frac{2\eta_1}{L_2^2}V^2(\eta_1)d\eta_1 \qquad (2.19)$$

Substitution of the value $d\eta_1 = U^2(z)dz$, following from the integrand (2.17), to (2.19) and replacement of product $V^2(\eta_1)U^2(z)$ by unity, according to (2.17), yields the equation:

$$\frac{dz}{L_2} = \frac{\sqrt{V(\eta_1)}dV(\eta_1)}{2\sqrt{V(\eta_1)-1}} = \frac{dU(z)}{2U^2(z)\sqrt{1-U(z)}} \tag{2.20}$$

where we have implicitly assumed that both U and V are real, and possess the same sign. Solution of Eq. (2.20) yields a new exactly solvable photonic barrier, expressed, unlike the previous models U = U(z), in a form of an inverse function z = z(U):

$$\frac{z}{L_2} = \frac{1}{2}\left[\frac{\sqrt{1-U(z)}}{U(z)} + \ln\left(\frac{1+\sqrt{1-U(z)}}{\sqrt{U(z)}}\right)\right] \tag{2.21}$$

Unlike (2.9), this barrier, shown in the Fig. 1b, is defined along all the half-axis $0 \leq z \triangleleft \infty$, so that $0 < U \leq 1$. Introducing, by analogy with (2.7) the variable

$$\eta_2(x) = \int_0^x V(\eta_1)d\eta_1 \tag{2.22}$$

where $V(\eta_1)$ is defined in (2.19), one obtains the solution of equation (2.16) for the heterogeneous barrier (2.21), expressed again in the form of a traveling wave

$$\theta = [U(z)]^{\frac{1}{2}} \exp[i(q\eta_2 - \omega t)] \tag{2.23}$$

Here the "wave vector" q is given by Eq. (2.14) with $p = L_2^{-1}$. The links between coordinates z and $\eta_{1,2}$ are given below in Eq. (2.26). Using (2.23), it can be directly shown that profile (2.21) provides an exactly solvable model for wave equation (2.16).

Thus, the transformation of Maxwell equations into a space of phase trajectories proves to be the useful tool for electromagnetics of heterogeneous dielectrics. It is remarkable that such an approach permits to reveal the exactly solvable models of photonic barriers, described by the inverse functions z = z(U). All these profiles U(z) yield generating functions for EM fields in heterogeneous media in the form of traveling waves in some phase space, which may (2.13). or may not (2.19) be obviously linked to the optical path, but in which phase accumulation is proportional to a space-like variable. Hence, the phase factors $\exp[i(q\eta - \omega t)]$ relate to harmonic wave, traveling along the phase coordinate $\eta$ with constant phase velocity $v_p = \omega/q$. This analogy with the traditional Fourier presentation of waves in homogeneous medium is shown below to simplify the analysis of dispersive properties of gradient films.

*2.2. Normal and anrmalous non-local dispersion of gradient photonic barriers.*

To visualize the dispersive properties of gradient media U(z) one has to inspect the wave numbers q and phase path lengths $\eta$ in the phase factors $\exp[i(q\eta - \omega t)]$ (2.13). This analysis reveals an important difference between concave and convex photonic barriers (Fig. 1a). In the case of a concave barrier (2.9) the value $p^2$ (2.11) is positive, and the wave number q in (2.14) can be expressed by means of some characteristic frequency $\Omega_1$

$$q = \frac{\omega}{c} n_0 N \; ; \; N = \sqrt{1 - \frac{\Omega_1^2}{\omega^2}} \; ; \; \Omega_1 = pc = \frac{2cy\sqrt{1+y^2}}{n_0 d} \qquad (2.24)$$

This spectral dependence $q(\omega)$ resembles a waveguide–like or plasma–like (normal) dispersion with a cutoff frequency $\Omega_1$. The phase path length (2.7) in this geometry

$$\eta(z) = \frac{L_2}{2\sqrt{1+y^2}} \ln\left(\frac{1 + \frac{zy_+}{L_2}}{1 - \frac{zy_-}{L_2}}\right) ; \quad y_\pm = \sqrt{1+y^2} \pm y ; \quad y_+ y_- = 1 \qquad (2.25)$$

is smaller than the distance z. Proceeding in a similar fashion with the model (2.19) and manipulating with condition (2.17), one obtains the explicit dependence $\eta_2(z)$, related in solution (2.22) to photonic barrier (2.21):

$$\frac{\eta_2}{L_2} = \frac{1}{2} \ln\left(\frac{1 + \frac{\eta_1}{L_2}}{1 - \frac{\eta_1}{L_2}}\right) ; \quad \frac{\eta_1}{L_2} = \sqrt{1 - U(z)} \qquad (2.26)$$

In contrast, the wave number for a wave traversing the convex photonic barrier (2.12) under the condition $p^2 < 0$ may be written by means of an another characteristic frequency $\Omega_2$

$$q = \frac{\omega n_0}{c} \sqrt{1 + \frac{\Omega_2^2}{\omega^2}} ; \quad \Omega_2 = \frac{2cy\sqrt{1-y^2}}{n_0 d} ; \quad y^2 < 1 \qquad (2.27)$$

This spectral dependence $q = q(\omega)$ corresponds to an anomalous dispersion. Coordinate $\eta$ for such barrier writes

$$\eta(z) = \frac{L_2}{\sqrt{1-y^2}} Arctg\left(\frac{\frac{z}{L_2}\sqrt{1-y^2}}{1-\frac{zy}{L_2}}\right) \qquad (2.28)$$

When the heterogeneity tends to zero ( $\Omega_1 \to 0, \Omega_2 \to 0$ ) the wave numbers q in (2.22) and (2.27) tend to the dispersionless limit $q = \omega n_0/c$.

Finally, in a special case $p^2 = 0$ that occurs when $L_2 = 2L_1$, the characteristic frequency $\Omega$ for decreasing function U(z) ( $s_1 = s_2 = +1$ ) is, according to (2.11), equal to 0, and the dispersive parameter N in (2.16) takes the value N = 1. The heterogeneous barrier (2.9), corresponding to this case

$$U(z) = \left(1 + \frac{z}{L_1}\right)^{-2} \qquad (2.29)$$

does not exhibit non-local dispersion (Fig. 1b).

Thus, in all these cases the generating functions are presented by waves, traveling in a dispersive phase space. We note that the characteristic frequencies $\Omega_1$ and $\Omega_2$ are determined by the heterogeneity-induced dispersion of photonic barrier (2.9). These frequencies are controlled by the heterogeneity scales $L_1$ and $L_2$ and barrier shape. The values $\Omega_1$ and $\Omega_2$ for the dielectric nanofilms belong to the visible and near infrared spectral ranges: e.g., for a film with thickness d = 80nm, refractive index $n_0$ = 1.8 and parameter $y^2$ = 0.25 one obtains for concave profile (2.24) $\Omega_1$ =2.355 $10^{15}$ rad $s^{-1}$ ( $\lambda = 800$ nm), while for the convex one (2.27) the value $\Omega_2$ is about 25% less: $\Omega_2$ = 1.805 $10^{15}$ rad $s^{-1}$ ( $\lambda = 1033$ nm).

To emphasize the formation of heterogeneity-induced dispersion (HID) in the dielectric layer, its material was assumed to be dispersionless, and the influence of surrounding media was neglected. However, when the layer at hand is assumed to be a part of resonant electromagnetic system, the heterogeneity of $\varepsilon(z)$ can change drastically the spectrum of this system. The examples of such spectral variations are discussed below.

*2.3. Comparison of local and non-local plasma-like dispersion.*

A simple example of manifestation of non-local dispersion in a resonant system can be given by spectra of eigenfrequencies of a cavity, filled by a heterogeneous dielectric. For simplicity, let us consider the electromagnetic cavity formed by two parallel, perfectly reflecting planes, spaced by distance d. In a model of cavity filled by homogeneous plasma with plasma frequency $\omega_p$, the lowest cutoff frequency $\Omega_c$ is known to be: $\Omega_c^2 = \omega_p^2 + (\pi/d)^2$ [45]. To illustrate the effect of a gradient layer, one can use again Fig. 1a, considering it as the cross section of cavity; the gradient dielectric layer between the planes z = 0 (v = 0) and z = d (v = 1) is characterized by concave profile U(z) (2.9). In this geometry the electric field $E_x$ inside the cavity may be written by means of substitution of generating function $\Psi$ (2.13) to (2.5):

$$E_x = \frac{i\omega A}{c\sqrt{U}}\sin(q\eta) \qquad (2.30)$$

To provide the values $E_x(z=0) = E_x(z=d) = 0$ the condition to be satisfied by the wave frequency $\omega$ is obtained from equation $q\eta_0 = m\pi$, where m = 1, 2, 3... and the phase path length $\eta_0$ for concave profile, related to the distance d, is found from (2.25):

$$\eta_0 = \frac{d}{2y\sqrt{1+y^2}}l_1; \quad l_1 = \ln\left(\frac{y_+}{y_-}\right) \qquad (2.31)$$

Using of the definition of wave number q (2.24) yields the spectrum of eigenfrequencies $\omega_c$ for the concave profile U(z)

$$\omega_c^2 = \Omega_1^2 \left[ \left( \frac{m\pi}{l_1} \right)^2 + 1 \right] \quad (2.32)$$

Here $\Omega_1$ is the characteristic frequency (2.24). Analogously, the cutoff frequency for the same cavity with convex profile U(z) (2.9) is expressed via the characteristic frequency $\Omega_2$ (2.27). By using the value $\eta_0 = \eta(d)$ obtained from (2.26) for this geometry

$$\eta_0 = \frac{d}{2y\sqrt{1-y^2}} l_2 \; ; \; l_2 = arctg\left( \frac{2y\sqrt{1-y^2}}{1-2y^2} \right) \quad (2.33)$$

one obtains the spectrum of eigenfrequencies for the convex profile U(z), that differs from that given by (2.32):

$$\omega_c^2 = \Omega_2^2 \left[ \left( \frac{\pi m}{l_2} \right)^2 - 1 \right] \quad (2.34)$$

When the heterogeneity of dielectric layer is vanishing $(y \to 0)$, both values $\omega_c$ reduce to the common limit, corresponding to the spectrum of the cavity filled by homogeneous dielectric $\omega_0 = \pi mc/n_0 d$. As compared with this "homogeneous" limit, the obtained eigenfrequencies $\omega_c$ are up-shifted (2.32) or down-shifted (2.34); thus, for the fundamental mode (m=1) and $y^2 = 1/3$ these formulae yield $\omega_c = 1.3\omega_0$ and $0.7\omega_0$ for concave and convex profiles U(z) respectively. Moreover, the difference between eigenfrequencies is increased (decreased) for concave (convex) profiles U(z).

Thus, the plane EM fields in a series of gradient media can be presented as amplitude-modulated waves, traveling along straight lines on some phase planes. These lines, as well as the constant phase velocity of transformed waves, are determined by the spatial profile $\varepsilon(z)$. Before using this analogy with Fourier optics for analysis of reflection–transmission problems for gradient layers, it is worth emphasizing the generality of this approach, mentioning some features, which are not discussed above:

(1) For simplicity we have considered a lossless medium. However, the absorption can also be included into this approach; if the absorption is characterized by some coordinate-independent parameter $\chi$ [46], one can replace $n_0$ by $n_0$ (1+i $\chi$). The spatially distributed absorption $\chi(z)$ can be modeled, e.g., by replacements in (2.9) the real characteristic lengths $L_1$ and $L_2$ by complex quantities $L_1 + iL_3$ and $L_2 + iL_4$. The effects of absorption in gradient layers will be shown in Section 4.

(2) The generating functions for EM fields in gradient barriers, obtained above, remain valid for arbitrary wavelengths and parameters of barriers (d, $n_0$ and $U_m$). This scalability property is useful for the parallel research in the different spectral ranges of gradient Electromagnetics, especially in optics and microwave physics.

(3) It is worth comparing the exact solutions of wave equations (2.6) and (2.16) with the approximate WKB solutions, based on a supposition of smallness of variations of profile U(z) with respect to the wavelengths. In this limit the characteristic frequencies $\Omega_{1,2}$, characterizing the non-local dispersion, are vanishing, factor N (2.14) tends to unity, and the exact solutions (2.13) and (2.22) are reduced to WKB solutions. It is remarkable that, by imposing the condition $\Omega=0$ without any WKB-like assumptions, we found the unique profile (2.29), for which the WKB solution coincides with the exact solution of the wave equation. Thus, the effects of non-local dispersion are lost in the framework of WKB approach (besides this profile). However, the gradient layers, discussed here, are characterized usually by subwavelength thicknesses, considerable variations of dielectric properties and strong non-local dispersion. WKB approach becomes invalid for these problems, and the above mentioned exact solutions and their generalizations are used below as the working tool for solving problems of gradient Electromagnetics.

## 3. Reflectance and transmittance of gradient photonic barriers.

This Section is devoted to the optical coating design problem, dealing with the search some spatial distribution of refractive index, which provides a fixed spectral reflectance in certain frequency bands. A practical approximate solution to this problem is given by so-called "digital algorithm" [47], based on merely splitting the arbitrarily shaped coating into pairs of thin homogeneous adjacent layers; each pair contains high-index ($n_h$) and low-index ($n_l$) layer with thicknesses $d_h$ and $d_l$, herein the optical thickness of each layer is assumed to be small: $n_h l_h \ll \lambda$, $n_l d_l \ll \lambda$. Multiplication of characteristic matrices [48] of two layers yields the parameters of one layer, optically equivalent to this pair; the thickness of such equivalent layer is $d_h + d_l$, its refractive index N has a value intermediate between $n_h$ and $n_l$. At the next step one considers the pairs of such equivalent layers and so on. In the limit of thin layers the equivalent index N is wavelength-independent, i.e. the resulting configuration is dispersionless.

An alternate method consists in approaching the continuous variation of the refractive index by that of a stack of different thin heterogeneous layers, characterized by variable profiles of dispersionless refraction index $n_m(z) = n_{0m}+(n_{dm} - n_{bm})(z/d_m)^g$ [49].

An opposite approach, based on the Green function technique, was developed in [50] for finding distributions of $\varepsilon(z)$ in the subsurface area of an heterogeneous dielectric half-space, providing its broadband antireflection properties. The numerical solutions in [50] illustrate the monotonic and oscillating profiles $\varepsilon(z)$ for "invisible" bulk media, described by Drude and Lorentz models, respectively. These models are known to include the local dispersive effects, connected, e.g., with resonant frequencies of polar dielectrics or plasma frequency in ionic crystals [51]. Introduction of periodicity into the refractive index depth profile of the coating provides the continuous index matching for broadband antireflection coatings [28].

Side by side with these artificial interfaces, the successes in nanotechnology has favored other applications of subwavelength dielectric films, such as, e.g., the gradient-index filters [22], photonic heterostructures (contiguous stacks) [52] and metal– dielectric structures [53]. The rugate optical filter, characterized by continuous sinusoidal index variations along the direction of substrate normal [54], exhibit a range of wavelengths in which the transmittance is nearly zero. Moreover, the intentional deviations from this periodicity, e.g., the inclusion of discontinuities into the index profile, such as adding of constant index layer or phase shifting of sinusoidal index variation, were shown to create a narrow band of wavelengths within a stop band, in which the transparency was significantly increased [27]. The total reflectivity for all the angles of incidence in the given spectral range (the omnidirectional reflection, ODR [55]) is exhibited by photonic heterostructures, consisting of alternating dielectric layers with high and low refractive indices [56] and different periods. Here the index contrast in the heterostructures with two periods can be much lower than those required in single periodic structures [57]. This property is especially important for the heterostructures intended for the visible range, since its difficult to find materials with a high refractive index contrast in this range [58]. The possibilities of creation of almost rectangular spectral windows are broadened due to enhanced coupling of EM waves and surface plasmons in structures, containing sandwiched metal and dielectric films [59].

On the contrary, the effects of heterogeneity-induced non-local dispersion are shown in this Chapter to provide a controlled reflectance spectra of subwavelength gradient layers. Using the models of concave and convex photonic barriers (2.9) and by expressing the fields inside these barriers as a sum of forward and backward waves (2.13)

$$\Psi = \frac{1}{\sqrt{U(z)}}\left[\exp(iq\eta) + Q\exp(-iq\eta)\right]\exp(-i\omega t) \qquad (3.1)$$

one obtains by means of continuity conditions on the interfaces of such barriers the reflection coefficients (Chapter 3.1). These generalized Fresnel formulae are valid for arbitrary thicknesses of barriers and depths of modulation of refractive indices. The reflectance of periodical sets of such gradient layers is found in chapter 3.2. Potential of highly dispersive reflectance of such sets for frequency-selective interfaces and antireflection coatings is considered in Chapter 3.3.

*3.1. Generalized Fresnel formulae for concave and convex photonic barriers.*

This chapter is devoted to reflectance and transmittance of photonic barriers with thickness d and symmetric profile of dielectric susceptibility U(z) (2.9): U(0)= U(d) = 1. To stress out the features of such barriers, stipulated by heterogeneity U(z), we first consider the model of a barrier without substrate. Let us assume that the wave is incidenting on the barrier interface z =0. To find the complex reflection coefficient R, one has to use the continuity conditions at interfaces z = 0 and z = d. Using the values of derivatives of profile (2.9)

$$\frac{1}{U^2}\frac{dU}{dz}\bigg|_{z=0} = -\frac{s_1}{L_1} \; ; \; \frac{1}{U^2}\frac{dU}{dz}\bigg|_{z=d} = \frac{s_1}{L_1} \qquad (3.2)$$

we can present the expressions for electric and magnetic components of the field (3.1) on the barrier interfaces:

$$E_x\big|_{z=0} = \frac{i\omega A}{c}(1+Q) \; ; \; E_x\big|_{z=d} = \frac{i\omega A}{c}\left[\exp(iq\eta_0) + Q\exp(-iq\eta_0)\right] \qquad (3.3)$$

$$H_y\big|_{z=0} = i\frac{\omega A}{c}\left\{-\frac{is_1\gamma}{2}(1+Q) + n_e(1-Q)\right\} \; ; \; n_e = n_0 N \; ; \; \gamma = \frac{c}{\omega L_1} \qquad (3.4)$$

$$H_y\big|_{z=d} = i\frac{\omega A}{c}\left\{\frac{is_1\gamma}{2}\left[\exp(iq\eta_0) + Q\exp(-iq\eta_0)\right] + n_e\left[\exp(iq\eta_0) - Q\exp(-iq\eta_0)\right]\right\} \qquad (3.5)$$

Here A is the amplitude of the refracted field, factors N and $\eta_0$ are determined for concave and convex profiles by formulae (2.22), (2.29) and (2.25), (2.31), respectively. For simplicity the factor $\exp(-i\omega t)$ is omitted hereafter. Parameter Q in expressions (3.3) – (3.5), describing the contribution of backward wave to the field inside the barrier, can be found for the continuity conditions on the interface z = d:

$$Q = -\exp(2iq\eta_0)\left[\frac{1 - \frac{is_1\gamma}{2} - n_e}{1 - \frac{is_1\gamma}{2} + n_e}\right] = Q_0 \qquad (3.6)$$

Substitution of $Q_0$ into (3.3) – (3.4) yields the expressions for the components of refracted field on the interface z = 0. Finally, the continuity conditions on this interface yield the reflection coefficient R [44], which reads as

$$R = \frac{\left(1 + \frac{\gamma^2}{4} - n_e^2\right)tg\delta - s_1\gamma n_e}{\left(1 - \frac{\gamma^2}{4} + n_e^2\right)tg\delta + s_1\gamma n_e + i(2n_e - s_1\gamma tg\delta)} \;;\; \delta = \frac{Nl}{u} \qquad (3.7)$$

Result (3.7) presents the generalization of Fresnel formula for reflection coefficient of a single layer for both concave ($s_1 = 1$) and convex ($s_1 = -1$) profiles U(z); parameter $\gamma$, defined in (3.4), can be written for concave ($\gamma_+$) and convex ($\gamma_-$) profiles as

$$(\gamma_{\pm}) = \frac{2un_0 y}{\sqrt{1 \pm y^2}} \tag{3.8}$$

To calculate the reflectivity of barriers with concave or convex profiles, one has to use the values N and l, defined by (2.22), (2.29) and (2.25), (2.31), respectively. In the limiting case when the heterogeneity is vanishing ($\gamma \to 0, y \to 0, N \to 1, l \to d$, $n_e \to n_0$), expression (3.7) is reduced to the well known Fresnel formula for homogeneous layers:

$$R = \frac{(1 - n_0^2) tg\delta_1}{(1 + n_0^2) tg\delta_1 + 2in_0} \; ; \; \delta_1 = \frac{\omega n_0 d}{c} \tag{3.9}$$

The non-local dispersion of gradient layers results in several peculiar features in reflectance–transmittance spectra of such layers - see Fig.2:

(1). A stack consisting of six convex gradient films, contains six points R = 0 in the visible and near IR ranges and six transmittance bands, surrounding these points. The total thickness of this stack, consider as a filter for IR waves, remains subwavelength.

(2). Unlike the homogeneous layers, characterized by equidistant frequencies $\omega_n$, providing the reflectionless propagation R($\omega_n$) = 0 (3.9), the widths of transmittance bands for gradient layers, particularly in the visible range, are unequal.

(3). Using gradient films allows to design a periodical system of contiguous films, characterized by equal values of refractive index $n_0$ and thickness d, with both equal or unequal profiles U(z).

*3.2. Optics of periodic gradient structures.*

Periodical system of gradient layers possess a considerable flexibility of reflection–transmission properties. The traditional multilayer structure consists of alternating homogeneous dielectric layers of two materials with high and low refractive indices $n_h$ and $n_l$ and layer thicknesses $d_h$ and $d_l$ respectively [54, 61]. Unlike the latter, the gradient layers forming a periodical structure can be diversified, side by side with the aforesaid values n and d, by profiles U(z). Even such a simple profile can broaden the variety of periodical structures. Owing to the flexibility of U(z), such structures can include alternating homogeneous and gradient layers as well as different combinations of adjacent gradient layers (Fig. 3). Since the layers with U(z), given by (2.9), are characterized by two free parameters, namely by the thickness d and the depth of spatial modulation $U_m$, we will consider below three combinations of such profiles:

(1) The systems, shown in Figs. 3a and 3b, containing only concave or only convex layers, fabricated from the same host material with equal values of $n_0$ and characterized by profiles U(z) with equal values of both free parameters, d and $U_m$. The reflection on the boundaries of adjacent layers is connected with the discontinuity of gradients: e.g., for the adjacent concave layers the values of grad U, expressed in normalized coordinates x = z/d, possess a jump on the boundary x = 1 from $dU/dx|_{1-0} = 4y^2$ to $dU/dx|_{1+0} = -4y^2$; herein the curvatures of both concave profiles on this boundary remain continuous: $K_1 = 8y^2(4y^2 + 1)(1+16y^4)^{-3/2}$. This geometry provides only one way to increase the total thickness of dispersive layer without changing its essential physical parameter, i.e., its characteristic frequency. By using the continuity conditions on each boundary between two adjacent layers, one can find the field in each layer; attributing the number m = 1 to the layer at the far side of the stack, one obtains a simple recursive relation for parameter $Q_m$ for m-th layer (m>=1):

$$Q_m = \exp[2i(m-1)q\eta_0]Q_0 \qquad (3.10)$$

where the value $Q_0$ is defined in (3.6). Proceeding in the same way as in the derivation of formula (3.7) for the reflection coefficient, we have found that (3.7) can be simply generalized for m layers after the replacement

$$\delta \to \delta_m = \frac{mNl}{u} \qquad (3.11)$$

Reflectance of a pairs (m = 2) of concave and convex photonic barriers, dependent on the internal discontinuity of grad U, is shown in Fig. 4.

A further increase of the amount of layers m results in significant distortions of the reflectance spectra. Thus, comparing the graphs of $|R|^2$ for convex barriers, containing m = 1, m = 2 and m = 6 layers (Fig.2), one can see the shifting of the points of $|R|^2_{min}$ and $|R|^2_{max}$ as well as the increase of the dispersion of $|R|^2$.

(2) The periodical set can be formed by alternating layers of concave and convex profiles U(z), their thicknesses d and refractive indices $n_0$ being equal - see Fig. 3c. To provide the smooth tangent of these profiles at the boundaries U = 1 (3.2), their heterogeneity scales $L_1$ must be equal, while the deviations of tops ($U_{max}$) and bottoms ($U_{min}$) of profile U(z) (2.10) from the boundary value U = 1 are unequal: $U_{max} - 1 \neq 1 - U_{min}$. Both profiles are characterized by the same value of parameter y, and the maximum and minimum values of U are linked by correlation

$$\frac{U_{max}}{U_{min}} = \sqrt{\frac{1+y^2}{1-y^2}} \qquad (3.12)$$

Unlike in the previous case, the gradients of both concave and convex profiles at the boundary x = 1 are equal (grad $U_1$ = grad $U_2$ = $4y^2$), and the internal reflectance arises due to discontinuity of curvatures of concave ($K_1$) and convex ($K_2$) profiles: $K_{1,2} = 8y^2(4y^2 \pm 1)(1 + 16y^4)^{-3/2}$.

(3) On the other hand, the periodical structure can be formed from another pair, characterized by equal deviations $U_{max} - 1 = 1 - U_{min}$ (see Fig. 3d). Herein the thicknesses of concave and convex parts, $d_1$ and $d_2$, described by different values of parameter y ($y_1$ and $y_2$), are linked by correlation

$$\frac{d_2}{d_1} = \left(\frac{y_2}{y_1}\right)^2 \tag{3.13}$$

The difference in reflectance for structures (3.12) and (3.13) is shown in Fig. 4. It is remarkable that this difference for gradient multilayers, having the same values of refractive index $n_0$ on both sides, arises from the difference in internal distributions of n(z).

These results were obtained for the model of gradient layers without substrate. To examine the influence of substrate on the reflectivity of such layers, one can generalize formula (3.6) by considering the substrate formed by homogeneous lossless layer of thickness D and refractive index $n_1$. Using of continuity conditions results in the replacement of unity in the numerator and denominator of (3.6) by expression :

$$1 \rightarrow \frac{n_1(1 - in_1 t_1)}{n_1 - it_1} \; ; \; t_1 = tg\left(\frac{\omega n_1 D}{c}\right) \tag{3.14}$$

In the case $n_1 = 1$ (layer-air interface) this generalized result is reduced back to (3.6). In the general case, all the basic features discussed above remain, with some quantitative changes in the position and width of resonances, that are treated in detail in ref [68]. Another interesting case is that of $t_1=0$, which means that the thickness of substrate layer contains an integer amount of halfwaves

$$n_1D = \frac{m\lambda}{2} \ ; m = 1, 2, 3... \tag{3.15}$$

In such a case, we obtain the formulae (3.6) and (3.7) again. Thus, the substrate obeying the condition (3.15) does not affect the reflectance of gradient layers.

*3.3. Frequenc- selective dielectric interfaces.*

The electromagnetic field inside the gradient layer is formed due to the interference of forward and backward waves. The electric component of transmitted field on the far side of barrier $E_t$ can be linked to the incident field $E_i$ by means of complex transmission coefficient T in a form $E_t = E_iT$. To find T one can substitute the value Q (3.6) to Eq. (3.3); presenting T in a form $T = |T|\exp(i\phi_t)$, we can write the amplitude and phase of T as

$$|T| = \sqrt{1-|R|^2} \ ; \ \phi_t = Arctg\left[\frac{\left(1-\frac{\gamma^2}{4}+n_e^2\right)tg\delta_m + s_1\gamma n_e}{2n_e - s_1\gamma tg\delta_m}\right] \tag{3.16}$$

Here the reflection coefficient R, defined in (3.7), can be written in a form $R = |R|\exp(i\phi_r)$, where the phase $\phi_r$ is:

$$\phi_r = Arctg\left[\frac{s_1\gamma tg\delta_m - 2n_e}{\left(1-\frac{\gamma^2}{4}+n_e^2\right)tg\delta_m + s_1\gamma n_e}\right] \qquad (3.17)$$

Comparison of (3.16) and (3.17) yields the correlation between the phases of reflected and transmitted waves:

$$\phi_t - \phi_r = \frac{\pi}{2} \qquad (3.18)$$

The velocity of energy transfer through the heterogeneous layer $v_g$ has to be found by means of energy flux (Pointing vector $\vec{P}$) and energy density W [59]. For normal incidence, there is only one component of Pointing vector, given by $P_z$:

$$v_g = \frac{P_z}{W} \; ; \; P_z = \frac{c}{4\pi}\text{Re}\left[\vec{E}\times\vec{H}^*\right] \; ; \; W = \frac{1}{8\pi}\left(\varepsilon|\vec{E}|^2 + |\vec{H}|^2\right) \qquad (3.19)$$

Substitution of (3.1) into (3.19) yields the expressions for $P_z$ and W:

$$P_z = \frac{cn_e^2|M|^2|E_i|^2}{\pi|\Delta|^2} \; ; \; W = \frac{n_e^2|M|^2|E_i|^2 \vartheta U}{4\pi|\Delta|^2} \qquad (3.20)$$

$$|M|^2 = \frac{(1+R)(1+R^*)}{(1+Q)(1+Q^*)} \; ; \; \Delta = 1 + n_e - \frac{is_1\gamma}{2} \qquad (3.21)$$

Here the reflection coefficient R and parameter Q are defined in Eqs. (3.7) and (3.6), where the dimensionless function $\vartheta$ is some coordinate-dependent function [44].

We note, that the energy flow $P_z$ (3.20) remains constant in each cross-section inside the barrier since the expression for $P_z$ is coordinate-independent while, in contrast, the energy density W proves to be coordinate-dependent: $W = W(z)$. Substitution of (3.20)–(3.21) into (3.19) yields the explicit formula for coordinate-dependent group velocity $v_g$. To visualize the effects of heterogeneity, it is convenient to present the values $v_g$, normalized by means of its values in homogeneous films ($v_{g0}$) with the same values of d and $n_0$:

$$v_g = \frac{4c}{U\vartheta} \; ; \; v_{g0} = \frac{2c}{1+n_0^2} \; ; \; V = \frac{v_g}{v_{g0}} \qquad (3.22)$$

These group velocities, which are presented in Fig. 5, illustrate the non-local dispersive properties of gradient photonic barriers: the velocity of energy transfer through such barriers can be increased (decreased) essentially due to concave (convex) profile U(z) inside the barrier. In this respect it is essential to note that, contrary to the refractive index profile, the group velocity profile is not symmetric (as it would if it was simply determined by the local value of the dielectric constant). This is an excellent illustration of the specificity of the heterogeneity-induced dispersion effect.

Thus, the reflectance-transmittance phenomena in gradient optics prove to be highly dispersive. This property can be used for design of frequency-selective interfaces and, in particular, the antireflection coatings in the given spectral ranges [60] – [61].

Besides these effective fequency-selective interfaces one can stress out some extra possibilities arising with the use of gradient layers:

(1) Flexible phase shifting, providing phase changes of the wave transmitted through the layer, satisfying the condition of complete transmission (reflectionless propagation) |T| = 1. The general expression for the phase of transmitted wave $\phi_t$ (3.16) that satisfies the condition R = 0 can be written in the form

$$tg\phi_t = \frac{\gamma s_1}{1 - \frac{\gamma^2}{4} - n_e^2} \qquad (3.23)$$

Unlike in the case of the nonreflecting homogeneous layer ($\gamma \to 0$, $tg\phi_t = 0$), the phase shift implied by gradient layer (3.23) can possess values $0 < \phi_t < \pi$.

(2) Rapid accumulation of dispersive effects in the subwavelength stacks of several similar films.

(3) Instead of using only the contrast between natural high and low refractive indices of adjacent homogeneous layers, gradient optics is dealing also with the technologically controlled contrast of gradients or curvatures of these indices - see Fig. 4; herein the host material and the thicknesses of all layers may remain unchanged. To optimize the amount of gradient layers m, providing a given reflectance of the multilayer in some fixed spectral range, one can merely use the generalized Fresnel formula (3.7), in which the phase $\delta_m$ is proportional to m (3.11).

(4) The dependence of transmittance of a gradient layer, the thickness and refractive index of the host material being given, may become useful for non-destructive optical control of the internal structure of the layer.

(5) Scalability of the obtained results, illustrated by optical effects in nanofilms, to other spectral ranges, e.g., to subwavelength gradient layers in microwave circuits (see Chapters 7b – 7d ).

**4. Non-attenuated tunneling of light through heterogeneous dielectric films.**

Tunneling of EM waves in opaque media with $\varepsilon < 0$ attracts nowadays a growing attention in optoelectronics and thin films physics [62]-[63], due to the peculiarities of energy transmission through such media. The research in this field focused mainly on the effects of frustrated total internal reflection (FTIR) and penetration of evanescent waves through photonic barriers of finite thickness. Spatial and polarization structures of evanescent waves in a homogeneous barrier with $\varepsilon < 0$ depend on the direction of their propagation with respect to the barrier's interface:

(1) The simplest example of tunneling through a photonic barrier can be illustrated by the normal incidence of wave with frequency $\omega$ on an homogeneous layer with free carriers, when the free carriers plasma frequency $\Omega_p$ is high enough: $\Omega_p > \omega$. The difference in wave polarization structure for this 1D problem is vanishing. In the case of a semiconductor with free carriers, $\varepsilon_L$ denoting the part of the dielectric susceptibility determined by the semiconductor crystal lattice, and d the layer thickness, one can write the power reflection $|R|^2$ and transmission $|T|^2$ coefficients in the tunneling regime in the form:

$$|R|^2 = 1 - |T|^2 \; ; \; |T|^2 = \frac{4n_e^2}{4n_e^2 + sh^2\vartheta(1+n_e^2)^2} \; ; \; n_e = \sqrt{\varepsilon_L}\, N_- \; ; \; N_-^2 = u^2 - 1$$

(4.1)

$$u = \frac{\Omega_p}{\omega} \; ; \; \vartheta = \frac{\omega n_e d}{c}$$

Thus, the transmittance decreases exponentially when the barrier thickness d is increasing: $|T|^2 \propto \exp(-2\vartheta)$; hence $|R|^2 \to 1$, and a non-attenuated (reflectionless) tunneling regime can not arise. The same occurs in the case of metallic layers.

(2) The tunneling effects for inclined incidence are considered frequently by means of a double-prism configuration, containing two right angle isosceles prisms of refractive index $n_p$, placed with their hypotenuses in close proximity, forming a vacuum gap of width d between them - see Fig. 6 [64]-[65]. Propagation into this gap can be viewed as a 2D problem of tunneling through a homogeneous photonic barrier for both S- and P- polarized waves, incidenting under the angle $\theta > \theta_c = Arc\sin(1/n_p)$. Total transmittance for these tunneling waves is also impossible.

(3) A special case of tunneling problem is associated with surface waves, traveling along the boundary of the medium and which are attenuated exponentially away from this boundary. Herein the media with negative dielectric susceptibility are known to support only TH surface wave [66].

In contrast, gradient media with $\mathrm{grad}\,\varepsilon < 0$, the dielectric susceptibility of host material being positive, possess the opposite properties, namely:

(1) The reflectionless tunneling is possible for both lossless and lossy concave photonic barriers.

(2) Heterogeneity–induced anisotropy, associated with the case of inclined incidence on gradient barrier, results in formation of evanescent S-wave and propagating P- wave in some spectral ranges; herein the reflectance spectra for S–wave can possess narrow peaks of reflectionless tunneling for large angles of incidence.

(3) Propagation of TE–polarized waves on the interface of gradient medium without free carriers is possible.

These properties as well as their potential for the design of subwavelength polarizers, frequency filters, antireflection coatings and frequency-selective interfaces are considered below.

*4.1. Energy transfer through gradient layers by evanescent modes.*

To examine the transmission of EM power in the tunneling regime (u > 1), one has to present the generating function for the evanescent mode inside the barrier, by analogy with (3.1), as a sum of forward and backward waves:

$$\Psi = \frac{1}{\sqrt{U(z)}}[\exp(-p\eta) + Q\exp(p\eta)]\exp(-i\omega t) \; ; \; p = \frac{\omega n_0 N_-}{c} \; ; \; N_- = \sqrt{u^2 - 1} \qquad (4.2)$$

The phase path length $\eta$ is determined in (2.29). Substitution of (4.2) into (2.5) yields the fields of evanescent modes. Following the scheme used in Section 3.1 for the analysis of the reflectance of a concave barrier with u ≤ 1, one obtains the reflection coefficient in the tunneling regime (u ≥ 1) for a stack containing m similar adjacent photonic barriers supported by a thick lossless substrate with refractive index $n_1$ in the form:

$$R = \frac{t\left(n_1 + \frac{\gamma^2}{4} + n_e^2\right) - \gamma n_e + i(n_1 - 1)\left(n_e - \frac{\gamma t}{2}\right)}{t\left(n_1 - \frac{\gamma^2}{4} - n_e^2\right) + \gamma n_e + i(n_1 + 1)\left(n_e - \frac{\gamma t}{2}\right)} \; ; \; t = th(mp\eta_0)$$

(4. 3)

$$n_e = n_0 N_- \; ; \; N_- = \sqrt{u^2 - 1}$$

The phase path length for concave barrier $\eta_0$ is determined in (2.29). In the special case $n_1 = 1$ (film-air interface) the formula (4.3) is reduced to a simple form, which can be independently obtained from (3.7), by replacements $N \to iN_-$, $q \to ip$, $tg\delta \to ith\delta$:

$$R = \frac{th(mp\eta_0)\left[1 + \frac{\gamma^2}{4} + n_e^2\right] - \gamma n_e}{th(mp\eta_0)\left[1 - \frac{\gamma^2}{4} - n_e^2\right] + \gamma n_e + i[2n_e - \gamma th(mp\eta_0)]} \quad (4.4)$$

The values of reflection coefficient R for the "critical" value u=1, defined as the limiting case for subcritical ($u \leq 1$) (3.7) and supercritical ($u \geq 1$) (4.4) regimes, coincide, providing a finite value for R in this case:

$$R\big|_{u=1} = \frac{ml_1\left(1 + \frac{\gamma^2}{4}\right) - \gamma n_0}{ml_1\left(1 - \frac{\gamma^2}{4}\right) + \gamma n_0 + i(2n_0 - ml_1\gamma)} \; ; \; l_1 = \ln\left(\frac{y_+}{y_-}\right) \quad (4.5)$$

Thus, generalized Fresnel formulae (3.7) and (4.4) describe the continuous variations of reflectivity of concave photonic barrier in all the spectral range $u \geq 0$. Photonic barriers in the tunneling regime are characterized usually by a high reflectivity: thus in a case $n_e = \gamma t/2$, eq. (4.3) may be written in the form:

$$R = \frac{n_1 t^2 - (1-t^2)n_e^2}{n_1 t^2 + (1-t^2)n_e^2} \quad (4.6)$$

Using a stack of gradient nanolayers can enhance the reflectance : a stack, containing 8 layers, characterized by concave profile (2.9) with parameters d=82 nm, $n_0$=1.8, $U_{min}$=0.75, deposited on a substrate with $n_1$=2, has a reflection coefficient, determined by Eq. (4.6), $|R(\lambda = 800 nm)|^2$=0.999; herein the total thickness without substrate of this effective mirror is smaller than the wavelength.

It is essential that the tunneling through rectangular barrier is always accompanied by some reflection (coefficient R (4.1) is never equal to zero), while the value R for concave barrier (4.) may reach the zero value. Such reflectionless tunneling constitutes an important peculiarity of gradient layers. Let us consider the model of contiguous gradient layers, supported by the substrate, whose thickness has been chosen to eliminate its influence according to (3.14). By setting the numerator of (4.4) to zero, one can write the condition R=0 in a form [67]

$$\text{th}(ml_1 x) = \frac{2xy}{\sqrt{1+y^2}} \left[ \frac{1}{n_0^2} + \frac{y^2}{1+y^2} + x^2 \left(1 - \frac{1}{n_0^2}\right) \right] \; ; \; x = \sqrt{1 - \frac{1}{u^2}} \qquad (4.7)$$

By solving Eq. (4.7) with respect to x one obtains the variable u and the cutoff frequency $\Omega_1 = \omega u$. Finally, making use of (2.22), one can calculate the layer thickness d. Thus, choosing the material with $n_0$=2.35 and the depth of modulation $U_{min}$=0.75 ($y^2$=1/3), we find that in a set of two adjacent barriers the condition $|T|$=1 is fulfilled when u=1.1. It means that, e.g., for the waves $\lambda$=800 nm (1055 nm), the reflectionless tunneling arises when d=65 (85) nm. It is remarkable that non-attenuative tunneling is obtained for these waves, even when the photonic barriers discussed are spaced by distance D, containing an integer amount of half waves (3.14). The transmittance spectrum for such waves is shown in Fig. 7.

Let us consider now the influence of absorption of gradient layers on the tunneling effects. Unlike the case (3.24), related with lossy substrates, we examine here the losses associated with the material of layers. To simplify the problem, let us assume that the imaginary part of complex dielectric susceptibility $\varepsilon = \varepsilon_1 + i\varepsilon_2$ ($\varepsilon_1 > 0$) is distributed inside the layer by the same function $U^2(z)$ as the real part [46]: this would be for instance the case if the gradient medium was created by mixing two components with varying concentrations.

$$\varepsilon = n_0^2(1+i\chi)U^2(z) \ ; \ n_0^2 = \varepsilon_1 \ ; \ \chi = \frac{\varepsilon_2}{\varepsilon_1} \tag{4.8}$$

Hereafter we neglect the influence of the substrate. By rewriting the expression for $N_-$ (4.3) for profile (2.9) in a generalized form

$$N_- = \sqrt{u^2 - 1 - i\chi} = a - ib \ ; \ a,b = \sqrt{\frac{1}{2}\left[\sqrt{(u^2-1)^2 + \chi^2} \pm (u^2-1)\right]} \tag{4.9}$$

substituting $N_-$ into the generating function $\Psi$ (4.2) and following to the standard procedure, one obtains the reflection coefficient for the wave tunneling through the stack of m lossy gradient layers:

$$R = \frac{S_t(1+G_\chi) - \gamma n_0(a-ib)}{\Delta_\chi} \ ; \ G_\chi = \frac{\gamma^2}{4} + n_0^2(a-ib)^2$$

$$\tag{4.10}$$

$$\Delta_\chi = S_t(1-G_\chi) + \gamma n_0(a-ib) + i[2n_0(a-ib) - \gamma S_t]$$

$$S_t = \frac{t_3 - it_4}{1 - it_3 t_4} \ ; \ t_3 = \text{th}\left(\frac{mal_1}{u}\right) \ ; \ t_4 = \text{tg}\left(\frac{mbl_1}{u}\right) \tag{4.11}$$

The transmission function for tunneling through the layers read as:

$$T = \frac{2in_0(a-ib)}{\Delta_\chi ch\left[\frac{ml_1(a-ib)}{u}\right]} \quad (4.12)$$

In a lossless limit $\chi \to 0$, $b \to 0, t_4 \to 0, a \to N_-$, and formulae (4.10)–(4.11) reduce to formula (4.4), describing the reflectance of lossless layers in the tunneling regime. Comparing these formulae, one can see that absorption can drastically weaken the reflectance : a stack, containing 5 lossless films ($n_0$=1.9, $\chi$=0, $U_{min}$=0.75), is characterized by $|R|^2$=0.95 for the normalized frequency u=1.25, while for the same frequency $|R|^2$=0.15, when $\chi$=0.5, and other parameters of the stack remain unchanged. Heterogeneity-induced dispersion can also provide the reflectionless tunneling through lossy films. To examine the feasibility of condition R=0 in this case, one can employ the conditions of nullification of real and imaginary parts of numerator in (4.10). Combing these conditions yields the equation which determines the parameters of lossy layers discussed:

$$\chi = \frac{u\left[bt_3(1+t_4^2) - at_4(1-t_3^2)\right]}{t_3^2 + t_4^2} \quad (4.13)$$

Considering the aforesaid set, containing 5 films with $n_0$=1.9, $U_{min}$=0.75, we find that the reflectionless regime arises for normalized frequency u=1.08, for the loss parameter $\chi = 0.35$. The calculation of reflectance for the same system of layers, but without absorption, yields the strong reflection: $|R|^2$=0.95; therefore, the absorption can result in a significant decrease of reflectance in the tunneling regime. It is remarkable that a similar analysis for homogeneous lossy layer results, instead of (4.13), to the condition of reflectionless propagation:

$\left(1-n_0^2\right)^2 + \left(\chi n_0^2\right)^2 = 0$, which cannot be satisfied by any non-zero values of $\chi$. Thus, in contrast to gradient layers, the reflectionless regime in homogeneous lossy layers is impossible.

For simplicity this consideration is restricted to normal incidence. Before going to the general case of an inclined incidence, it is worth summarizing the salient features of the discussed phenomena:

(1) The regime of reflectionless tunneling, due to interference of forward- and backward-propagating evanescent modes in gradient layer, can arise in layers manufactured from both conservative and lossy materials.

(2) The velocity of energy transfer, $v_g$, by waves tunneling through the photonic barrier can be calculated by means of general formulae (3.19)-(3.22). An example of this velocity, shown in Fig. 5c, illustrates its non-monotonic coordinate dependence; herein the velocity $v_g$ can exceed by 30–40% the constant velocity $v_{g0}$ of energy transfer through the homogeneous layer with the same values of $n_0$ and d. Thus, the gradient photonic barrier can provide a non-attenuated accelerated energy flow; however, the velocity of such energy transfer by evanescent modes always remains subluminal, $v_g < c$.

(3) The phase of the wave tunneling through the stack of gradient layers in the non-attenuated regime, can be found from (4.4):

$$\phi_t = Arctg\left[\frac{\gamma}{1-\frac{\gamma^2}{4}+n_e^2}\right] \qquad (4.14)$$

Thus, in the abovementioned example of reflectionless tunneling through lossless films (m=2, $n_0$=2.35, $U_{min}$=0.75, d=65 nm, u=1.1) Eq. (4.7) yields $\phi_t = 1.38$ rad; thus, the wave envelope zero-crossing point is traversing the distance 2d during the subluminal time t=$\phi_t/\omega$=1.35 $t_0$, where $t_0$=2d/c relates to a free space propagation. Herein, the phase shift $\phi_0$, accumulated by this free space propagation, would be $\phi_0$= $2d\omega/c$= 1.02 rad, which is smaller than the abovementioned tunneling phase shift 1.38 rad. Here we arrived to the result already reported previously in [69-70] for other photonic barriers saying that, despite the fact that the phase shift seems to indicate a superluminal behavior, the signal velocity as well as the energy transfer velocity in the course of tunneling through photonic barrier are subluminal.

*4.2. Polarization–dependent tunneling (heterogeneity–induced anisotropy of gradient photonic barriers).*

This chapter is devoted to the reflectance-transmittance problems for waves, incidenting on the gradient layer under an arbitrary angle $\theta$. Here, unlike in the case of normal incidence, the waves have a different polarization structure and are described by different equations. Denoting the normal to the layer as z-axis and choosing the projection of the wave vector on the layer interface as y-axis, one can describe the polarization structure of an S-wave by means of its electric component $E_x$ and magnetic components $H_y$ and $H_z$. The layer is assumed to be lossless and non-magnetic, and one can write the Maxwell equations linking these components in the form:

$$\frac{\partial E_x}{\partial z} = -\frac{1}{c}\frac{\partial H_y}{\partial t} \;;\quad \frac{\partial E_x}{\partial y} = \frac{1}{c}\frac{\partial H_z}{\partial t} \;;\quad \frac{\partial H_z}{\partial y} - \frac{\partial H_y}{\partial z} = \frac{\varepsilon(z)}{c}\frac{\partial E_x}{\partial t} \qquad (4.15)$$

$$div(\varepsilon\overline{E}) = 0 \;;\; div(\overline{H}) = 0 \qquad (4.16)$$

Components of P-wave ($H_x$, $E_y$ and $E_z$) are also linked by Eq. (4.16); but the Eqs. (4.15) have to be replaced by:

$$\frac{\partial H_x}{\partial z} = \frac{\varepsilon(z)}{c}\frac{\partial E_y}{\partial t} \; ; \quad \frac{\partial H_x}{\partial y} = -\frac{\varepsilon(z)}{c}\frac{\partial E_z}{\partial t} \; ; \quad \frac{\partial E_z}{\partial y} - \frac{\partial E_y}{\partial z} = -\frac{1}{c}\frac{\partial H_x}{\partial t} \qquad (4.17)$$

The effects of heterogeneity-induced dispersion for inclined incidence can be found from the set (4.15) – (4.17) by means of exactly solvable models of $\varepsilon(z)$ given by Eq. (1), suitable for both S- and P- polarizations. The Rayleigh profile U(z) (1.1) and exponential profile (1.2) provide exact solutions of the set (4.15) –(4.17) for monotonic variations of U(z) [71]; however, these models are not suitable for description of smoothly varying concave photonic barriers that we plan to discuss. Thus, the reflection–refraction problem for the set of equations (4.15)-(4.17) is considered below with the basic method described in Section 2. Let us express the field components in Maxwell equations by means of the following auxiliary, polarization-dependent functions $\Psi_s$ and $\Psi_p$:

$$\text{S – polarization: } E_x = -\frac{1}{c}\frac{d\Psi_s}{dt} \; ; \quad H_y = \frac{d\Psi_s}{dz} \; ; \quad H_z = -\frac{d\Psi_s}{dy} \qquad (4.18)$$

$$\text{P – polarization: } H_x = \frac{1}{c}\frac{d\Psi_p}{dt} \; ; \quad E_y = \frac{1}{\varepsilon(z)}\frac{d\Psi_p}{dz} \; ; \quad E_z = -\frac{1}{\varepsilon(z)}\frac{d\Psi_p}{dy} \qquad (4.19)$$

The system (4.15) – (4.17) then reduces to two equations, governing S-and P-waves respectively. Omitting for simplicity the factor $\exp[i(k_y y - \omega t)]$, these equations can be written as:

$$\frac{\partial^2 \Psi_s}{\partial z^2} + \left( \frac{\omega^2 n_0^2 U^2}{c^2} - k_y^2 \right) \Psi_s = 0 , \qquad k_y = \frac{\omega}{c} \sin\theta \qquad (4.20)$$

$$\frac{\partial^2 \Psi_p}{\partial z^2} + \left( \frac{\omega^2 n_0^2 U^2}{c^2} - k_y^2 \right) \Psi_p = \frac{2}{U} \frac{dU}{dz} \frac{\partial \Psi_p}{\partial z} \qquad (4.21)$$

By introducing the new variable $\eta$ and new functions $f_s$ and $f_p$:

$$\eta = \int_0^z U(z_1) dz_1 ; \quad f_s = \frac{\Psi_s}{\sqrt{U(z)}} ; \qquad f_p = \Psi_p \sqrt{U(z)} \qquad (4.22)$$

one can present Eqs. (4.20) – (4.21) for S- and P – waves in the form:

$$\frac{d^2 f_s}{d\eta^2} + f_s \left( \Lambda - \frac{U_{\eta\eta}}{2U} + \frac{U_\eta^2}{4U^2} \right) = 0 \qquad (4.23)$$

$$\frac{d^2 f_p}{d\eta^2} + f_p \left( \Lambda + \frac{U_{\eta\eta}}{2U} - \frac{3U_\eta^2}{4U^2} \right) = 0 \qquad (4.24)$$

where $\Lambda = \frac{\omega^2 n_0^2}{c^2} - \frac{k_y^2}{U^2}$; $U_\eta = \frac{dU}{d\eta}$; $U_{\eta\eta} = \frac{d^2 U}{d\eta^2}$. Eqs. (4.23) – (4.24) are valid for arbitrary profiles of photonic barriers $U(z)$ and all angles of incidence $\theta$, including in particular the two well-known exactly solvable models mentioned above (Rayleigh and exponential profiles). The fact that some polarization effects will depend on the barrier profile is evident even without solving these equations. In the case of the widely used Rayleigh profile $U_R$, which can be rewritten in $\eta$-space by means of (4.22) as $U_R(\eta) = \exp(-\eta/L)$, both equations (4.23) and (4.24) coincide:

$$\frac{d^2 f_{s,p}}{d\eta^2} + f_{s,p}\left(\frac{\omega n_0}{c}\right)^2\left(1 - \frac{\omega_{cr}^2}{\omega^2} - \frac{\sin^2\theta}{n_0^2}\exp\left(\frac{2\eta}{L}\right)\right) = 0 \qquad (4.25)$$

Eq. (4.25) shows that, due to heterogeneity-induced dispersion, characterized by a cutoff frequency $\omega_{cr} = c/2n_0 L$, the tunneling regime arises in the Rayleigh barrier for both S- and P-waves simultaneously. In the case of the exponential profile, rewritten in $\eta$-space as $U = 1 - \eta/L$, no heterogeneity-induced cutoff frequency can be defined from Eqs. (4.23)-(4.24). In contrast to these traditionally used asymmetrical profiles, we consider here the exactly solvable symmetrical concave barrier $U(z)$, formed inside the dielectric film (thickness $d$) by variation of $U$ from the value $U = 1$ in the centre of barrier ($z = 0$) up to the values $U_m$ at the interfaces $z = \pm d/2$ - see Fig. 8 :

$$U = \frac{1}{\cos\left(\frac{z}{L}\right)}; \qquad U_m = \frac{1}{m}; \qquad m = \cos\left(\frac{d}{2L}\right) \qquad (4.26)$$

Note that this model is characterized by one free parameter (the length-scale $L$), however it can be used to study many features associated with this type of potential as we will show in Chapter 4.2. In fact, it represents a particular case of a more general potential with two adjustable parameters which, however, leads to much more cumbersome calculations and thus is not developed here. Substitution of (4.26) into (4.22) yields the variable $\eta$ and profile U($\eta$) [72]:

$$\eta = L \ln\left[\frac{1+tg(z/2L)}{1-tg(z/2L)}\right]; \quad U = \frac{1}{\cos\left(\frac{z}{L}\right)} = ch\xi; \quad \xi = \frac{\eta}{L} \quad (4.27)$$

Using Eq. (4.27) one can rewrite Eqs. (4.23)-(4.24) for S- and P-waves inside the barrier, given by Eq. (4.26), into the similar form:

$$\frac{d^2 f_{s,p}}{d\xi^2} + f_{s,p}\left(q^2 - \frac{T_{s,p}}{ch^2\xi}\right) = 0 \quad (4.28)$$

$$q^2 = \left(\frac{\omega n_0 L}{c}\right)^2 \left(1 - \frac{\Omega^2}{\omega^2}\right) \quad (4.29)$$

$$T_s = (k_y L)^2 + \frac{1}{4}; \quad T_p = (k_y L)^2 - \frac{3}{4}$$

where the characteristic frequency $\Omega$, writes(4.26):

$$\Omega = \frac{c}{2n_0 L} = \frac{c}{n_0 d} Arc\cos(m) \quad (4.30)$$

The sign of parameter $q^2$, given by Eq.(4.29), changes in $\omega = \Omega$. Let us consider here. In the low frequency range $\omega < \Omega$, $q^2 < 0$, and the expression in brackets in Eq.(4.28) for an S-wave is always negative : a low frequency S-wave, incidenting on the barrier (4.26) under an arbitrary angle $\theta$, is traversing this barrier as an evanescent wave. Inversely, the same expression in Eq. (4.28), for a P-wave and for realistic values of the modulation of the refractive index (4.26) in the layer ($U_m = m^{-1} \leq 1.3$–$1.4$) remains positive and implies a traveling mode regime for a P-wave.

Thus, unlike in homogeneous rectangular barrier, where tunneling of EM waves is determined by a condition common to both S- and P-polarizations, tunneling through the concave barrier (4.26) proves to be polarization-dependent. This heterogeneity-induced anisotropy can result in a significant difference in reflectivity and transmittivity of such barrier for S- and P-waves, their frequencies and angles of incidence on the film being equal. This difference is illustrated below by means of exactly solvable model (4.26).

In the case of one layer without substrate, embedded in free space, introducing the dimensionless quantities

$$u = \Omega/\omega > 1 \, ; \qquad N = \sqrt{1 - u^{-2}} < 1 \qquad (4.31)$$

and transforming eq.(4.28) by means of new variable $\eta$ and new function $W$

$$v = \frac{1 - th\xi}{2} \, ; \quad f_{s,p} = (ch\xi)^{\frac{N}{2}} W_{s,p} \qquad (4.32)$$

one obtains for this function the hypergeometric equation

$$v(1-v)\frac{d^2 W}{dv^2} + [\gamma - v(1 + \alpha + \beta)]\frac{dW}{dv} - \alpha\beta W = 0 \, ; \quad \gamma = 1 - \frac{N}{2} \qquad (4.33)$$

Although Eq.(4.33) is valid for both polarizations, the values of parameters $\alpha$ and $\beta$ have to be specified for each wave; definition $\gamma$ (4.33) is valid for both waves. In the case of an S-wave :

$$\alpha_s, \beta_s = \frac{1}{2}\left[1 - N \pm \frac{i\sin\theta\sqrt{1-N^2}}{n_0}\right] \qquad (4.34)$$

Since $\alpha_s + \beta_s + 1 = 2\gamma$, two linearly-independent solutions of Eq.(4.33) are given by the hypergeometric functions $F(\alpha_s, \beta_s, \gamma, v)$ and $F(\alpha_s, \beta_s, \gamma, 1-v)$, denoted below for compactness as $F(v)$ and $F(1-v)$; moreover, due to condition $\mathrm{Re}(\alpha_s + \beta_s - \gamma) < 0$ the series, presenting these functions, are converging absolutely. The general solution of Eq.(4.33) is

$$W = F(v) + Q_s F(1-v) \qquad (4.35)$$

Here $F(v)$ and $F(1-v)$ can be considered as forward and backward waves (meaning more exactly in a tunneling case, evanescent and antievanescent waves), while the factor $Q_s$ is associated with the contribution of backward wave to the total field. Using the variables, defined in Eqs. (4.22) and (4.31), and expressing the factor $\mathrm{ch}\,\xi$ in terms of $\cos(z/L)$ according to Eq.(4.27) the generating function $\Psi_s$ takes the form:

$$\Psi_s = \left[\cos\left(\frac{z}{L}\right)\right]^{\frac{1-N}{2}} W \qquad (4.36)$$

Substitution of (4.36) into (4.18) yields the explicit expressions for components of S-wave inside the film and the continuity conditions allow to determine the complex reflection coefficient for this wave $R_s$. By proceeding in a similar fashion, one can find the generating function for P- waves $\Psi_p$ and reflection coefficient $R_p$ [72].

The spectral properties of transmittance for S- and P- waves are presented in Figs. 9, 10 and 12 on the ($|T|^2$ - $\gamma$) plane, providing a uniform scale on horizontal axis of graphs. Herein the comparison of Figs. 9a – 9b, drawn for one gradient layer, shows that the increase of depth of modulation of refractive index $m$ can result in dramatic changes of transmittance-reflection spectra of gradient film for S-waves. A narrow asymmetrical peak of transmittance due to reflectionless tunneling for S-waves arises near by the point $u = 1$, $\gamma = 1$. This peak is contiguous with a narrow area of high dispersion of transmittance coefficient with almost vertical tangent to the graph $|T_s(\gamma)|^2$. The value of $|T_p|^2$ in this range remains almost constant, approximately 87 %. When the angle of incidence is decreased, the existence of the peak is unaffected while transmittance $|T_p|^2$ tends to 100% - see Fig.9b. In contrast, in the low frequency spectral range, close to the point $\gamma = 0.5$ - see Fig.9b, the reflection of S- waves is varying negligibly, close to 100%, while the transmittance of P- waves tends to zero; herein the frequency dispersion of $|T_p|^2$ in this range is profound. Using of a pair of contiguous similar layers, as shown on Fig. 9, can result in the splitting of peak of transmittance for S-waves for two closely-spaced narrow-banded peaks with $|T|^2 = 1$ near by the frequency u = 1 – see Fig. 11. The same tendency is observed in the transmittance of a gradient film supported by a thick substrate, also in this case the peak $|T|^2 = 1$ is broadened – see Fig. 12.

Considering the transmittance regime, related to some values $\gamma$, m, $\theta$ and $n_0$ and using the characteristic frequency $\Omega$, given by Eq. (4.30), one can choose the thickness of layer d, providing the formation of such regime for any given polarization and wavelength $\lambda$ by means of expression, following from Eqs.(4.30) and (4.31):

$$\frac{d}{\lambda} = \frac{\sqrt{1-4(1-\gamma)^2} \, Arc\cos(m)}{2\pi n_0} \qquad (4.37)$$

The high contrast between the transmittance of gradient layers for S- and P-waves may be interesting for polarizing systems, operating under large angles of incidence. These layers may be rather thin: e.g., a polarizing screen, providing for transmitted waves the ratio $|T_p|^2/|T_s|^2 < 0.05$, is characterized, satisfying the conditions of Fig. 9b (u = 3.2), by the ratio $d/\lambda = 0.025$ (eq. 4.37). Such miniaturized scale d is a remarkable feature of the anisotropic gradient nanolayers considered.

Thus, the heterogeneity-induced anisotropy, occurring in the course of inclined propagation of polarized waves through the gradient layer (4.26), provides new features of reflectionless tunneling of waves, associated with this geometry:

(1) Unlike the homogeneous layer, providing the 100% - transmittance for P-polarized waves, incidenting under the Brewster angle, the gradient layer (4.26) can provide complete transmittance for S – polarized waves in a wide range of angles of incidence.

(2) The non-attenuative tunneling of S-wave can be characterized by narrow closely located peaks $|T|^2 = 1$ (Fig.11) with the halfwidths $\Delta\omega/\omega \leq 0.01$.

(3) Such narrow-banded filtration of S-wave arising near by the critical frequency $\omega \approx \Omega$ may become useful for the design of high contrast spectral filters and polarizers, operating in the area of large angles of incidence ($\theta \approx 45° - 75°$).

It is worth noting that the model $U(z) = [\cos(z/L)]^{-1}$ and the analysis of reflectance of such films, performed above, can be used for the search of new exactly solvable model U(z). To do this according to approach, described in Chapter 2.1, let us rewrite the wave equation (4.20) for the normal incidence ($k_y = 0$) in $\eta_1$- space in a form (2.18); here the barrier $V(\eta_1)$ is:

$$V(\eta_1) = \frac{1}{\cos\left(\dfrac{\eta_1}{L}\right)} \qquad (4.38)$$

Considering (4.38) as a transform of some unknown barrier U(z) into $\eta_1$ -space, one can find U(z) from (2.16) by inverse transform

$$U^2(z) = \frac{1}{1+\left(\dfrac{z}{L}\right)^2} \ ; \ \frac{\eta_1}{L} = Arctg\left(\frac{z}{L}\right) \qquad (4.39)$$

The symmetric barrier (4.39), occupying the layer ($-z_0$, $z_0$), corresponds in $\eta_1$ - space to profile (4.38) with parameter m = $\left(\sqrt{1+(z_0/L)^2}\right)^{-1}$. Thus, we obtained a new exactly solvable model of photonic barrier – the Cauchy profile (4.39), shown in Fig. 1b.

*4.3. Gradient optics of surface waves.*

Until now we were dealing with waves tunneling along the gradient of dielectric susceptibility (Chapter 4.1) or under some angle $\theta$ to this direction (Chapter 4.2). In the special case $\theta = \pi/2$ it is useful to consider the waves on the interface of heterogeneous medium. These surface waves are confined close to interface, while their spatial structure depends on the profiles of refractive indices in the subsurface area of the bordering media. Such profiles provide the continuous transition of bulk values of dielectric susceptibility to their surface values both in some natural media, e.g., anchoring layer in liquid crystals [7], and in specially fabricated optoelectronic materials, such as, e.g., nanocluster-doped glasses [73], ion-implanted dielectrics [74] or dye-doped polymers [75]. To point out the peculiarities of surface waves, supported by smoothly varying profiles $\varepsilon(z)$, it is instructive to compare them with surface electromagnetic waves on the sharp boundary between two homogeneous dielectrics. Polarization and spectra of such waves, traveling in the y-direction along the boundary (plane z = 0), are well known [67]:

(1) The wave field contains components $E_y$, $E_z$, $H_x$ (TH- polarization); TE- polarized surface wave (components $H_y$, $H_z$, $E_x$) is not supported by such system, although such waves were described for other media, e.g., for sandwiched structures [76], photorefractive materials [77], and antiferromagnetics [78].

(2) TM(P-polarized) surface wave can propagate along the boundary of two homogeneous media, if their dielectric susceptibilities obey to condition $\varepsilon_1 + \varepsilon_2 < 0$. It means that, in any case, one medium must have $\varepsilon < 0$; such situation occurs usually for solid state plasma in metals and dielectrics [79].

(3) The spectrum of surface waves at the air-plasma interface with plasma frequency $\Omega_p$ possesses an upper bound: $\omega < \Omega_p/\sqrt{2}$. However, unlike these TM-waves, the smooth variation of profile $\varepsilon(z)$ near by the boundary can provide another structure of surface wave. This structure can be considered in the framework of exactly solvable model of $\varepsilon(z) = n_0^2 U^2(z)$ for the dielectric, filling the half-space z >= 0 (Fig.13):

$$U^2(z) = 1 - \frac{1}{g} + \frac{W^2(z)}{g} \; ; \; W(z) = \left(1 + \frac{z}{L}\right)^{-1} \qquad (4.40)$$

Here g is a free parameter; function (4.40) describes the "saturation" of dielectric susceptibility in the depth of medium (z >> L), where the refraction index takes its "bulk" value $n_v$:

$$n(z \gg L) = n_0 \sqrt{1 - 1/g} = n_v \qquad (4.41)$$

The values g<0 (g>1) correspond to the case $n_0 < n_v$ ($n_0 > n_v$), respectively.

TE (S-polarized) surface wave can propagate along the boundary between the medium (4.40) and air (W = 1). The components of wave field $E_x$, $H_y$ и $H_z$ can be expressed by means of (4.18) via the generating function $\Psi$, governed by wave equation (4.20). Presenting the function $\Psi$ in a form [80]

$$\Psi = A\sqrt{W} f(u) \exp[i(k_y y - \omega t)]; \quad u = 1 + z/L \qquad (4.42)$$

one can find from (4.20) that function f(u) obeys Bessel equation

$$\frac{d^2 f}{du^2} + \frac{1}{u}\frac{df}{du} + f\left(p^2 - \frac{s^2}{u^2}\right) = 0 \qquad (4.43)$$

$$p^2 = \left(\frac{\omega}{c}\right)^2 (n_v^2 - b^2) = -p_1^2; \quad s^2 = \frac{1}{4}\left(1 - \frac{\omega^2}{\Omega_c^2}\right); \quad b = \frac{ck_y}{\omega}; \quad \Omega_c = \frac{c}{2L\sqrt{n_0^2 - n_v^2}} \qquad (4.44)$$

The solutions of Eq. (4.43), that decrease as the coordinate z increases, are given by the modified Bessel functions $K_s(p_1 L u)$. These solutions, arising, when $p^2 < 0$ ($p_1^2 > 0$) and $s^2 > 0$, describe the spatial structure of the field, localized near by the plane z=0; herein $n_v^2 < b^2$.

Let us stress out that $n_0 > n_v > 0$; the refractive index is decreasing in the subsurface layer, remaining positive in the depth of medium. Thus, the field localization near by the interface of dielectric is determined by, instead of traditional condition $\varepsilon < 0$ [67], another condition: $\text{grad}\,\varepsilon < 0$. Generating function in the air ($z \leq 0$) can be written as

$$\Psi_1 = B\exp[z/l + i(k_y y - \omega t)] \; ; \; k_y^2 = \frac{\omega^2}{c^2} + \frac{1}{l^2} \tag{4.45}$$

Here l is the characteristic e-folding length of the wave in the air ($z \leq 0$), A and B are the normalization constants of solutions (4.42) and (4.45) respectively. The continuity conditions on the interface z=0 yield the link between the constants A and B and spatial scales l and L:

$$B = AK_s(p_1L) \tag{4.46}$$

$$L/l = \tfrac{1}{2} + p_1L[dK_s/dz](K_s)^{-1} \tag{4.47}$$

By factorizing the function $K_s(p_1L)$ in the range of small values of argument $[(p_1L)^2 \ll 1]$ in a power series and taking into account the first term of this series, one can rewrite (4.47) in a form [81]:

$$K_s(p_1L) = \frac{\Gamma(s)}{2}\left(\frac{2}{p_1L}\right)^s \; ; \; L/l = \frac{1}{2} - s + .. \tag{4.48}$$

here $\Gamma(s)$ is the gamma – function. Condition $L/l > 0$ brings the limitation of index of Bessel function $K_s$: $0 < s < \tfrac{1}{2}$. The dispersive equation of the surface wave considered can be found by means of correlation between the scale l and frequency $\omega$, following from (4.45): $b^2 = 1 + (c/l\omega)^2$. Herein $b = ck_y/\omega > 1$, thus, the wavenumber of the surface wave exceeds the value $k_y$ for the wave of the same frequency $\omega$ in free space z<0. Such a surface wave can be excited due to the increase of projection $k_y$ of the wave, incidenting from the half–space $z < 0$. To provide this increase, one can use the optical scheme, resembling FTIR in the configuration of

Fig. 6: the wave is passing through the triangle prism with refractive index $n_p$, located on the interface of gradient medium. In this geometry the dimensionless parameter b given by Eq. (4.44) is defined as $b = n_p \sin\theta$ ; thus, the condition $b > n_v$, which is in need for field decrease in the depth of medium, can be fulfilled. Expressing the e-folding length l in (4.48) via parameter b, one can obtain the dispersive equation of surface TE wave:

$$\frac{b^2 - 1}{n_0^2 - n_v^2} = \frac{1}{\left(u - \sqrt{u^2 - 1}\right)^2} \; ; \; u = \frac{\Omega_c}{\omega} \qquad (4.49)$$

Here $u \geq 1$ is the normalized frequency of the surface wave, the critical frequency $\Omega_c$ is defined in (4.44). The spectral range of surface waves considered is : $1 \geq u \geq u_c$; here $u_c$ is defined as:

$$u_c = \frac{n_0^2 - 1}{2\sqrt{(n_0^2 - n_v^2)(n_v^2 - 1)}} \qquad (4.50)$$

Thus, in a case $n_0 = 1.84$, $n_v = 1.38$, corresponding to 25% decrease of refractive index in the depth of medium, one has $u_c = 1.03$. The wavelengths $\lambda$, related to this normalized spectral range $1 \geq u \geq u_c$, is very sensitive to the characteristic thickness of transition layer L; according to definition of critical frequency $\Omega_c$ (4.44) and (4.49) the layers with L = 30 – 45 nm can support TE surface waves in the blue end of the visible spectrum : 458 nm < $\lambda$ < 472 nm, $\Delta\lambda$ = 14 nm, when L = 30 nm, while the value L = 45 nm corresponds red end: 688 nm < $\lambda$ < 710 nm, $\Delta\lambda$ = 22 nm ("colored" surface waves). The increase of L up to 55 nm shifts the spectral range of surface TE waves to a near IR range: 840nm < $\lambda$ < 865 nm, $\Delta\lambda$ = 25 nm.

These narrow spectral ranges, determined by Eq. (4.50), can be broadened due to increase of difference between the surface and bulk values of refractive index n. Thus, considering the medium (4.41) with $n_0 = 2$, $n_v = 1.84$ and L = 30 nm, we obtain the spectral range 545 nm < $\lambda$ < 595 nm, $\Delta\lambda$ = 40 nm, which is broadened and red-shifted as compared with the aforesaid range, related to $n_0 = 1.84$, $n_v = 1.38$, L = 30 nm.

The same situation arises when the next term of factorization of function $K_s$ (4.48) is taken into account:

$$\frac{L}{l} = \frac{1}{2} - s + \frac{(p_1 L)^2}{2(1-s)} - ... \qquad (4.51)$$

The corrections to the values b, obtained by means of approximation (4.50), do not exceed a few percents.

The TE waves on the boundary of gradient media are distinguished from TM waves on a sharp interface between two homogeneous media by an another mechanism of field penetration inside the media: P-polarized TM waves are tunneling, as was mentioned above, in the medium where $\varepsilon < 0$, while tunneling of S-polarized TE waves is possible, if $d\varepsilon/dz < 0$, even if $\varepsilon > 0$. The transition layer with characteristic scale L is in need for existence of TE waves; when this layer is broadening (L $\to \infty$), the critical frequency $\Omega_c$ tends to zero, and equation (4.49) is reduced to the trivial condition of total internal reflection of incidenting waves $n_p \sin\theta = 1$.

It is worth outlining some peculiarities of surface TE waves, stipulated by tunneling through the thin ($L < \lambda$) subsurface layer of lossy dielectric susceptibility, namely:

(1) TE surface waves can propagate along the interfaces of solids, which do not contain free carriers. This property broadens essentially the list of materials making possible the use of these waves.

(2) High values of $\Omega_c$, determined by the non-local dispersion in the gradient layer, permit one to broaden the spectrum of surface TE waves to the near IR and even visible spectral range.

(3) Surface TE waves, unlike TM ones, exist in the narrow spectral ranges $\Delta\lambda \leq 0.1\ \lambda$.

## 5. Electromagnetics of heterogeneous spherical structures.

Scattering of linearly-polarized electromagnetic plane waves by spherically symmetrical objects of natural and artificial origin, also called Mie scattering, has been considered by many authors. These researches, pioneered by Rayleigh in his explanation of the color of the sky by scattering of light on small water droplets, had obtained about a half of century ago the second wind due to attention to the radar backscattering and targets identification [82]. The ability to produce micrometer-size droplets in a controlled manner and the growing applications of composites, containing such droplets, led nowadays to a burst of renewed interest into the problems of interaction of EM waves with curvilinear surfaces for the different values of important parameter – the ratio of wavelength $\lambda$ to the characteristic scales of the scatterer [83-87]. However, the theoretical analysis of Mie scattering, based on solutions of wave equation in the spherical coordinates,

$$\left(\nabla^2 + k^2\right)\Psi = 0 \tag{5.1}$$

faces a lot of difficulties even in the simplest geometry of homogeneous spherical dielectric scatterers. These difficulties lie in a complicated angular pattern due to interference of many partial waves arising, in their turn, due to modal decomposition of the incident plane wave. Unlike the reflectance of plane gradient layer, determined by non-local dispersion of curvilinear profiles $\varepsilon(z)$, the salient feature of curvilinear layers is their "geometrical" dispersion, arising in reflectance of even homogeneous curved interface. This dispersive effect can be simply illustrated, without solving Eq. (5.1), by means of reflection-refraction phenomena near the angle of total internal reflection (TIR) on the curved surface of dispersionless homogeneous dielectric. Owing to this effect some rays, predicted by Fresnel laws to be totally reflected from the interface, are reflected only partially, providing dispersive losses.

To examine this curvature based losses let us consider an EM wave, incidenting from the optically denser medium on the curved interface between two dielectric lossless media with slightly different refractive indices n₁ and n₂ (n₁ > n₂); defining the complement $\theta_c$ of TIR angle.

$$\theta_c^2 = 1 - \left(\frac{n_2}{n_1}\right)^2 \tag{5.2}$$

one can see from Eq. (5.2), that $\theta_c \ll 1$ ($\theta_c > 0$). The power transmission coefficient $|T|^2$ in this geometry, found in [88], reads as

$$|T|^2 \frac{1}{\pi\theta_c}\left(\frac{2}{k\rho}\right)^{\frac{1}{3}} \left|Ai\left[l \exp\left(\frac{2\pi}{3}\right)\right]\right|^{-2} ; \; l = \left(\frac{k\rho}{2}\right)^{\frac{2}{3}}\left(\theta_c^2 - \theta_1^2\right) \tag{5.3}$$

Here $\theta_1$ is the complement of an angle of incidence $\theta$ ($\theta_1 = \frac{\pi}{2} - \theta$), $\rho$ is the radius of curvature in the plane of incidence, k = $\omega$ n₁/c is the wave vector in the denser medium, and $Ai$ is the Airy function.

Let us examine the grazing incidence of rays onto the curved interface, when the angle $\theta_1$ is small, but finite: $1 \gg \theta_1 \gg (k\rho)^{-\left(\frac{2}{3}\right)}$. When the ray is incident with $\theta_1 > \theta_c$, geometric optics presents the Fresnel value of power transmission coefficient $|T|^2$:

$$|T|^2 = 4\sqrt{\left(\frac{\theta_1}{\theta_c}\right)^2 - 1} \qquad (5.4)$$

However, the asymptotic forms of Airy function [81]

$$\lim|Ai(x)|_{x<0,|x|>1} = \frac{1}{2}|x|^{-\frac{1}{4}} \; ; \; \lim|Ai(x)|_{x>>1} = \frac{1}{2}|x|^{\frac{1}{4}}\exp\left(-\frac{2}{3}x^{\frac{3}{2}}\right) \qquad (5.5)$$

show that Eq. (5.3) can be reduced to Eq. (5.4) only in the case of large negative values of x in Eq. (5.5), when

$$\theta_1^2 - \theta_c^2 >> (k\rho)^{-\frac{2}{3}} \qquad (5.6)$$

When the ray is incident with $\theta_1 < \theta_c$, Fresnel value (5.4) is zero, while the value, given by Eq. (5.3), is finite. This effect indicates some radiation energy losses, due to leaky waves, penetrating into the medium $n_2$. A simple expression for these losses can be obtained for leaky rays, which propagate not too close to the TIR angle: $\theta_c^2 - \theta_1^2 >> (k\rho)^{-\frac{2}{3}}$. The asymptotics (5.5) in this case yield

$$|T|^2 = 4\sqrt{1 - \left(\frac{\theta_1}{\theta_c}\right)^2}\exp\left[-\frac{2}{3}k\rho(\theta_c^2 - \theta_1^2)^{\frac{3}{2}}\right] \qquad (5.7)$$

These values of power transmission coefficient depend exponentially on the radius of curvature. The losses, arising due to formation of leaky waves in the vicinity of TIR angle, may become important for energy transfer in bent optical fibers. Thus, tunneling through cylindrical barriers in the optical waveguides with small bending radius can result in formation of parasitic modes, resembling Rayleigh acoustical "whispering gallery" effects. For the fiber with a core diameter d, external radius of cladding $r_0$, refractive indices of core and shell $n_1$ and $n_2$ respectively the condition of generation of such mode is: $d/r_0 > (n_1/n_2 - 1)$ [89].

This analysis deals with the reflection on the curved interface of a scatterer whose sizes are not supposed to be restricted. To examine the problem of waves scattering on a sphere with finite radius, one has to use the wave equation (5.1). This equation in spherical coordinates is known to be separable into ordinary differential equations in $r, \theta$ and $\phi$, and its solution can be written as a superposition of spherical modes, presented by the products of radial- and angular-dependent factors:

$$\Psi = \sum_{l=0}^{\infty} \sum_{m=0}^{\infty} R_{lm}(r) Y_{lm}(\theta, \phi) \qquad (5.8)$$

For a few types of radial distributions of refractive index n(r), which describe the dependence of index on the distance r from the centre of spherical sample, it is possible to obtain exact analytical solutions of Maxwell equations in terms of known functions. These solutions are examined below for the model of 3D photonic bandgap materials, consisting of stratified sphere with the step-like variations of refractive index between the neighboring spheres (Chapter 5.1) and the core-shell model of the sphere with power law distribution of dielectric susceptibility $\varepsilon(r)$ in the shell (Chapter 5.2). The effect of electromagnetic cloaking, arising due to shell produced from metamaterial with continuously varying tensor distributions of $\varepsilon > 0$ and $\mu > 0$ is illustrated too. Another heterogeneous shell, designed from metamaterial with negative values of $\varepsilon$ or $\mu$ or both of them, is shown to provide a potential for perfect gradient lens (Chapter 5.3).

*5.1. Geometrical dispersion and photonic bandgaps in Mie scattering.*

Mie scattering has long been a subject of both experimental and theoretical interest [83], [90]. Current interest lies mainly in interaction of light with large droplets (ka >>1), where a is the droplet radius and k is the wave number in vacuum. In this case the scattering of light exhibits sharp peaks, related to morphology-dependent resonances (MDR). The high values of quality factors Q (Q $\approx 10^6 - 10^7$) of these resonances provide their small widths, implying long storage times and strong optical feedback for photons [91].

The rational design of concentric dielectric multilayers exploits the analogy with stratified planar systems. This analogy permits to treat these multilayers as potential models of three-dimensional (3D) photonic bandgap (PBG) structures. The detailed analysis of interaction of light with these concentric systems in the framework of the full apparatus of Mie scattering theory is cumbersome algebraically and computationally, especially for ka >> 1 [92]. Therefore, only some features of these processes in concentrically stratified spheres, useful for analysis of spherical layers with radially symmetrical profiles of dielectric susceptibility $\varepsilon(r)$, will be pointed out below.

Let us consider the simplified version of Mie theory for a plane wave $\vec{E}_0$, propagating in z-direction and linearly polarized in x-direction, scattered by a homogeneous lossless dielectric sphere of radius a. The internal ($\vec{E}_i$) and scattered ($\vec{E}_{sc}$) fields can be written as the superposition of vector spherical harmonics [93]:

$$\vec{E}_i = \sum_{p=1}^{2}\sum_{n=1}^{\infty} q_{np} b_{np} \vec{N}_{np}^{(1)} \; ; \; \vec{E}_{sc} = \sum_{p=1}^{2}\sum_{n=1}^{\infty} q_{np} c_{np} \vec{N}_{np}^{(2)} \; ; \; q_{np} = -\frac{i^{n+p}(2n+1)}{n(n+1)} E_0 \qquad (5.9)$$

Here the subscript p refers to either TM (p = 1) or TE (p = 2) partial-wave polarization; the superscripts 1 and 2 indicate, that the radial parts of vector spherical harmonic are governed by the spherical Bessel function and Hankel function of the first kind, respectively. Applying the continuity condition on the surface of the sphere (r = a) yields the coefficients for the scattered and internal fields; in particular, the coefficient $c_{np}$ for scattered wave is:

$$c_{np} = \frac{m^{2-p}\psi_n(mx)\psi_n'(x) - m^{p-1}\psi_n(x)\psi_n'(mx)}{m^{2-p}\psi_n(mx)\xi_n'(x) - m^{p-1}\xi_n(x)\psi_n'(mx)} \quad (5.10)$$

Here $\psi$ and $\xi$ are Ricatti-Bessel and Ricatti-Hankel function, respectively, x is the size parameter x=ka, and m is the relative refractive index

$$m = \frac{n_{part}}{n_{med}} \quad (5.11)$$

where $n_{part}$ and $n_{med}$ are the refractive indices of the particle and surrounding medium. In the limit x >> 1 the numerator of (5.10) is periodic in x (or $1/\lambda$) as the Fresnel reflection coefficient for a thin plane film, and the various values of $c_{np}$ are in phase. Superposition of these periodic oscillations gives rise to an interference structure of the scattered wave as a function of x. Roots of denominator of $c_{np}$ provide the morphology-dependent resonances in scattering. Using the asymptotic expressions for functions $\psi$ and $\xi$ yields the frequency spacing between the interference maxima [90]

$$\Delta\omega = \frac{\pi c}{a n_{med}(m-1)} \quad (5.12)$$

This analysis can be generalized for a N-layer concentrically stratified sphere [94]. Although the coefficients $b_{np}$ and $c_{np}$ (5.9) are analogous to Fresnel transmission and reflection coefficients for plane geometry, the important difference between plane and spherical geometries has to be emphasized: unlike a plane wave, incident on a plane structure, which has only one reflection coefficient, a plane wave, incident on a spherical structure, has a multitude of scattering (reflection) coefficients as a result of modal decomposition of the incident wave on the partial waves.

The concentric stratification provides a modification of the interference structure of the scattered wave. Thus, for a periodic structure with an odd numbers of layers, the spacing between the interference maxima is given by [90]:

$$\Delta\omega = \frac{\pi c}{d(N-1)\Delta n} \tag{5.13}$$

where $\Delta n$ is the refractive index difference between the successive layers and d is the thickness of the even-numbered layers. This effect is distinguished from that associated with the existence of one-dimensional photonic band gap observed in planar layers, where the frequency spacing would be $\Delta\omega = \pi c/2nd$. The origin of this difference is that for small particles a portion of the incident wave goes around the structure [95], while planar multilayers are assumed to be infinite, occluding all the incident light. Side by side with these bulk effects it is worth to examine the manifestation of geometrical dispersion in the multimode spectra of whispering gallery modes, which travel on the homogeneous sphere close to its circumference. This problem attracts attention due to unusual phenomena in the IR optics of microspheres, connected with high values of their Q-factors ($Q \approx 10^7 - 10^8$) [96].

Analysis of spatial structure of these modes can be reduced to solving a scalar wave equation in the spherical coordinates:

$$(\nabla^2 + k^2)\Psi = 0 \qquad (5.14)$$

where the generating function $\Psi$ is determined from

$$\vec{E} = \nabla \times (\vec{r}\Psi) \qquad (5.15)$$

In Eq. (5.14) the wave vector k is assumed to be a function of frequency $\omega$ and radial coordinate r. This equation is separable into ordinary differential equations in r, $\theta$ and $\phi$

$$\Psi = R(r)\Theta(\theta)\Phi(\phi) \qquad (5.16)$$

Following [97], one can assume that the wave is localized in a narrow ring area around the equator of the sphere, $\theta = \pi/2$,

$$\Phi(\phi) = \exp[i(m\phi - \omega t)] \ ; \ |m| = 1; 2; 3... \qquad (5.17)$$

For whispering gallery modes traveling along the equator, it is convenient to approximate the solutions for $\Theta(\theta)$ by Hermite-Gaussian functions $H_p(\theta_1)$, where $\theta_1 = \pi/2 - \theta, \theta_1 \ll 1$, p = l - |m|, l = 0; 1; 2... As to radial function R(r) – let us introduce a new coordinate $r_1$ = r – a, so that $r_1 \ll$ a. Using the normalized radial variable

$$\rho = \left(\frac{2k^2}{a}\right)^{-\frac{2}{3}} \left[\frac{2k^2 r_1}{a} - k^2 + \frac{l(l+1)}{a^2}\right] \tag{5.18}$$

equation for R can be rewritten in a dimensionless form

$$\frac{d^2 R}{d\rho^2} - \rho R = 0 \tag{5.19}$$

with a proper solution given by Airy function R = Ai($\rho$). These Airy-Hermite-Gaussian whispering gallery modes were observed experimentally in [98]. Using the continuity conditions at r = a, one can obtain the dispersive relation for $q^{th}$ mode [99]:

$$(ka)^2 = l(l+1) + \varsigma_q \left(2k^2 a^2\right)^{\frac{2}{3}} \tag{5.20}$$

where $\varsigma_q$ are zeros of Airy function ($\varsigma_1 = 2.34$; $\varsigma_2 = 4.09$; $\varsigma_3 = 5.52$ ...). The radius related to $q^{th}$ mode maximum is decreasing with the increase of the mode number q : the higher order modes travel closer to the center of the sphere than low order ones.

Thus, the geometrical dispersion, arising due to sphericity of the scatterer, gives rise to multimode wave structures; this peculiarity distinguishes both bulk and surface waves in/on the sphere from those ones in the plane geometry. These modal structures, formed owing to decomposition of incidenting plane wave into partial waves, provide to each partial wave a possibility to have its own photonic bandgap. The variety of such mode effects is extended due to gradient multilayers, especially, as shown below, when these multilayers are formed by properly designed metamaterials.

*5.2. Optics of core–shell gradient spherical layers.*

The electromagnetics of dielectric spheres, whose refractive index n(r) is varying continuously along the radius r from the centre (r = 0) up to the interface (r = a), was examined analytically only for few models of n(r) [100]. Solutions of Maxwell equations in these cases is expressed in terms of two scalar potential functions $F_1$ and $F_2$, which describe respectively the TE waves ($\vec{r}.\vec{E} = 0$) and TM waves ($\vec{r}.\vec{H} = 0$). The models of such continuous profiles are presented, e.g., by Cauchy distribution n(r) and generalized parabolic distribution $n^2(r)$ [101]:

$$n(r) = \frac{n_0}{1+\varsigma r^2 a^{-2}} \; ; \; n^2(r) = m^2[1- \varsigma r^2 a^{-2}] \; ; \; \varsigma > 0 \; ; \; 0 \leq r \leq a \qquad (5.21)$$

In the special case $\varsigma = 1$ Cauchy distribution n(r) is reduced to well-known "Maxwell fish-eye" profile, and parabolic distribution $n^2(r)$ for $m^2 = 2$ and $\varsigma = 1/2$ reduces to a spherical Luneberg lens. The field structures in these cases are described by complicated combinations of hypergeometric functions. A useful approximation to the exactly solvable model is given by linear profile $\varepsilon = a + br$ [102].

However, these profiles of n(r) do not readily cover an important case, characterized by a small gradient in the index near the centre of sphere and a rapid drop near the boundary. Such a dependence on radial distance is conventionally represented by the model of uniform inner core, surrounded by a spherically symmetric inhomogeneous outer shell, (core-shell model), for which n(r) varies according to a power law:

$$n(s) = n_0 \; (0 \leq s \leq x_0) \; ; \; n(s) = gs^p \; ; \; s = kr \; ; \; (x_0 \leq s \leq x) \; ; \; x_0 = ka_0 \qquad (5.22)$$

In a special case p = - 1 the wave equations governing the radial functions R(r) in solution of (5.8) for the model (5.22), reduce to simple Euler homogeneous equations. By introducing

$$q = \sqrt{\left(l+\frac{1}{2}\right)^2 - g^2} \tag{5.23}$$

and assuming l+1/2 >g, one can find the radial functions R(r) for TE and TM waves in the gradient shell

$$R_{TE} = \sqrt{s}\left(s^q + Q_{TE}s^{-q}\right) ; \quad R_{TM} = \frac{s^q + Q_{TM}s^{-q}}{\sqrt{s}} \tag{5.24}$$

In an opposite case (l+1/2 <g, q=iq$_1$) the functions R$_{TE}$ and R$_{TM}$ read as:

$$R_{TE} = \sqrt{s}\left[\cos(q_1 \ln s) + Q_{TE}\sin(q_1 \ln s)\right]$$

$$R_{TM} = \frac{\cos(q_1 \ln s) + Q_{TM}\sin(q_1 \ln s)}{\sqrt{s}} \tag{5.25}$$

This presentation of fields in model (5.22), in the case p = -1, in terms of elementary functions simplifies the analysis of the wave scattering problems for core-shell gradient structures. Comparison of the angular scattering patterns on cores coated by gradient shells (5.22) shows that, subject to the slope of profile n(s) determined by the power p in (5.22), these patterns become essentially different, especially for the large scattering angles [103]. Herein the forward scattering in this geometry was found to be the least dependent on the profile n(s), while the backscattering proved to be very sensitive to the variations of this profile. Moreover, the shadow behind the illuminated sphere can be affected by a portion of incident wave, rounding the sphere [95].

An intriguing application of gradient shells is connected with the reduction of backscattering and weakening of shadow in the transmitted field. These properties provide the possibility to some body, coated by appropriately designed shell, to have in a narrow spectral range the optical parameters of free space. Interacting with such coated body, the incidenting wave will remain undistorted, and, thus, this body will remain invisible in the given spectral range for the external observer. This unprecedented flexibility in manipulating EM waves gives rise nowadays to a growing interest in the new field of the transformation optics based on the use of metamaterials [104]- [107].

The theoretical fundamentals of this so called electromagnetic cloak are based on the invariance of Maxwell equations, written in Minkowski form, in the coordinate transformation; herein only the components of time-independent dielectric susceptibility and magnetic permeability tensors are affected by these transformations [108] – [109].

Referring the reader, interested in a rigorous mathematical treatment of these problems, to the original works [110] – [112], we will restrict the present analysis to examples illustrating the possibility to bring the optical parameters of core-shell system nearer to those of free space. Let us consider the spherically symmetric coordinate transform, squeezing the space from the sphere $0 \leq r \leq b$ to a shell $a \leq r_1 \leq b$ [113] :

$$r_1 = \frac{b-a}{b} r + a \qquad (5.26)$$

Herein, when $r = b$, then $r_1 = b$; outside this domain the identity mapping is assumed. Thus, the area $r \geq b$ remains undisturbed, and there is no discontinuity of transformation on the boundary of the shell $r = r_1 = b$.

Let us consider the position vector $\vec{x}$, characterized by components $x^i$ in the original three-dimensional Cartesian free space, and components $x^\alpha$ in the transformed coordinate system, linked by Jacobian transformation matrix $\Lambda_i^\alpha$, which reads for the transform (5.26) as

$$\Lambda_i^\alpha = \frac{\partial x^\alpha}{\partial x^i} = \frac{r_1}{r}\delta_i^\alpha - \frac{ax^j x^k \delta_\alpha^j \delta_{ki}}{r^3} \tag{5.27}$$

The tensors of dielectric susceptibility $\varepsilon^{ij}$ and magnetic permeability $\mu^{ij}$ of the original space are known to transform as [109]

$$\varepsilon^{\alpha\beta} = \left|\det(\Lambda_i^\alpha)\right|^{-1} \Lambda_i^\alpha \Lambda_j^\beta \varepsilon^{ij} \tag{5.28}$$

Assuming, that the original medium is free space, its relative dielectric susceptibility and magnetic permittivity are both equal to unity. Calculating the determinant of $\Lambda_i^\alpha$ (5.27)

$$\det(\Lambda_i^\alpha) = \frac{b-a}{b}\left(\frac{r_1}{r}\right)^2 \tag{5.29}$$

and substituting (5.29) to (5.28), one obtains the material properties, which are needed for the gradient spherical shell considered [113]:

$$\varepsilon^{\alpha\beta} = \mu^{\alpha\beta} = \frac{b}{b-a}\left(\delta^{\alpha\beta} - \frac{2ar - a^2}{r^4} x^\alpha x^\beta\right) \tag{5.30}$$

Note that the electromagnetic response of the shell, described by tensors $\varepsilon^{\alpha\beta}$ and $\mu^{\alpha\beta}$, is anisotropic. Proceeding in a similar fashion one can design a cloak of cylindrical symmetry. The coordinate transform, used in this case, has to compress a cylindrical region $0 \leq r \leq b$ to a concentric cylindrical shell $a \leq r_1 \leq b$ (5.26). This mapping can also be described by (5.26), however, the radii a and b confine now a cylindrical shell along the axis z. The results are simplified in the cases where the electric (magnetic) component of incidenting TE (TM) polarized wave is directed along the z axis [114]. Thus, in a case of TM illumination this transformation results in the following requirements for the non-zero components of anisotropic susceptibility $\varepsilon$ in the cloaking shell:

$$\varepsilon_{rr} = \left(\frac{b}{b-a}\right)^2 \left(\frac{r_1 - a}{r_1}\right)^2; \quad \varepsilon_{\theta\theta} = \left(\frac{b}{b-a}\right)^2 \qquad (5.31)$$

The magnetic permeability in this geometry is characterized by only one component of tensor $\mu$ ($\mu_{zz} = 1$). The radial distribution of $\varepsilon_{rr}$ is varying from the value 0 at the inner boundary of the shell ($r_1 = a$) up to $\varepsilon_{rr} = 1$ at the outer interface $r_1 = b$. Thus, at the outer interface the value $\varepsilon_{rr}$ is matched with the value of relative dielectric susceptibility of free space.

To provide the desired gradients of tensor components in (5.30) and (5.31) as the functions of radius, a very complicated metamaterial structure is needed. Artificial dielectrics with positive values of $\varepsilon_r$ less than unity, were designed in [115–116] by means of metal wires of subwavelength size, embedded in the radial direction in a host dielectric medium. The desired radial gradient of magnetic permeability was associated in [117] with the variable dimensions of a series of split-ring resonators (SRR), providing the proper magnetic polarizability of this structure in a given spectral range given by [118]

$$\mu = 1 - \frac{F\omega^2}{\omega^2 - \omega_0^2} \qquad (5.32)$$

Here the resonant frequency $\omega_0$ and dimensionless parameter F (0<F<1) depend on the geometrical scales of the structure. To approximate the cloak as a continuous material, a wavelength–structure ratio $\lambda/a \geq 10$ is desirable [115]. Herein, the shell-type cloaks can work only in a narrow frequency range, since the metamaterials with n < 1 are dispersive.

The effectiveness of transformation based cloak was examined by full wave numerical simulations [119]. The wave fronts were shown to flow around the cloaked target, exhibiting an essential reduction of distortions of scattered fronts, although such distortions were not completely suppressed due to incomplete matching of cloak with the free space. However, despite the aforesaid limitations, these first realizations of cloaking by means of gradient shells are of great potential interest.

*5.3. Gradient lenses, containing left handed metamaterials(LHM).*

Combination of properties of gradient spherical shells and unusual features of metamaterials with negative $\varepsilon$ and $\mu$ stimulate the efforts to design a "perfect lens", whose resolution is not restricted by the classical diffraction limit. These efforts are inspired by the model of slab, made from material with $\varepsilon = \mu = -1$ [105], characterized by negative refraction index n < 0. These so-called negative index materials (NIM) [120] were considered during several decades as an academic curiosity until recently, when the first samples of NIM were fabricated. These samples were based on the interleaving arrays of metallic nanowires for negative $\varepsilon$ and split-ring resonators (SRR), providing, according to (5.32), negative values of $\mu$ for GHz waves [121]

$$\omega < \omega_{cr} = \frac{\omega_0}{\sqrt{1-F}} \qquad (5.33)$$

Such slab would act as a lens but, being invariant in the transverse directions, would not provide any magnification in the image. To create a magnification, this transverse invariance has to be broken by means of curved surfaces.

Following [122], one can consider the lossless NRM spherical shell, confined by radii b > a, and the source, located inside the shell at the point with radius $a_0$ < a. This sphere is assumed to be embedded in a right-handed dielectric medium ($\varepsilon > 0, \mu > 0$). Let us examine the power distributions $\varepsilon(r) = \alpha r^p$ and $\mu = \beta r^q$ in the gradient shell. If the powers p and q obey to condition

$$p + q = -2 \qquad (5.34)$$

the power form solution of wave equation (5.14) can be obtained for both TE and TM polarized waves, irradiated by the source. Thus, e.g., the radial electric component of TM wave can be written in this case as

$$E_r = \sum_{l,m} \left[ n_+ A_{lm} r^{n_+ - 1} + n_- B_{lm} r^{n_- - 1} \right] Y_{lm}(\theta, \phi) \qquad (5.35)$$

$$n_{\pm} = \frac{1}{2}\left[ -(p+1) \pm \sqrt{(p+1)^2 + 4l(l+1) - 4\alpha\beta\omega^2 c^{-2}} \right] \qquad (5.36)$$

Similar solutions can be also obtained for TE modes. Such "power" solutions are valid for any values of parameters p and q, obeying to condition (5.34). In a special case p = q = - 1 these solutions were shown to describe the perfect lens model [122], providing the near–field magnification in the image, formed by a factor $(b/a)^2 > 1$.

This near-field limit is important for the Rayleigh (quasistatic) approximation, when all the length scales of the problem are much smaller than the wavelength. In this case the wave equation (5.14) is reduced to Laplace equation; its power solutions define the profiles of $\varepsilon$ and $\mu$ providing the perfect lens effect for TE or to TM modes alone. Thus, a spherical shell with $\varepsilon = -C/r^2, \mu =$ const, satisfying (5.34), can result in a perfect lens for TM modes, while the case $\varepsilon =$ const, $\mu = -C_1/r^2$, yields the same effect for TE modes [122]. A similar lens-like property was found in [123 – 124] for the cylindrical annulus with $\varepsilon = -1$.

Side by side with near-field magnification problems this analysis of shells with heterogeneous $\varepsilon$ is shown below to be useful for the design of composite materials for optoelectronics consisting of coated microspheres.

## 6. Nonlinear subwavelenght optics of heterogeneous structures.

Nonlinear electromagnetics of heterogeneous media presents nowadays one of the dominating frontlines in optoelectronics, gradient optics and plasma-assisted technologies. The interplay of heterogeneity and nonlinearity provides a richness of promising applications, hampering, on the other hand, the theoretical analysis of physical fundamentals of these phenomena. Therefore exactly solvable models can be rarely found in these researches, especially for two- and three-dimensional configurations. One can mention only some exceptional cases such as, e.g., the model of self-focusing of axisymmetrical light beam, traveling across the lens-like Kerr medium, which was reduced in [125] owing to the special coordinate transform to a customary non-linear Schrödinger equation (NLSE). Herein the similar problem for an another geometry-self-focusing of wave beam, traveling in heterogeneous Kerr medium along the gradient of refractive index, e.g., in the heterogeneous plasma, remains the subject of numerous computer simulations [126]. Compared with these macroscopic problems, admitting often WKB-like approximations, the nonlinear optics of subwavelength heterogeneous layers is even more complicated, since such simplifications of basic equations become invalid.

Nonlinear optical properties of heterogeneous media have drawn a special attention due to successes in fabrication of composite materials, made from small grains of semiconductor or metal particles, dispersed in a dielectric matrix. These materials possess the possibility to increase the local electric field inside the particle due to difference between the dielectric properties of the particle and the host material. Moreover, the metal coated inclusions can enhance the local field near the surface plasmon resonance. Both of these effects can provide enormous enhancements of optical nonlinearity of composite material due to optimization of volume fractions of constituents as well as the nanometer scale size and shape of inclusions. To find the largest nonlinear effects one has to analyze the following problems:

(1). Single particle properties. The nonlinear dielectric function of inclusion $\varepsilon_s$ is taken to be the saturable form of two-level system [127]

$$\varepsilon_s = \varepsilon_\infty + \frac{\beta(i+\delta)}{1+\delta^2 + |E|^2/I_{sat}} \; ; \; \delta = \frac{\omega - \omega_0}{\Gamma} \qquad (6.1)$$

Here $\omega_0$ and $\Gamma$ are resonant frequency and linewidth, respectively $\beta$ is the dipole oscillation strength, $I_{sat}$ is the saturation intensity. This model can be used for calculation of dipole moments for dielectric inclusions of different geometry, such as circular cylinders, elliptical cylindres, spheres, concentric spheres [128]. The dipole moments of one parameter family of stratified confocal ellipsoidal layers $x^2/a^2 + y^2/b^2 + z^2/c^2 = \aleph^2$ ($\aleph \leq 1$) and potentials of dielectric ellipsoids with three-dimensional Ferrers distributions of $\varepsilon(x,y,z)$, containing 3 free positive parameters ($\alpha, \beta, \gamma$),

$$\varepsilon(x,y,z) = \varepsilon_i \left(\frac{x}{a}\right)^\alpha \left(\frac{y}{b}\right)^\beta \left(\frac{z}{c}\right)^\gamma \qquad (6.2)$$

that are expressed in [129] via complicated elliptical functions. It has been known since long ago that anisotropic shaping of amorphous silicate nanoparticles can be induced by MeV ion irradiation [130]. This effect has been since then the subject of many studies, and among them on Ag nanocrystals embedded in a silica glass [131] (as well as on other metallic nanoparticles : Co, Au), and on photonic-bandgap crystals made of spherical $SiO_2$ and ZnS –core- $SiO_2$-shell nanoparticles [132].

The strong enhancement of nonlinear response of nanoparticles can be achieved by means of plasmon resonance in the compound microstructures, containing nonlinear semiconductor core and metal shell or, vice versa, metal core and non-linear shell. The particles are much smaller than the wavelength, but still large enough to be considered as a bulk material, i.e., not smaller than a few nanometers in diameter.

(2). Effective-medium model. If each of the inclusions was treated as a dipole, their contributions summed up, and the net polarization found for the medium, one could determine the effective dielectric function of composite material $\varepsilon_{eff}$. Considering the incidenting wavelength $\lambda$ to be much larger than the characteristic size of microparticles d, one can calculate the polarization of one microparticle by means of quasistationary Rayleigh approximation. In this limit the polarization is described by Laplace equation; whose solutions yields the expression for function $\varepsilon_{eff}$ via the dielectric functions of microstructure material $\varepsilon_s$ and the host medium $\varepsilon_m$ [133]

$$(1-p)\frac{\varepsilon_m - \varepsilon_{eff}}{2\varepsilon_{eff} + \varepsilon_m} + p\left(\frac{\varepsilon_s - \varepsilon_{eff}}{2\varepsilon_{eff} + \varepsilon_s}\right) = 0 \qquad (6.3)$$

where p is the volume fraction of inclusions. In the problems discussed below, the values of p are often small (p <<1) and, thus, factor (1- p) in Eq. (6.3) can be replaced in such cases by unity.

(3). Nonlinear wave propagation in composite material. Analysis of power dependent reflectance and transmittance of composite layers, designed often as the subwavelength elements in optical circuitry, has to be performed in the framework of full wave equations; to obtain a quantitative result numerical methods are usually in need.

The nonlinear effects considered in this Section, displayed in both threshold, e.g., bistability, and thresholdless regimes, such as the waves conversion or mechanical pressure of EM field, can also arise in homogeneous media. However, as it is shown below, the heterogeneity of dielectric properties of medium enriches these effects by a multitude of new features, promising for optoelectronic devices.

*6.1. Enhanced bistability of graded metallo-dielectric films.*

Metallo-dielectric composites attract attention due to possibility of enormous enhancements of their nonlinear properties. These systems may be of practical importance in design of new optoelectronic materials and signal processing devices, since their nonlinear response is controlled by the volume fraction of small metallic or semiconductor particles, dispersed in a dielectric matrix, as well as by their shapes. For the composites, containing the metal coated particles, a strong enhancement of the local field can be provided by the surface plasmon resonance.

This resonant effect, reported for composite medium, consisting of silver nanoparticles, embedded in a silica matrix [134], stimulated a series of researches of strong nonlinearity of coated microspheres [135]. These researches revealed a diversity of manifestations of surface plasmon resonances, influenced by the structure and geometry of particles, such as, e.g., spherical and ellipsoidal particles [136], thickness of shell [137], metallic core and dielectic shell [138] or, vice versa, dielectric core and metallic shell [139].

The anisotropically shaped dielectric (metallic) particles, embedded into a metallic (dielectric) component with a compositional or shape-dependent graded profile, e.g., colloidal ellipsoids with continuously varying shapes [140], can provide new properties for these materials. To illustrate these properties let us consider a metallo-dielectric composite layer with a variation of volume fraction of anisotropic particles along the z-axis, normal to the layer; herein the effective dielectric susceptibility of this film is coordinate-dependent $\varepsilon_{eff} = \varepsilon_{eff}(z)$. By denoting the volume fraction of dielectric inclusions, embedded into the metallic component, as p(z), (p << 1) one can obtain $\varepsilon_{eff}(z)$ from the first kind of Maxwell-Garnett approximation [141]:

$$\frac{\varepsilon_{eff}(z)-\varepsilon_1}{L_z^{(2)}\varepsilon_{eff}(z)+\left[1-L_z^{(2)}\right]\varepsilon_1} = p(z)\frac{\varepsilon_2-\varepsilon_1}{L_z^{(2)}\varepsilon_2+\left[1-L_z^{(2)}\right]\varepsilon_1} \qquad (6.4)$$

where $L_z^{(2)}$ is the depolarization factor of dielectric particles along z-axis; $\varepsilon_2$ ($\varepsilon_1$) stands for the dielectric constant of dielectric (metallic) particles. Alternatively, the profile $\varepsilon_{eff}(z)$ for the metallic particles, embedded into the dielectric host, can be determined by means of Eq. (6) using the replacements $\varepsilon_1 \Leftrightarrow \varepsilon_2$, $L_z^{(2)} \to L_z^{(1)}$ and p(z) $\to$ 1 – p(z); where $L_z^{(1)}$ is the depolarization factor of metallic particle along z-axis. The values of $L_z$ <1/3, = 1/3 or > 1/3 indicate that the inclusions have the form of prolate spheroid, sphere or oblate spheroid respectively.

The prolate spheroidal metallic nanoparticles can be formed in a dielectric host by means of MeV ion bombardment, which is providing simultaneously the orientation of these particles along the direction of ion beam [131]. This graded metallo-dielectric composite is characterized by plasmon resonance peak and broad plasmon band. A large enhancement of nonlinearity occurs, when the electric component of optical wave is parallel to the direction of gradient of p(z), coinciding, in its turn, with the direction of ion beam, forming the structure [142].

To illustrate the salient features of coated compound particles, let us consider the coated spheres with a spherical metal core with radius $a_1$ and dielectric constant $\varepsilon_c$, surrounded by concentric spherical nonlinear dielectric shell of radius $a_2 > a_1$, dielectric constant $\varepsilon_s$ and nonlinear susceptibility $\chi_s$; when the spheres are distributed randomly in a host dielectric of $\varepsilon_m$. The dielectric function of composite structures is described usually by complicated formulae even in a linear limit [143]. Thus, introducing the dimensionless parameter y so, that $a_1 = a_2 y^{1/3}$, one can present the effective linear response of a dielectric, containing small volume fraction p << 1 of coated spheres, embedded into a host medium, in a form [144]:

$$\varepsilon_{eff} = \varepsilon_m \left[ 1 + 3p \frac{\varepsilon_s - \varepsilon_m + xy(\varepsilon_m + 2\varepsilon_s)}{\varepsilon_s + 2\varepsilon_m + 2xy(\varepsilon_s - \varepsilon_m)} \right] ; x = \frac{\varepsilon_c - \varepsilon_s}{\varepsilon_c + 2\varepsilon_s} \qquad (6.5)$$

Here x is the dipolar factor, relating the core and shell materials. The effective nonlinear response $\chi_{eff}$, obtained in the aforesaid Rayleigh approximation, is proportional to the volume fraction of spheres p and nonlinear susceptibility of their shells $\chi_s$:

$$\chi_{eff} = p\chi_s F(\varepsilon_c, \varepsilon_s, \varepsilon_m) \qquad (6.6)$$

where F is a complicated fractional algebraic function [138].

Using formulae (6.5) and (6.6), one can optimize the composite parameters, providing the maximum value of nonlinear response. For a metallic core the value $\varepsilon_c$ is complex with a negative real part. A good candidate for core material is silver, since the small imaginary part of its dielectric constant results in sharpening of the surface plasmon resonance and helps lower the bistability threshold. Taking, e.g., for the wavelength $\lambda = 386.7$ nm the value $\varepsilon_c = -10 + 0.037i$, $\varepsilon_s = 2$, $\varepsilon_m = 64$, p = 0.02, one can examine the dependence of nonlinear response $\chi_{eff}$ (6.6) on the shell thickness, determined by parameter y [144]. This thickness determines the conditions of excitement of plasmon resonance in the core-shell structure. Both the real and imaginary parts of $\chi_{eff}$ were shown to possess a sharp resonant maximum under the conditions discussed at y = 0.52 [138]. Herein the large host dielectric constant $\varepsilon_m$ = 64 proves to be essential for strong nonlinear response; such a large value of $\varepsilon_m$ may be realized by use of the CuCl host material near by the biexcitonic transition [145].

The power-dependent transmittance of a film, made from this composite material, can be investigated by means of nonlinear wave equation, governing the electric component of the EM wave. The numerical solution of this equation for the normal incidence

$$\frac{d^2 E}{dz^2} + \frac{\omega^2}{c^2}\left(\varepsilon_{eff} + \chi_{eff}|E|^2\right)E = 0 \qquad (6.7)$$

reveals the bistable behavior of reflectance and transmittance near by the aforesaid maxima of nonlinear Kerr response for the film of subwavelength thickness d = $0.4\lambda$, where $\lambda = 2\pi c/\omega\sqrt{\varepsilon_m}$. Thus, the hysteretic jumps of transmittance, increasing from $|T|^2$ = 0.3 up to $|T|^2$ = 0.8 and decreasing from $|T|^2$ = 1 down to $|T|^2$ = 0.2, were shown to occur due to variations of radiation intensity in a narrow range around the threshold of bistability value $|E_c|^2$ [138]. Owing to local amplification of the field near by surface plasmon resonance this value $|E_c|^2$ for composite structure, containing the coated spheres, proves to be about four orders of magnitude smaller than the bistability threshold for the homogeneous dielectric film [146].

Side by side with the dielectric parameters of composite constituents, the crucial influence on its nonlinearity is stipulated by the distribution of inclusions and their shapes. Unlike the spherically shaped inclusions, the resonant plasmon bands for anisotropically shaped metallic nanoparticles are split for orientations along major and minor axes [147]; moreover, such particles can show a reduced plasmon relaxation time [148]. In contrast to the model discussed above (metallic core and the dielectric shell), analysis [139] is devoted to the dielectric spheroidal core and confocal metallic shell, characterized by an arbitrary ratio of the minor to major axes. Analysis of dielectric function $\varepsilon_{eff}$ for oblate spheroids with symmetry axis aligned along the external electric field shows that for nonlinear core material (CdS), coated by silver shell and embedded into silica glass, this function exhibits an intrinsic bistability and discontinuous jump from the lower branch of dependence $\varepsilon_{eff}$ upon the wave intensity I to its upper branch. The switching threshold for this bistability was found to be as small as 12 W/cm$^2$. This figure represents a reduction about six orders of magnitude over the threshold intensity of uncoated silver particles.

Formation of bistability in a more complicated three-dimensional configuration may be illustrated by doubly doped monomode optical fiber with circular cross section. One of the dopants is assumed to be homogeneously distributed over the fiber core, meanwhile the second has an inhomogeneous radial distribution over the cross section $0 \leq r \leq a$. The nonlinear part of electric displacement $D_{NL}$ for inhomogeneous core may be written as [149]

$$D_{NL} = \left[\varepsilon_2^{(1)} + \alpha \varepsilon_2^{(2)} f(r)\right] \frac{|E|^2 E}{1 + |E|^2 / I_s} \qquad (6.8)$$

Here $\varepsilon_2^{(1)}$ and $\varepsilon_2^{(2)}$ are the nonlinear permittivities for the first and second dopant, respectively, $I_s$ is the saturation intensity, which is supposed, for simplicity, to be equal for both dopants, f(r) stands for the radial dependence of concentration of the second dopant

$$f(r) = \frac{r^2}{r_0^2} \ (0 \leq r < a) \ ; \ f = 0 \ (r \geq a) \qquad (6.9)$$

$\alpha$ stands for its concentration, relative to the first. It is important that the second dopant is assumed to be defocusing. Substitution of (6.8) into Maxwell equations, using of the customary slowly varying envelope approximation and averaging the nonlinearity (6.9) over the fiber cross section yields the evolution equation, governing the pulse propagation in this doubly doped fiber. The numerical analysis of this equation, similar to the nonlinear Schrödinger equation, shows that the pulse energy W may become a multivalued function of propagation constant $k_n$, satisfying to the criteria of bistability [133]. Herein the lower and upper branches of S-like nonlinear dependence $W = W(k_n)$ corresponds to the bistable soliton regime, while the intermediate branch corresponds to the cnoidal waves [149].

These power threshold phenomena provide the physical basis for design of miniaturized flexible all-optical switches and power limiters. The flexible choice of nonlinearity and heterogeneity parameters is shown below to be promising also for using in thresholdless small scale power-dependent systems.

*6.2. Conversion of waves in subwavelength dielectric layers.*

The rapid development of laser optics over the past decades gave rise to a growing interest into nonlinear dynamics of EM waves in continuous media. Side by side with the cubic $\left(\chi^{(3)}\right)$ nonlinearities appearing, e.g., in soliton phenomena, the attention was focused on quadratic $\left(\chi^{(2)}\right)$ nonlinearities providing, in particular, three-wave mixing, optical rectification and second harmonic generation (SHG). The tendencies of such processes are determined by the dispersive properties of interacting waves. Initially $\chi^{(2)}$- effects in merging and decay of waves were treated in the framework of a simple 1D model of phase-matched wave processes in an unbound dielectric medium [150]. Unlike these classical results we will consider some peculiarities of wave conversion phenomena, associated with heterogeneous media.

The recent investigation of frequency doubling has shown that the effectiveness of SHG in a photonic bandgap material, doped with $\chi^{(2)}$ medium, may be essentially larger than the SHG effectiveness stipulated by an equal length of phase-matched bulk homogeneous material [151]. In a periodic structure [152] the pump pulse propagates through the band gap device, formed by dielectric layers with alternating high and low values of refractive index. Herein the strong confinement of both pump and SH wave, occurring near the photonic band edge, where the density of EM modes is large, results in the enhancement of SHG effectiveness by 2–3 orders of magnitude [153]. Such systems, providing the replacement of mm- or cm- long phase-matched devices by layered structures of only few micrometers in length, are promising for Raman-type lasers and frequency up- and down-conversion schemes. Moreover, the power-dependent effects in periodic structures, such as the modulational instabilities in optical fibers with periodical variations of nonlinearity or dispersion [154], are in the focus of studies of stability of long distance transmission system, influenced by amplifiers or dispersion management. Unlike this periodic modulation, the self-action of waves in media with non–periodic (linear and exponential) continuous modulation of refractive index in the direction of wave propagation was examined in [155]. The dynamics of waves in nonlinear media with an imprinted transversal modulation of refractive index is described in [156].

A traditional object for researches of nonlinear polarization effects is connected with electron gas in solid plasmas. The heterogeneity-induced nonlinear phenomena open new trends in these researches, such as the enhancement of optical nonlinearity of graded metal films was reported in [157]. Semiconductors of $A^{III} - B^{V}$ groups with gradient subsurface distributions of free carriers were shown to support the propagation of S-polarized surface waves [158], admitting four-waves conversion and formation of both bright and dark infrared solitons of these waves [159].

Another peculiar mechanism of SHG in subsurface area can occur, when p-polarized light is incidenting obliquely on the boundary of centrosymmetric semi-infinite medium [160]. The medium is assumed to consist of electric dipoles, whose density is varying continuously across the surface, tending to a constant value in the bulk. The origin of nonlinearity in this model is the spatial variation of normal component of electric field across each dipole. This variation gives a small contribution of order $\lambda/a$ in the bulk, where $\lambda$ is the size of each polarizable entity and $\lambda$ is the wavelength. However, the normal component of p-polarized electric field is varying rapidly in the subsurface area in a distance smaller than $\lambda$, resembling its formal discontinuity on an abrupt boundary. This variation creates a sizeable surface nonlinear polarization.

To examine this effect of the field heterogeneity, let us consider a single charge e with mass m at a distance z from its equilibrium position $z_0$, to which it is bound by a harmonic force with resonant frequency $\omega_0$. Considering, for simplicity the one-dimensional model, one can describe the motion of this charge, governed by spatially varying normal component of electric field E(z,t), by equation

$$m\frac{d^2z}{dt^2} = -eE(z,t) - m\omega_0^2 z \qquad (6.10)$$

where the dissipative term and the Lorentz force are neglected. The field has to be evaluated at the actual electron position $z_0 + z$ ($|z| \ll |z_0|$):

$$E(z,t) = E(z_0,t) + z\frac{dE}{dz} + ... \qquad (6.11)$$

The higher order terms of Taylor expansion (6.11) have no effect in the quadratic nonlinear response discussed. Expanding the solution of Eq. (6.10) in powers of $E = E_\omega \exp(-i\omega t)$ and putting $z = z_1 + z_2 + ...$, $z_1 = \text{Re}[z_1\exp(-i\omega t)]$, one obtains from Eq. (6.10) the linear approximation of the induced electric dipole moment $p_\omega = -ez_1$:

$$p_\omega = \alpha(\omega)E_\omega; \quad \alpha(\omega) = \frac{e^2/m}{D(\omega)} \; ; \; D(\omega) = \omega_0^2 - \omega^2 \tag{6.12}$$

Proceeding in a similar fashion, one can find from Eq. (6.10) the second-order displacement $z_2$; the second-order dipole moment, oscillating with the double frequency $p_{2\omega} = -ez_2$, may be written as:

$$p_{2\omega} = -\frac{1}{2e}\alpha(\omega)\alpha(2\omega)\nabla E_\omega^2 \tag{6.13}$$

We note that in the geometry discussed there is an electric quadrupole moment $Q_{2\omega} = -ez_1^2$, oscillating with frequency $2\omega$:

$$Q_{2\omega} = -\frac{1}{e}\alpha^2(\omega)(E_\omega)^2 \tag{6.14}$$

Introducing the coordinate-dependent density of dipoles per unit volume n(z), and making use of (6.13) and (6.14), we can find the second-order polarization per unit volume

$$P_{2\omega} = np_{2\omega} - \frac{1}{2}\nabla(nQ_{2\omega}) \tag{6.15}$$

This polarization, which is distinct from zero only in a thin subsurface layer of centrosymmetric media, permits one to find the efficiency of SHG induced by p-polarized wave. Referring the reader for the details of cumbersome calculations to the article [160], we note that this effect is particularly interesting in the spectral range $\Omega_p^2 + \omega_0^2 > \omega^2 > \omega_0^2$, where the bulk material is opaque ($\Omega_p^2 = 4\pi e^2 n_s /m$, $n_s$ is the bulk value of electron density). In this range the reflectance of waves with frequency $2\omega$, generated at the surface, becomes an important tool for non-destructive surface control [161]. The microscopic crystalline structure was ignored in this analysis; however, the SHG effects from surface crystallized glasses were observed for glasses, containing $LiNbO_3$ [162] and $LiTaO_3$ [163].

The nonlinear wave phenomena considered above are dependent on the rapid variations of fields over subwavelength distances. Mechanical manifestations of the same phenomena are examined below.

*6.3. Optomechanical effects in heterogeneous optical fields.*

An external laser illumination is known to create forces between closely located dielectric or metallic spheres due to interactions between the induced dipole moments of spheres, when the attractive or repulsive action of these forces depends upon the relative phase shift of induced dipoles [164-166]. Physically similar effects in trapping [167], manipulating [168] and propelling [169] of nanoparticles were also reported. Moreover, the mechanical forces can arise due to internal illumination of transparent samples, e.g., due to interaction of two coupled parallel optical waveguides, carrying one trapped mode; a fraction of energy of this mode is traveling in the air between waveguides. The spatial heterogeneity of the mode field gives the rise to attracting and repulsive forces, acting on waveguides. This optomechanical interaction was examined in [170] for the configuration composed of two parallel silicon strip waveguides with length L, separated by distance d; each waveguide has a square-shaped cross section with size a. The eigenfrequency of trapped mode $\omega$ depends upon the distance d, and small adiabatic variation of d (d =$d_0$ +x, |x| << $d_0$) results in variations of $\omega$ and, respectively, in variation of mode energy per unit length U=N$\hbar\omega$. Here N is the amount of light quanta per unit length, and $\hbar$ the Plank's constant. Expressing U via the total power transmitted through the coupled waveguides P and group velocity $v_g$ as $P = Uv_g/L$, one can present the force of waveguides interaction F per unit length as:

$$F = -\frac{dU}{dx} = -\frac{1}{\omega}\frac{d\omega}{dx}\frac{P}{v_g} \qquad (6.16)$$

Negative (positive) values of F correspond to attractive (repulsive) forces and the character of the force depends upon the relative separation $d_0/a$ and upon the symmetry of trapped mode [170].

Analysis of the interaction of parallel waveguides, resting on a substrate, with a free standing sections of length L, reveals the possibility of controlled small deflection w ( w << L) of waveguides due to this attractive force. This deflection is governed by bending equation of an elastic rod:

$$GI\frac{d^4w}{dx^4} = -F(w) \qquad (6.17)$$

Here G is Young modulus, I is the moment of inertia of waveguide cross–section; for the square-shaped cross section (size a) $I = a^4/12$.

Consideration of Si waveguides ($d_0$ = 46,5 nm, a = 310 nm, L = 30 $\mu m$, G = 169 GPa) for P = 100mW and $\lambda$ = 1550 nm brings the maximum value of deflection at the centre of free–standing sections $w_m$=20 nm [170].

A perspective to enhance this effect is connected with the use of mechanical resonances in ocillations of coupled rods [171]. Another chance to enhance the force F is shown by $1/v_g$ dependence of F given by Eq. (6.16), the power P being given [172]. From this viewpoint it is interesting to refer to the article [173], describing the mechanical interaction between an EM field and thin semiconductor films near by the electron resonance.

Another manifestation of optomechanical force can be connected with a peculiar mechanism of nonlinearity of dipole dielectrics, independent upon the nonlinear polarizability of dipoles. To visualize this mechanism, let us consider a simple model: a monochromatic plane wave $E_0$ is incidenting normally on the interface of a lossless dielectric homogeneous layer with thickness d and refractive index n. The power reflection coefficient $|R|^2$ was given in (3.9) :

$$|R|^2 = \frac{t^2(n^2-1)^2}{t^2(n^2+1)^2 + 4n^2} \; ; \; t = tg\left(\frac{n\omega d}{c}\right) \tag{6.18}$$

The energy density W inside the layer can be expressed via the incident energy flux $P = c|E_0|^2/4\pi$, as

$$W = \left(\frac{2n^2(1+n^2)}{4n^2 + t^2(1+n^2)^2}\right)\frac{P}{c} \tag{6.19}$$

The maximum value W for the given flux P is attained under the condition $t = 0$ (in this case the reflection is vanishing, $R = 0$) :

$$W_m = \left(\frac{1+n^2}{2}\right)\frac{P}{c} \tag{6.20}$$

The value of W for the incident beam is $W_i = P/c$. In this case ($R = 0$) there is no external force of light pressure acting on the layer; however there are internal forces, produced by standing wave inside the layer. The density of energy is known to determine the field pressure. Herein the difference between $W_m$ and $W_i$ gives the pressure F, acting normally on the interfaces

$$F = \left(\frac{n^2-1}{2}\right)\frac{P}{c} \qquad (6.21)$$

In a case, e.g., of polar dielectrics (n > 1), the internal pressure F tends to increase the thickness d; its relative variation $\delta$ is defined by Young module G : $\delta = F/G$. This deformation results in the decrease of the density of dipoles inside the layer and, respectively, in decrease of its polarizability per unit volume $\chi$: $\Delta\chi = -\chi\delta$. The change of dielectric susceptibility $\Delta\varepsilon$, linked with the polarizability ($\varepsilon = 1 + 4\pi\chi$), can be written by means of Eqs. (6.20) – (6.21):

$$\Delta\varepsilon = 4\pi\Delta\chi = -4\pi\chi\left(\frac{n^2-1}{2}\right)\frac{P}{cG} \qquad (6.22)$$

Using $\Delta\varepsilon = 2n\Delta n$, $\Delta n = n_2 P$, where $n_2$ is the nonlinear part of refractive index, one obtains the value $n_2$, produced by optomechanical effect

$$n_2 = \frac{-\pi(n^2-1)^2}{ncG} \qquad (6.23)$$

Let us compare this nonlinearity with the Kerr nonlinearity described e.g., for glass by the value $n_2 = 3 \cdot 10^{-16}$ cm$^2$/W; taking for glass n = 1.5, Young's module G = 50– 80 GPa, one obtains $n_2 = -(1.4–2.2) \cdot 10^{-16}$ cm$^2$/W. To emphasize the physical basis of the effect discussed, we were treating an homogeneous layer. In the case of a gradient layer, characterized by profile U(z), the spatial distribution of forces inside the layer depends upon the gradient dU/dz. Thus, for the convex profile U(z) the additional force F = - dU/dz, providing an additional stretch of layer, tends to increase the optomechanical nonlinearity.

The salient features of this effect, distinct from the other mechanisms of nonlinearity, are:

(1). Media n >1 possess a "defocusing" opto-mechanical nonlinearity; its magnitude may become comparable with Kerr nonlinearity.

(2). The nonlinearity in question is relevant to any material, including those which do not support a cubic Kerr effect.

(3). The smaller is the value of Young module G, the stronger will be the optomechanical nonlinearity. Thus, one has to deal with the problem of finding the suitable soft materials; one can seek reasonable candidates between plastics or polymers. Search of deformable materials with high refractive index n, like, e.g., CuCl (n $\approx$ 8) may also turn out to be useful.

Considering here the EM wave pressure, we are dealing, in fact, with the part of this pressure, averaged in time with respect to the frequency $\omega$. Herein the nonlinear variations of polarization with frequency $2\omega$ could provide a source of frequency doubling; however, this problem still remains opened.

## 7. Conclusion: some more about the evanescent modes in gradient structures.

In summary, it should be noted, that the problems of non-local dispersion of electromagnetic waves, implied by the heterogeneity of dielectric media, have been considered here for simple geometries, when the gradient of refractive index of plane photonic barrier is normal to the barrier interface; the direction of wave propagation can be parallel, oblique or normal to this gradient. The unified approach to this problem, based on the use of new variables – the phase coordinates – allowed the construction of exact analytical solutions of Maxwell equations for broad classes of coordinate - dependent continuous distributions of dielectric susceptibility; herein the spatial structure of continuous wave (CW) fields is visualized in some cases by means of elementary functions. The flexibility of these models stems from the existence of several free parameters, characterizing the heterogeneity scale lengths.

Tunneling is known to be one of the fundamental phenomena in the dynamics of waves of different physical nature. The formal approach, developed so far for optical problems, can be pursued for the other EM fields and, in particular, for tunneling modes in other spectral ranges. Dynamics of evanescent optical modes in gradient films, arising in media with real values of refractive index n, but negative spatial derivatives of n, was shown to distinguish itself by important peculiarities, e.g., by non-attenuative energy transfer through such films. This property, following from exact analytical solutions of Maxwell equations, can be scaled to other spectral ranges of EM waves, in particular, to microwaves, forming the parallel branch of gradient optics - the gradient radiooptics. Some examples of such scaling are illustrated below.

*7.1. Layered structures vs. split- ring resonators.*

An important achievement of electrodynamics of metamaterials is the creation of artificial media with negative magnetic permeability $\mu < 0$ at high frequencies of EM wave. This intriguing property may be provided, in particular, by a periodical array of split–ring microresonators (SRR), embedded in a dielectric matrix [118]; the value of $\mu$ for such a structure in the lossless limit is given by formula (5.32) in which the resonant frequency $\omega_0$ and dimensionless parameter F ($0 < F < 1$) depend upon the geometric parameters of the structure. In the low-frequency range $\omega < \Omega_m = \omega_0 \sqrt{1-F}$ the magnetic field, induced in SRR, and the external magnetic field have the opposite directions; this configuration relates to the condition $\mu < 0$. Such systems, developed for microwave [104], THz [117] and infrared [174] ranges, open new possibilities for spectral and directional control of radiation.

In this conjunction one can outline the investigation on multilayer structures, intended for formation of unusual direction diagrams of radiative systems, based on the excitation of surface polaritons on the interfaces of neighboring layers, with the condition $\varepsilon\mu < 0$ being fulfilled in each of them. In a simplest case of double layer medium TE-polaritons are excited on the boundary of contiguous layers 1 ($\varepsilon_1 < 0, \mu_1 > 0$) and 2 ($\varepsilon_2 > 0, \mu_2 < 0$) [175]. Layer 1 is fabricated from a material with free carriers, e.g., from a metal ($\varepsilon_1 = 1 - \Omega_p^2/\omega^2$, $\Omega_p$ is plasma frequency) and layer 2 from the metamaterial (5.32). The dispersive equation for such waves, induced by the plane electromagnetic wave, incidenting on the layer 1 under an angle $\theta$, may be written as [176]:

$$\frac{\mu_2}{l_1} + \frac{\mu_1}{l_2} = 0 \tag{7.1}$$

Here $l_{1,2}$ are the e-folding attenuation lengths for waves in media 1 and 2; since $l_{1,2} > 0$, eq. (7.1) can have solutions if one of the values $\mu_{1,2}$ is negative. Substitution of values $l_{1,2}$ for homogeneous layers

$$l_{1,2} = \frac{c}{\omega\sqrt{\sin^2\theta - (\varepsilon_{1,2})(\mu_{1,2})}} \quad (7.2)$$

into Eq. (7.1) yields in a simple case $\mu_1 = 1$, $\varepsilon_2 = n_0^2$ the dispersive equation for surface TE wave:

$$\left(1 - \frac{F\omega^2}{\omega^2 - \omega_0^2}\right)\sqrt{\frac{\Omega_p^2}{\omega^2} - \cos^2\theta} + \sqrt{\sin^2\theta - n_0^2\left(1 - \frac{F\omega^2}{\omega^2 - \omega_0^2}\right)} = 0 \quad (7.3)$$

Equation (7.3) has a non–zero root even in a case $\theta = 0$; thus, surface TE polaritons can be excited even by the normal incidence of EM waves on the two–layer structure, containing the layer with $\mu < 0$.

It is remarkable that two-layer structures can support the propagation of surface TE waves in a case where the metamaterial with $\mu < 0$ is replaced by the gradient layer of non-magnetic dielectric $(\mu = 1)$, if its dielectric susceptibility, while remaining positive, is decreasing in the depth of this layer according to profile (4.40). Following the analysis of surface waves on the air–gradient dielectric interface (chapter 4.3) and substituting the attenuation length (7.2) into the condition of continuity of components $E_y$ (4.48), one obtains the dispersive equation for TE surface waves in the gradient two-layer structure:

$$\frac{L\omega}{c}\sqrt{\frac{\Omega_p^2}{\omega^2}-\cos^2\theta}=\frac{1}{2}\left(1-\sqrt{1-\frac{\omega^2}{\Omega_c^2}}\right) \qquad (7.4)$$

This structure with $\mu_2 = 1$, as well as the structure with $\mu_2 < 0$, admits the excitation of TE surface waves on the boundary between layers in a case of normal incidence $\theta = 0$ ($k_y = 0$); thus, for the critical frequency of transition layer (4.40) $\Omega_c = 3 \cdot 10^{15}$ rad s$^{-1}$, corresponding to the layer parameters L = 100 nm, $n_0$ = 1.65, $n_v$ = 1.57, and plasma frequency $\Omega_p = 3.185 \cdot 10^{15}$ rad s$^{-1}$, one can find from (7.4) the frequency of excited TE waves $\omega = 2.945 \cdot 10^{15}$ rad s$^{-1}$. This frequency relates to the wave of visible range $\lambda$ = 640 nm, herein the characteristic thickness of transition layer remains subwavelength: $L < \lambda$.

Thus, the gradient dielectric layers can possess some properties of SRR arrays with $\mu < 0$. The high-frequency magnetism, obtained by means of SRR, was reported for GHz [104] and THz [117] ranges. However, the direct scaling of these systems to visible optics would require a structure with critical features controlled at a level of about 10 nm, which is technologically challenging. The nanofabricated media with $\mu < 0$ for visible frequencies [177] contains electromagnetically coupled pairs of gold dots. The subwavelength scales of the gradient structures (7.4) as well as the absence of expensive gold details attract attention to the potential of gradient components in metamaterials, possessing a negative magnetic response at high frequencies.

*7.2. Tunneling phenomena in transmission lines with continuously varying parameters.*

The model of transmission line (TL) is used traditionally for description of wave processes of different physical nature, e.g., in electronics, plasma physics, communications theory [33], [109]. A recent development of this model is connected with the lines with continuously distributed parameters, characterized by capacity C or inductance L per unit length. For a strip line with strips width a and spacing between strips d, the quantities C and L are [33]:

$$C = \frac{\varepsilon_0 \varepsilon a}{d} \; ; \; L = \frac{\mu_0 \mu d}{a} \qquad (7.5)$$

Here $\varepsilon_0 = 8.85 \cdot 10^{-12} \, F/m, \mu_0 = 4\pi \cdot 10^{-7} \, G/m$ - dielectric susceptibility and magnetic permittivity of vacuum, $\varepsilon$ and $\mu$ - the relative values of these parameters for the medium. If the quantities C and L keep the constant values $C_0$ and $L_0$ along the line, the wave velocity $v_0$ and impedance of the line $Z_0$ are:

$$v_0 = \frac{1}{\sqrt{L_0 C_0}} \; ; \; Z_0 = \sqrt{\frac{L_0}{C_0}} \qquad (7.6)$$

Spatial distributions of current I and voltage V along the lossless TL with continuously distributed parameters is known to be governed by the pair of equations:

$$\frac{\partial V}{\partial z} + L \frac{\partial I}{\partial t} = 0 \; ; \; \frac{\partial I}{\partial z} + C \frac{\partial V}{\partial t} = 0 \qquad (7.7)$$

Let us consider in the framework of eq. (7.7) the case where the distributions of capacity or/and inductance are heterogeneous. Starting from the simple case where, in some segment of TL ($0 \leq z \leq d$), the distribution of C is coordinate-dependent, while the value L remains invariable, we examine the case:

$$C(z) = C_i U^2(z) \; ; \; U(0) = U(d) = 1 \; ; \; L = L_0 \qquad (7.8)$$

Introducing the generating function $\Psi$ by correlations

$$V = -L_0 \frac{\partial \Psi}{\partial t} \quad ; \quad I = \frac{\partial \Psi}{\partial z} \qquad (7.9)$$

one obtains an equation, coinciding after the replacement $c/n_0 \to v_0$ with the wave equation (2.6). Herein the presentation (7.9) for V and I are analogous to presentation (2.5) for electric and magnetic components of EM field.

Continuing this analogy, one can examine a strip line with concave profile of capacity (7.8), formed, e.g., by smooth variation of $\varepsilon(z)$, the sizes a and d being invariable: in this case the heterogeneous segment of strip line will be characterized by cutoff frequency $\Omega$, dependent upon the parameters of profile U(z) and velocity $v_0$:

$$\Omega = \frac{2v_0 y\sqrt{1+y^2}}{d} \qquad (7.10)$$

Out of this segment the line is homogeneous, with the parameters $C_0$ and $L_0$. Tunneling of current and voltage waves through the discussed segment of TL arises in the subcritical regime $\omega < \Omega$. The evanescent modes are described by generating function $\Psi$ (4.2) with q=$\omega$ N./$v_0$, and the reflection coefficient of this segment with modulated capacity as well as the condition of reflectionless tunneling are given by formulae (4.4) and (4.7) respectively, after the replacements

$$\gamma \to \frac{2\beta yu}{\sqrt{1+y^2}} \quad ; \quad \beta = \sqrt{\frac{C_0}{C_i}} \quad ; \quad n_0^2 \to C_0/C_i \qquad (7.11)$$

Thus, tunneling of current and voltage waves in the TL with continuously distributed capacity proves to be similar to the tunneling of light in gradient optics of subwavelength layers.

This analysis covers the case where the capacity of TL is coordinate-dependent, while its inductance remains spatially unchanged. An opposite case C = const, L = L(z) can be treated analogously, and it is interesting to consider the more complicated case, when both capacity and inductance are coordinate-dependent. This problem is described by Eq. (7.7) with $C=C_i U^2(z)$, $L=L_i F^2(z)$. In this case only few exactly solvable models of heterogeneous TL were discussed; a well known example of such model is presented by "exponential horns", where both C(z) and L(z) are proportional to exp(z/l), so that the ratio C/L remains constant.

However, using the approach developed in the chapter 2.1, one can treat a more flexible problem, where the distributions of capacity and inductance along the TL are different. Introducing the generating function $\Psi$ by correlations

$$I = -\frac{\partial \Psi}{\partial t} \; ; \; V = \frac{1}{C_0 U^2(z)} \frac{\partial \Psi}{\partial z} \qquad (7.12)$$

one obtains the equation, governing the generating function $\Psi$

$$\frac{\partial^2 \Psi}{\partial z^2} + \frac{\omega^2 U^2(z) F^2(z)}{v_0^2} \Psi = \frac{2}{U} \frac{dU}{dz} \frac{\partial \Psi}{\partial z} \qquad (7.13)$$

The latter equation is similar to Eq. (2.16); introducing the new variable $\varsigma$, by analogy with (2.17), one can rewrite Eq. (7.13) as

$$\varsigma = \int_0^z U^2(z_1)dz_1 \; ; \; \frac{d^2\Psi}{d\varsigma^2} + \frac{\omega^2}{v_0^2} W^2(\varsigma)\Psi = 0 \; ; \; W^2(\varsigma) = \frac{F^2(z)}{U^2(z)} \tag{7.14}$$

Let us assume that the function $W(\varsigma)$ is one of the exactly solvable models discussed in Chapter 2; taking, e.g., the model (2.9), let us put there, for simplicity, $s_2 = 0$, $W(\varsigma) = (1 + \varsigma/l)^{-1}$. Using the formal similarity of Eq. (7.14) and (2.6), one can write the solution of (7.14) in a form, resembling (3.1) ;

$$\Psi = \frac{1}{\sqrt{W(\varsigma)}}\left[\exp(iq\xi) + Q\exp(-iq\xi)\right] \tag{7.15}$$

$$\xi = \int_0^\varsigma W(\varsigma_1)d\varsigma_1 = \int_0^z U(z_1)F(z_1)dz_1 \; ; \; q = \frac{\omega}{v_0}\sqrt{1 - \frac{\Omega^2}{\omega^2}} \; ; \; \Omega = \frac{v_0}{2l} \tag{7.16}$$

Until now the functions $U(z)$ and $F(z)$, as well as the scale l, connected with the characteristic scales of these functions, remain unknown. Choosing one of these functions, e.g. distribution of inductance $F(z)$, freely, we find from (7.14) the variable $\varsigma$ ; then, the link between functions $F(z)$ and $U(z)$, postulated in (7.14), yields the function $U(z)$. Let us consider, e.g., $F(z) = (1 + z/l_1)^{-1}$; following to this scheme of calculation, we find:

$$\varsigma = \frac{z}{1 + \frac{z}{l_1}} \; ; \; U(z) = W[\varsigma(z)]F(z) = \frac{1}{1 + z\left(\frac{1}{l_1} + \frac{1}{l}\right)} \tag{7.17}$$

Thus the capacity has to be distributed according to (7.17) as $U(z) = (1 + z/l_2)^{-1}$ and comparison of these distributions yields the unknown scale l:

$$l = \frac{l_1 l_2}{l_1 - l_2} \tag{7.18}$$

Finally, the variable $\xi$, calculated by substitution of F(z) and U(z) to (7.16), is

$$\xi = l \ln\left[1 + \frac{z(l_1 - l_2)}{l_2(z + l_1)}\right] \tag{7.19}$$

Now the wave (7.15) in the transmission line with continuously distributed inductance F(z) and capacity U(z) is defined; herein the characteristic scales of these distributions, $l_1$ and $l_2$, are the free parameters of this line. The reflectance/transmittance of this line can be found by the standard procedures.

Thus, the continuous variations of capacity/inductance distributions along transmission line can provide new properties of such lines:

(1). Frequency dispersion, controlled by distributions of capacity and inductance.

(2). Flexible possibility of matching of homogeneous and heterogeneous segments of TL.

(3). Control of the phase of wave tunneling through the transmission line without loss of power, when the phase shift depends not on the phase path length, but on the parameters of the heterogeneous segment.

Despite the formal similarity of these phenomena with the effects of gradient optics an important physical difference between such systems has to be emphasized: the refractive index of film material obeys to condition $n_0 > 1$, while the value of corresponding parameter $[C_0/C_i]^{1/2}$ in formulae (7.11) for TL can be both more or less than unity. This property provides the possibilities of flexible modeling of tunneling effects in transmission lines with continuously distributed capacity and inductance.

*7.3. Reflectionless tunneling of microwaves through the heterogeneous layer in waveguide.*

Transmittance spectra of microwave waveguides can be essentially modified by means of a thin gradient layer, located normally to the waveguide axis z. Unlike the spectral distortions, produced by varying the waveguides width [178], such modification is based on the non–local dispersion of a wave barrier formed by gradient layer; herein the waveguide cross section remains unchanged. This dispersion results in an intriguing peculiarities of propagation of trapped modes, tunneling through the barrier with concave profile of dielectric susceptibility. The underlying physics of these phenomena, treated as a frustrated total internal reflection (FTIR), can be simply visualized for $TE_{01}$ fundamental mode in the rectangular waveguide with sides a and b, a > b, dammed at the segment $0 \leq z \leq d$ by barrier with profile $\varepsilon(z)$ (2.1), where the function $U^2(z)$ is:

$$U^2(z) = 1 - \frac{1}{g} + \frac{W^2(z)}{g} \; ; \; W(z) = [\cos(z/L) + M \sin(z/L)]^{-1} \quad (7.20)$$

Three free parameters of this symmetrical profile (the dimensionless values g>1, M>0 and the scale L) are linked with the layer thickness d, mimimum value of refractive index $n_m$ and scale L of the profile $U^2(z)$ near by the interfaces z = 0 and z = d by formulae:

$$L = \frac{d}{Arc\sin\left(\frac{2M}{1+M^2}\right)} \; ; \; n_m = n_0 \sqrt{1 - \frac{M^2}{g(1+M^2)}} \quad (7.21)$$

The polarization structure of $TE_{01}$ mode is characterized by transversal components $E_x$, $H_y$ and longitudinal component $H_z$. The components $E_x$ and $H_y$ can be expressed via the generating function $\Psi$ (2.5), $H_z = -\dfrac{\partial \Psi}{\partial y}$; herein the equation $\operatorname{div}\vec{H} = 0$ is fulfilled automatically, and function $\Psi$ is governed by equation

$$\frac{\partial^2 \Psi}{\partial z^2} + \frac{\partial^2 \Psi}{\partial y^2} + \frac{(\omega n_0)^2 U^2}{c^2}\Psi = 0 \qquad (7.22)$$

Outside the barrier ($z \leq 0$ and $z \geq d$) one has $n_0 = 1$, $U = 1$; the function $\Psi$ for these parts of waveguide is known:

$$\Psi = A\sin\left(\frac{\pi y}{b}\right)\exp[i(\beta z - \omega t)] \; ; \; \beta = \sqrt{\left(\frac{\omega}{c}\right)^2 - (k_\perp)^2} \; ; \; k_\perp = \frac{\pi}{b} \qquad (7.23)$$

Let us consider the frequencies $\omega > \omega_{cr} = ck_\perp$. To find the solution of Eq. (7.22), it is convenient to introduce the new variable u:

$$u = \ln\left[\frac{1 + m_+ tg(z/2L)}{m_+ - tg(z/2L)}\right]; \; m_+ = \sqrt{1+M^2} + M \; ; \; W(z) = \frac{chu}{\sqrt{1+M^2}} \qquad (7.24)$$

With the help of variable u, the solution of Eq. (7.24) can be written as

$$\Psi = B\sqrt{\cos(z/L) + M\sin(z/L)}\sin\left(\frac{\pi y}{b}\right)F(u)\exp(-i\omega t) \qquad (7.25)$$

Quantities A and B in (7.23) and (7.25) are the normalization constants. Function F(u) in (7.25) is determined by equations:

$$\frac{d^2F}{du^2} + F\left(q^2 - \frac{T}{ch^2 u}\right) = 0 \tag{7.26}$$

$$q^2 = \left(\frac{\omega n_0 L}{c}\right)^2 \frac{1}{g(1+M^2)} - \frac{1}{4}; \quad T = L^2\left[\left(\frac{\omega n_0}{c}\right)^2\left(1-\frac{1}{g}\right) - k_\perp^2\right] - \frac{1}{4} \tag{7.27}$$

It is worth pointing out that the model (7.20) is the generalization of model (4.26), used in chapter 4 for analysis of polarization effects in gradient optics of thin photonic barriers. Namely, the model (4.26) is the limiting case, following from (7.20), when $M = 0$, $g = 1$; respectively, in this case the expressions for $q^2$ and $T_s$ (4.29), as well as for $\eta$ (4.27), follow from (7.27) and (7.24) respectively.

Equation (7.26) is well known in quantum mechanics; its solutions are given by hypergeometric functions. Let us consider, for simplicity, the limiting case $T = 0$; in this limit Eq. (7.26) reduces to a form

$$\frac{d^2F}{du^2} + q^2 F = 0 \tag{7.28}$$

When $q^2 < 0$, Eq. (7.28) describes the tunneling mode. Such a situation occurs, e.g., for $TE_{01}$ mode with frequency 10 GHz ($\omega = 2\pi 10^{10}\, rads^{-1}$) in the rectangular waveguide with sides a = 1 cm, b = 2 cm and the parameters of gradient barrier $n_0$ =1.86, g =2.25, M= 1.7, $n_m$ = 1.52, d = 0.22 cm; $q^2$ = - 0.177. Introducing the parameter $p^2 = -q^2 > 0$, one can present the solution of Eq. (7.28) as a sum of forward and backward waves, resembling the solution (4.2):

$$F = \exp(-pu) + Q\exp(pu); \quad p = \sqrt{-q^2} \tag{7.29}$$

$$Q = \exp(-2pu_0)\left(\frac{2p\sqrt{1+M^2} + M + 2i\beta L}{2p\sqrt{1+M^2} - M - 2i\beta L}\right); \quad u_0 = \ln(m_+) \tag{7.30}$$

Using the continuity conditions on the boundaries of gradient layer $z = 0$ ($u = -u_0$) and $z = d$ ($u = u_0$), one can find that the reflection coefficient for $TE_{01}$ mode from this layer can be obtained from (4.3), setting there $n_1 = 1$. Generalizing this result for m contiguous layers and using formulae (3.10) with the value q, given by (7.27), we find the reflection coefficient for the system of m similar gradient layers in the waveguide:

$$R = \frac{t_m\left[(\beta L)^2 + p^2(1+M^2) + M^2/4\right] - Mp\sqrt{1+M^2}}{t_m\left[(\beta L)^2 - p^2(1+M^2) - M^2/4\right] + Mp\sqrt{1+M^2} + i\beta L\left(2p\sqrt{1+M^2} - Mt_m\right)}$$

$$\tag{7.31}$$

$$t_m = th(2pmu_0)$$

The condition of reflectionless tunneling $R = 0$ (nullification of denominator in (7.31)) can be fulfilled, e.g., for a diaphragm consisting of two layers (m = 2) with parameters given above (M=1.7, g = 2.25, $n_0$ =1.86, $n_m$ =1.52, d = 0.22 cm). Varying some parameters of these layers ($n_0$=1.9, $n_m$=1.55, d=0.2 cm) and leaving the values M and g unchanged, we find that the condition for reflectionless tunneling $R = 0$ is fulfilled in this case for the diaphragm consisting of three layers. These cases relate to 100% transmittance of barrier in the tunneling regime, provided by evanescent modes; the phase of transmitted wave, defined by expression, following from (7.31),

$$\phi_t = Arctg\left\{\frac{t_m\left[(\beta L)^2 - p^2(1+M^2) - M^2/4\right] + Mp\sqrt{1+M^2}}{\beta L\left(2p\sqrt{1+M^2} - Mt_m\right)}\right\} \qquad (7.32)$$

possesses values close to $\pi/2$. Thus, such gradient diaphragm can be considered as a peculiar phase shifter, leaving the wave power unchanged. The total thickness of diaphragm D, containing both two or three gradient layers, remains subwavelength D < $2\pi/\beta$.

While discussing the total transmittance of waveguides with gradient barriers one can mention the measurement of transmittance spectra of guided waves, tunneling through an array of split-ring resonators inside the waveguide [179]. These spectra show the possibility of total transmittance of tunneling modes, implied by plasmon transfer of EM energy. A similar effect for photons, tunneling through the layered LH metamaterial structure, was examined in [180]. Unlike this, the aforesaid effect illustrates the perspectives of total transmission of microwaves through gradient layers, fabricated from usual nonmagnetic materials with positive values of dielectric susceptibility.

*7.4. Hartman paradox in gradient optics.*

Side by side with the above mentioned perspectives of using gradient structures in optoelectronics the phenomena of non-attenuative tunneling of EM waves through such structures may become useful for an old academical problem of great principal importance, connected with the velocity of tunneling. This problem attracted attention as long ago as in 1928, after the prominent paper of G. Gamow, devoted to the nuclear alpha - decay [181]. In this paper the probability of penetration of alpha–particle with energy E through the potential barrier with maximum $U_0$, so, that E < $U_0$, was determined namely by tunneling. Three years later the attempt of Condon [182] to calculate in the framework of tunneling theory the velocity or the flight-time of the particle in the area E < $U_0$, revealed a principal problem: how should we define these amounts in the "classically forbidden" zone, where the impulse of particle would have an imaginary values? A year later MacColl reported about the absence of any delay of a wave packet moving inside the barrier [183]. This question, considered, on the background of successes of quantum mechanics, as an "academical trick", remained opened during more than three decades.

A new burst of interest into this problem had sparked due to result of Hartman [184], where the tunneling time of the particle with energy E through a barrier was defined via the phase of complex transmission function of the barrier $T = |T|\exp(i\phi)$ by means of general formula [185]:

$$\tau_p = \hbar \frac{\partial \phi}{\partial E} \qquad (7.33)$$

Using the well known expression for the transmission function for a particle with mass m, tunneling through the rectangular potential barrier with maximum $U_0$ and width d,

$$|T| \propto \exp(-\aleph d) \;;\; \phi = Arctg\left[\frac{(2-U_0/E)th(\aleph d)}{2\sqrt{U_0/E-1}}\right] \;;\; \aleph = \frac{\sqrt{2m(U_0-E)}}{\hbar} \qquad (7.34)$$

and considering the wide barrier ($\aleph d \gg 1$), Hartman found from (7.33) and (7.34) a simple formula:

$$\tau_p = \frac{\hbar}{\sqrt{E(U_0-E)}} \qquad (7.35)$$

This result showed an unexpected property of the "phase time" of tunneling $\tau_p$: the time $\tau_p$ proved to be independent upon the tunneling path length; this path being long enough, the velocity of particle v could reach superluminal values v > c. This conclusion was entitled in the literature as "Hartman paradox".

An attempt to redefine the phase time, linking it with both phase and amplitude of transmission function by means of generalized formula

$$\tau_B = \sqrt{\tau_p^2 + \tau_T^2} \; ; \; \tau_T = \frac{\partial \ln|T|}{\partial \omega} \quad (7.36)$$

was performed in [186]. According to an another approach the velocity of tunneling has to coincide with the group velocity of energy transfer $v_g$ [187], determined by division of energy flow on energy density (3.19); in this approach the tunneling time in the bi-prism configuration (Fig. 6) $t_d = d/v_g$ proves to be subluminal. Hartman paradox, derived without any new hypotheses from the standard formulae (7.34), which can be found in many textbooks, stimulated a hot discussion; this controversial situation had acquired a new sense of urgency, leading to conflicting theoretical works (see [188–192] and the literature therein). However, the direct measurement of times of tunneling electron transitions through the quantum barriers proved to be a problem difficult to solve, and thus, the idea to check the Hartman's conclusion in the classical effects of tunneling of EM waves through the macroscopic photonic barriers became popular. This idea was based on the formal equivalence of time-independent Schrödinger equation and the wave equation: tunneling of particles through the forbidden zone could be compared with the propagation of EM wave through the opaque dispersive layer.

The THz technique was used as a convenient experimental basis for these researches, since the spatial-temporal scales of observed effects – centimeters and nanoseconds – can be measured easily, by comparison with the relevant small scales in the optical tunneling. Thus, the measurements of lateral shift of rays in the total internal reflection (Goos–Hanchen shift), performed for a beam of THz microwaves [65], were considered in the favor of concept of subluminal tunneling velocity. Unlike this, while measuring the delay time for $TE_{01}$ mode, tunneling through the "undersized" segment in waveguide, some authors [193] had emphasized the proximity of their results to those given by Eq. (7.36). However, these experiments are hampered by the exponential weakening of tunneling wave and its partial scattering on the butt-ends of undersized segment.

On the other hand, both these sources of losses can be excluded due to the reflectionless tunneling of $TE_{01}$ mode through the gradient barrier. To compare and contrast the FTIR effects in waveguides, caused by the narrowing of its cross section and gradient diaphragm, one has to stress out some peculiarities of wave barriers under discussion:

(1). The reflectionless barrier can be considered as a phase shifter for the transmitted wave without its attenuation.

(2). The results of measurement of propagation time through the undersized waveguide were reported in [193] to tend to the values $\tau_B$ (7.36), dependent upon the derivative $\frac{\partial \ln|T|}{\partial \omega}$; however, in a case of reflectionless tunneling ($|T| = 1$) this derivative vanishes, and this limit shows the possibility to check this theory of FTIR.

(3). The phase shift $\phi_t$ (7.32), accumulated by the mode tunneling in waveguide through the diaphragm with total thickness D, can exceed the phase shift of the same mode, accumulated at the same distance D in the waveguide without diaphragm $\Phi = \beta D$; thus, in the above mentioned example, when R = 0 for two layers (m = 2, D = 2d) the phase shift for evanescent $TE_{01}$ mode is $\phi_t = 1.49$ rad, meanwhile $\Phi = 0.61$ rad. Exciting $TE_{01}$ mode in the system of two similar parallel waveguides, one of which contains the gradient diaphragm for FTIR, one can expect, that some phase shift $\Delta\phi$ between the modes, passing through this system, will occur; in the example discussed $\Delta\phi = 0.88$ rad. Calculating from the measured value $\Delta\phi$ the phase $\phi_t$ and comparing the time of this phase formation ($t_1 = \Phi/\omega$) with the phase time $\tau_3 = \dfrac{\partial \phi_t}{\partial \omega}$ and the time $t_g$, determined by the group velocity of the evanescent mode (3.19), one can compare the effectiveness of competing theories of FTIR.

*7.5. Next steps: phase effects, random heterogeneties, spac- time analogies ...*

Side by side with the abovementioned eventual generalizations of models of heterogeneous photonic barriers to GHz range discussed above one can mention the some topics in gradient optics which, though being "hot" ones, are however insufficiently elaborated:

(1). Phase effects : the phases of waves, reflected from and transmitted through the gradient barriers, are sensitive to the gradient of refractive index profile inside the barrier. This property attracts attention to the compositionally graded ($Ba_xSr_{1-x}$)$TiO_3$ (BST) thin films (0 < x <1), regarded as a promising candidate for applications in electrically controllable microwave filters [194] and phase shifters [195]. Thus, in tunable phase shifters [195] for 10 GHz wave the composition of layer, varied from $BaTiO_3$ (x = 1) on one side up to $SrTiO_3$ (x = 0) on the other side, provided phase shifts of about 20° – 70° for a film's thickness of d = 330 nm.

(2) The random heterogeneties are of great practical interest for optics of very thin films, where the creation of controlled profile of refractive index becomes technologically difficult. One can mention the analysis of such effects for non-regular wires [196], surface plasmon polaritons in random nanostructures [197] and depolarization of EM field in randomly heterogeneous media [198].

(3). There is a remarkable space-time analogy between the wave propagation in gradient stationary media and homogeneous non-stationary media [199]. Thus, the Maxwell equation for time-dependent dielectric susceptibility $\varepsilon(t) = n_0^2 U_1^2(t)$ can be obtained from equation (2.6) due to replacement U(z) → $U_1(t)^{-1}$, $L_{1,2}$ → $t_{1,2}$ and, respectively, the solution of this non-stationary equation follows from stationary solution (2.13) due to replacements in the phase factor $\eta \to z$, $t \to \tau$, where the characteristic non-stationarity-induced frequency Ω is found due to the aforesaid replacements in (2.11), with variable $\tau$ is defined by analogy with (2.7): $\tau = \int dt / U_1(t)$; herein this newly obtained phase factor reads as exp[i( qz - $\omega\tau$ )]. An important problem is posed by combined analysis of effects, caused by spatial-temporal variations of dielectric parameters, e.g., reflectance/transmittance of non-stationary gradient media [200].

This approach may be continued in a problem of diffraction of localized wave beams and pulses in heterogeneous media [201]. Moreover, the series of actual tasks in dynamics of ultrashort single-cycle pulses with gradient layers still remains unsolved [202].

Opening the avenues to design materials with electromagnetic properties unattainable in nature, gradient optics is developing at the frontier between electromagnetics and nanotechnology. Penetration of physical concepts and analytical methods of this field to the theories of waves of another physical nature, especially to quantum mechanics, is progressing rapidly. Referring to the recent Nobel Lecture of Prof. T. Haensch one can expect, that "the biggest surprise in these endeavors would be if we found no surprise".

**Acknowledgements.**

The authors are much obliged to Prof. T. Arecchi, S. Haroche, A. Migus, L. Stenflo for their interest into these topics and numerous discussions. One of the authors (A.S.) thanks the Calouste Gulbenkian Foundation and Prof. V.V. Konotop for their hospitality in the Lisbon University, where this work had begun. Work of V. K. was partially supported by COST P11 Action, and that of A.S. and G.P. by NATO grant n° PST.CLG 980331

**Figure Captions :**

Figure 1: (a) Concave and convex profiles U(z), given by Eq.(2.9), are indicated by full and dashed lines, respectively; (b) profile (1) given by Eq. (2.27) (full), (2) given by an inverse dependence z = z(U) (2.20) (dotted).) and (3), presenting the Cauchy profile (4.39) (dot-dash), are plotted vs the normalized coordinate z/L.

Figure 2 : Reflection coefficient $|R|^2$ for gradient photonic barriers (2.9) ($n_0$ = 1.8, y = 0.45, d = 120 nm) in visible and near IR ranges (normal incidence). Reflectance of the single layer with concave profile U(z) and one, two and six layers with convex profiles U(z) are represented by curves 1, 2, 3 and 4, respectively.

Figure 3 : Pairs of adjacent identical concave (a) and convex (b) profiles of gradient barriers U(z) (2.9), illustrating the discontinuities of gradient and curvature of U(z) on the internal boundary z = d, where U = 1. discontinuities of adjacent grad U in pairs of concave (3a) and convex profile(b) ,both with $y^2$ = 1/3. (c), (d) pairs of adjacent smoothly tangent concave and convex profiles characterized, unlike (a,b), by continuity of gradU and discontinuity of curvatures of $U_1$ and $U_2$ on the border z =d. (c) ($y_1^2 = y_2^2 = 1/3$, $d_1 = d_2$) shows the case of unequal deviations of $U_{min}$ and $U_{max}$ from the boundary value U = 1. 3; (d) ($y_1^2 = 1/3$, $y_2^2 = 0.2025$, $d_2 = (y_2^2/y_1^2)$, corresponds to the case of equal deviations of $U_{min}$ and $U_{max}$ from U = 1.

Figure 4. : Reflectance spectra for the pairs of contiguous gradient photonic barriers ($n_0$ = 2) with profiles U(z), shown on Fig.3. Fig. 4a corresponds to the pairs depicted on Fig. 3a – 3b (reflection from the discontinuity of gradients of profile U(z)). Fig. 4b corresponds to the configurations, shown on Fig. 3c – 3d (reflection from the discontinuity of curvatures of profiles U(z)).

Figure 5 : Group velocity $v_g$ and normalized group velocity V (3.22), plotted as a function of coordinate z in three different symmetric barriers U(z), $n_0$ = 1.8: (a) convex profile, d = 100 nm, $U_{max}$ = 1.25; (b) concave profile, d = 100 nm, $U_{min}$ = 0.75; (c) concave profile, d = 80 nm, $U_{min}$ = 0.75.

Figure 6 : Frustrated total internal reflection of wave 1 and Goos-Hanchen shift l for reflected (2) and tunneling (3) waves in the double-prism configuration.

Figure 7 : Transmittance spectra for waves tunneling through 1 (curve 1) and 2 (curve 2) gradient films (2.9) ($n_0$ = 2.35, $y^2$ = 1/3); the normalized frequency u = 1.1 relates to the reflectionless tunneling (non-attenuated transmittance) through 2 films.

Figure 8 : Set of gradient layers, supported by homogeneous substrate, thickness of each layer and substrate are d and D, respectively; profile of refractive index inside each layer is given in (4.26).

Figure 9 : Spectra of transmittance of a single layer (4.26) ($n_0$ = 1.4, m = 0.75) for inclined incidence of S- and P- waves, plotted vs the frequency-dependent parameter $\gamma$ (4.33); Fig. 9(a) and 9(b) correspond to the angles of incidence $\theta = 75°$ and $\theta = 65°$, respectively.

Figure 10 : Spectra of transmittance of a pair of similar layers (4.26) ($n_0 = 1.4$, $\theta = 75°$) for S – and P- waves; Fig 10a and 10b relate to the depth of refractive index modulation m = 0.95 and m = 0.86 respectively, frequency-dependent parameter $\gamma$ is defined in (4.33).

Figure 11 : Narrow-banded peaks of reflectionless tunneling of S- waves through the pair of gradient layers (4.26) ($n_0 = 1.4$, m = 0.75, $\theta = 65°$); the transmission coefficients $|T|^2$ are plotted vs the normalized frequency u.

Figure 12 : Influence of a thick substrate (n = 2.32) on transmittance spectra of S- and P- polarized waves through one layer gradient photonic barrier (4.26) ($n_0=1.4$, m = 0.86, $\theta = 65°$); curve $S_0$ shows the transmittance of the same barrier without substrate.

Figure13 : Profile of normalized dielectric susceptibility $U^2(z)$, supporting the propagation of surface TE- wave along the boundary of gradient dielectric medium without free carriers.

Figure 14 : Scheme of waveguide, containing the gradient layer with thickness d. The reflectionless tunneling of $TE_{01}$ mode through the layer is provided by concave profile of dielectric susceptibility $U^2$ (7.20).

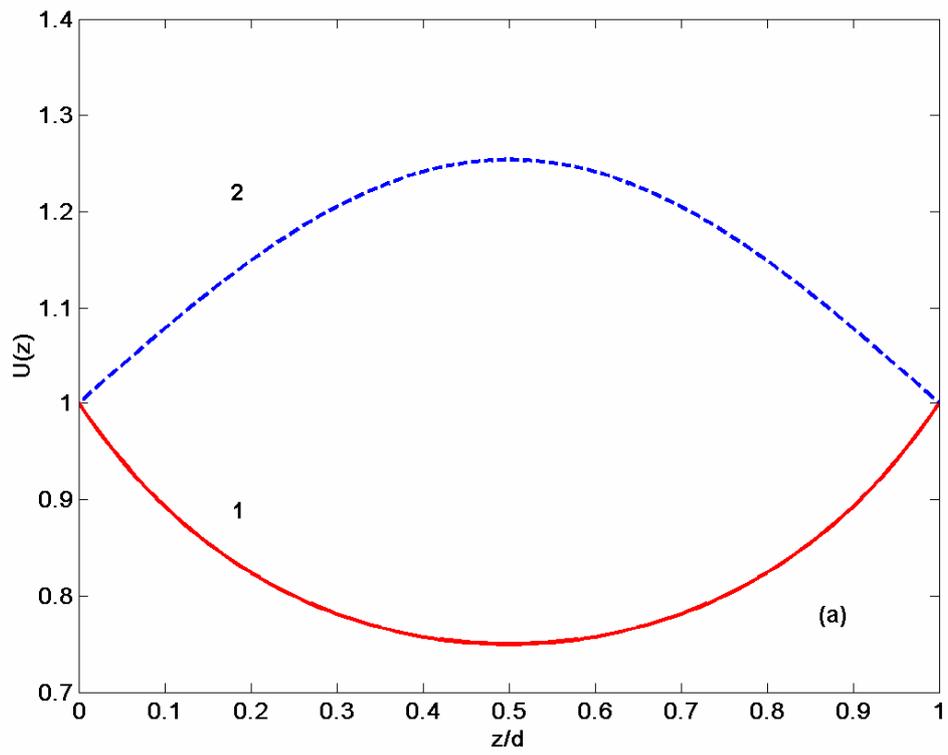

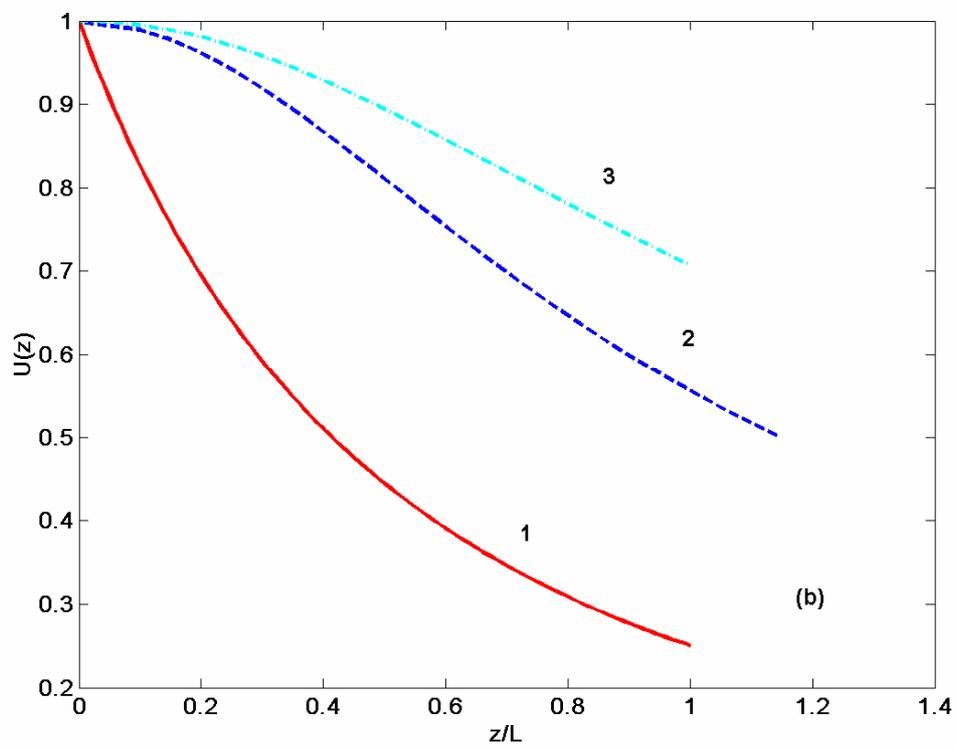

Shvartsburg et al. Figure 1

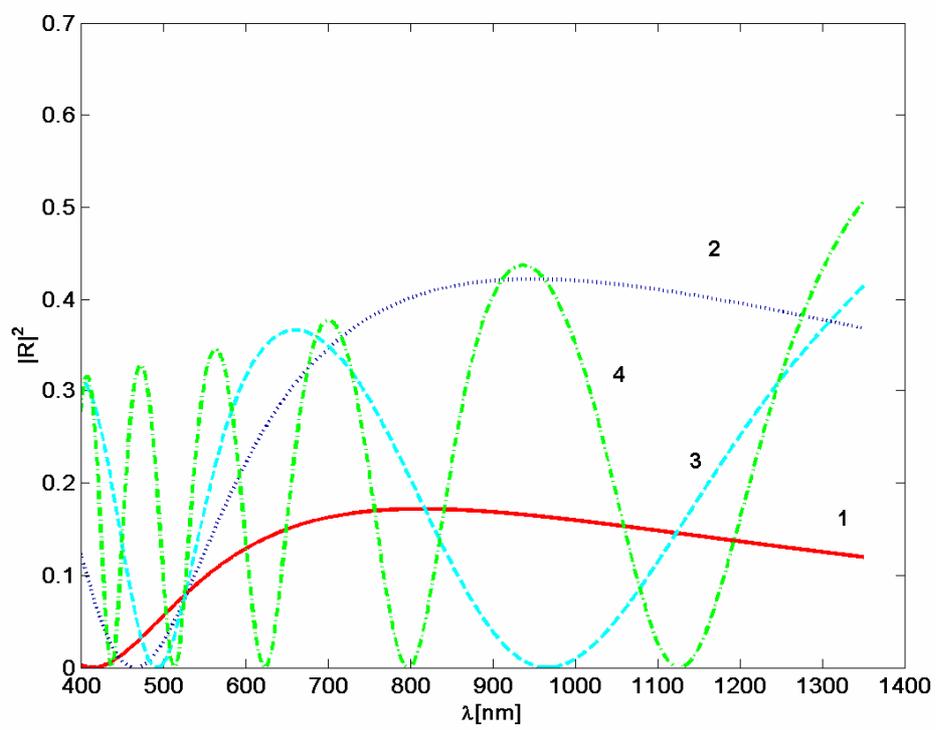

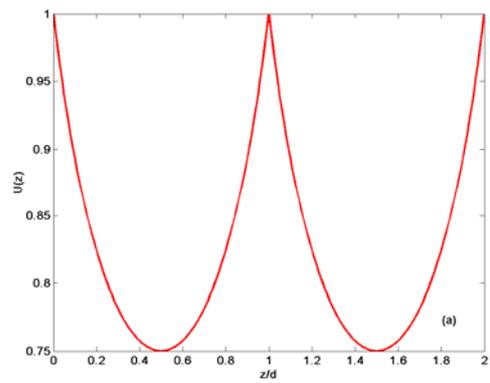
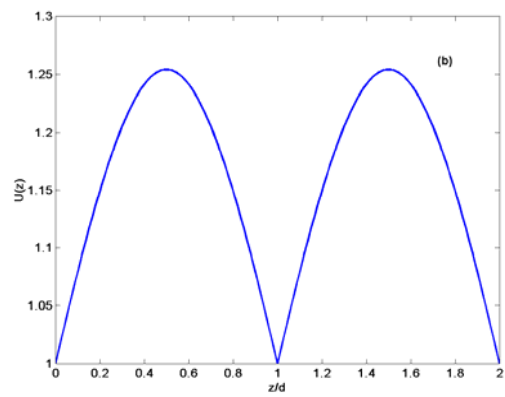
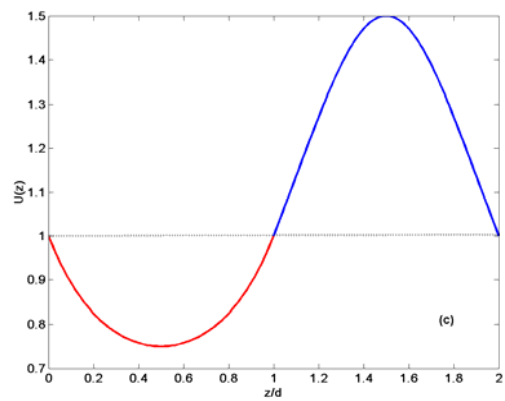
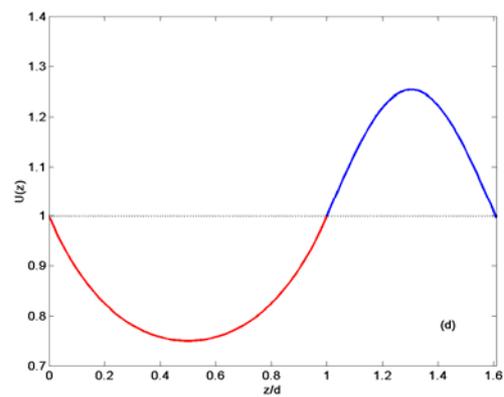

Shvartsburg et al. : Figure 3

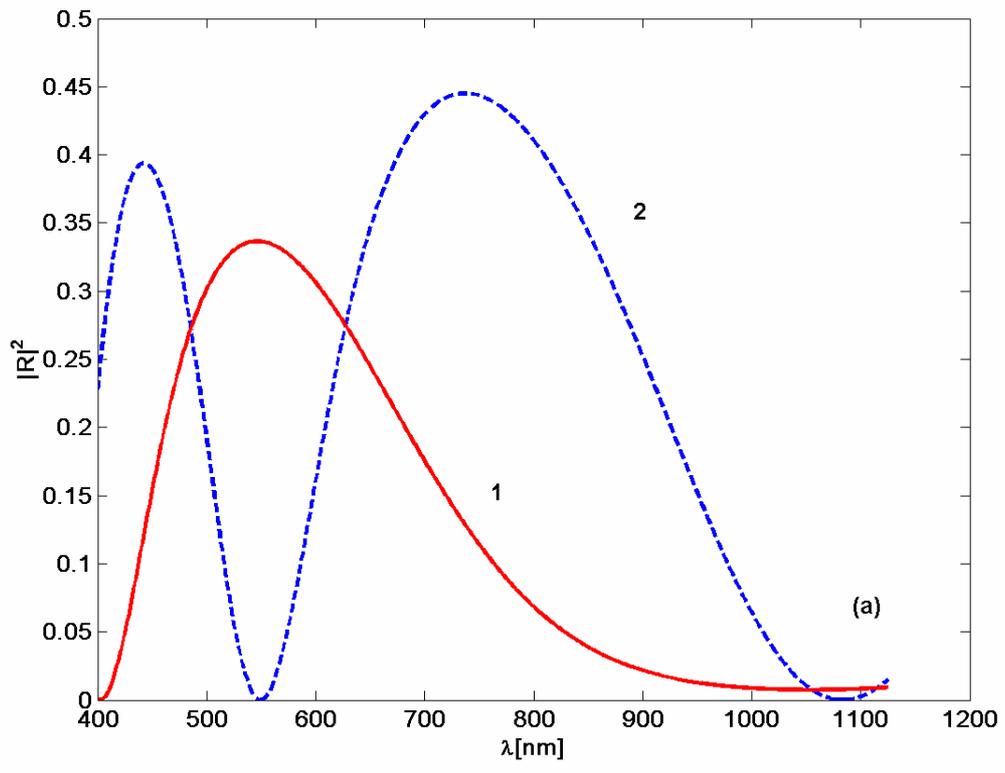

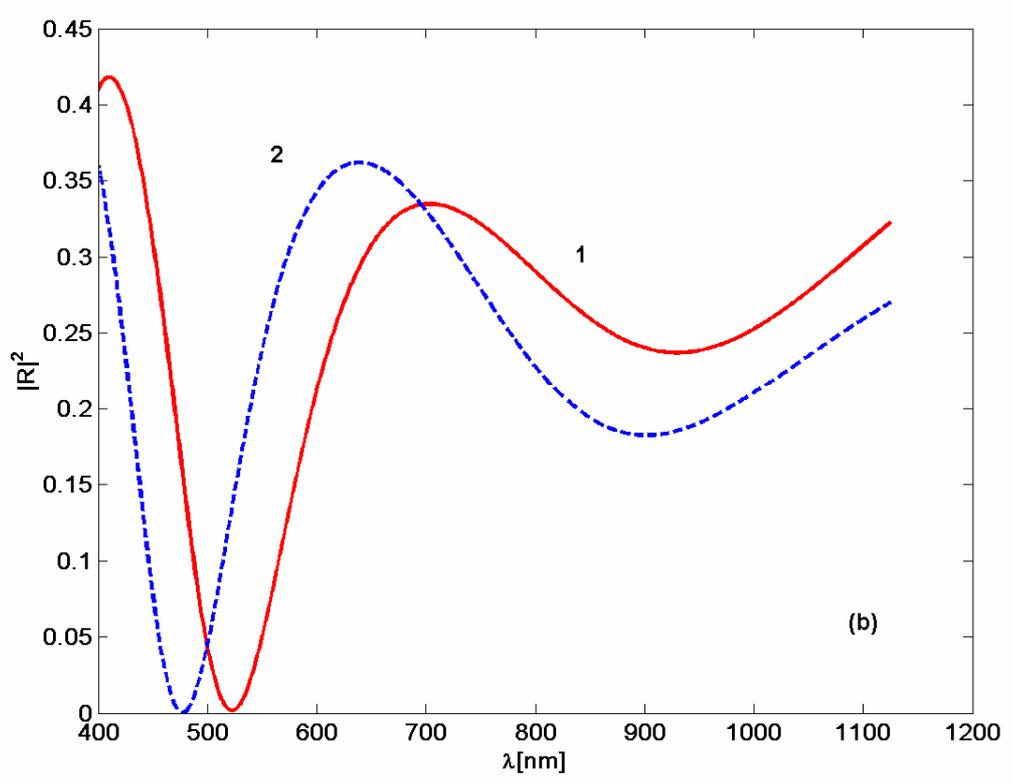

Shvartsburg et al. : Figure 4

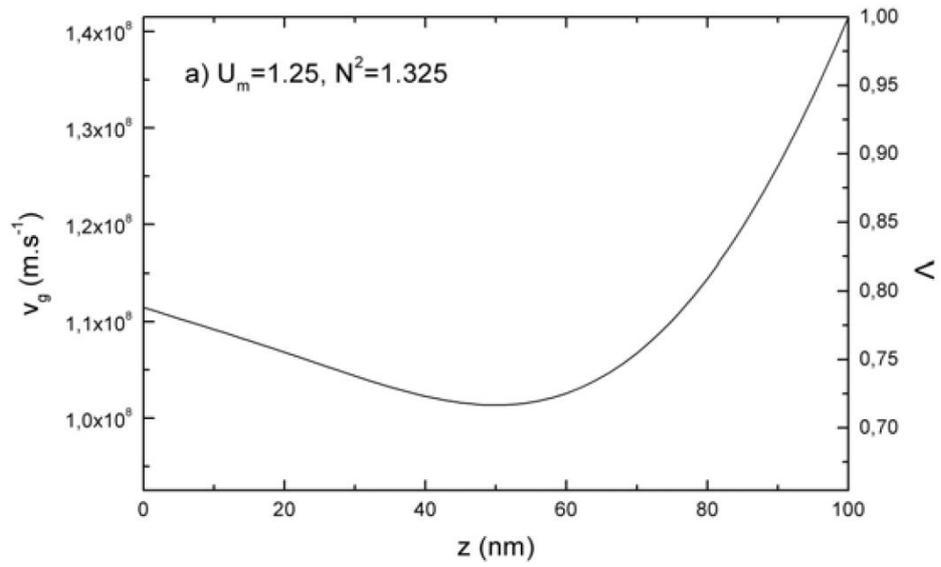

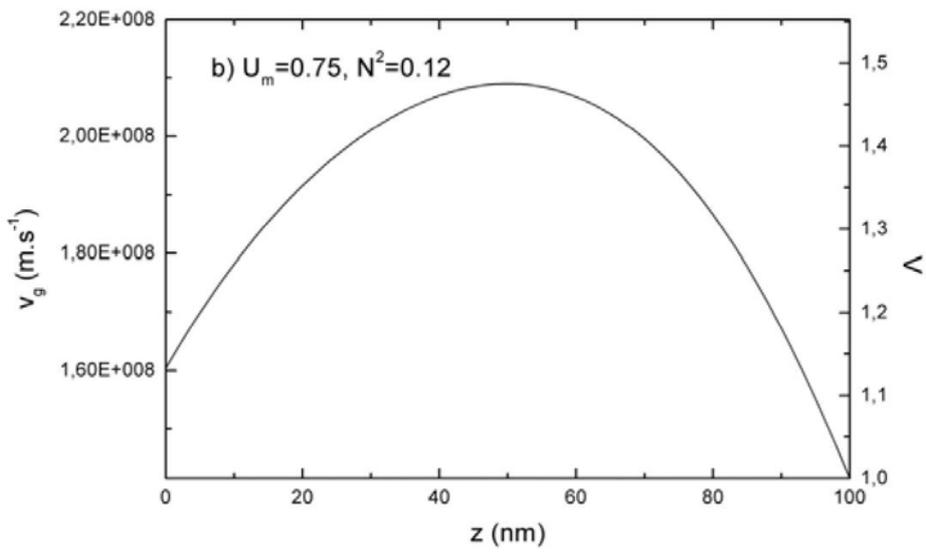

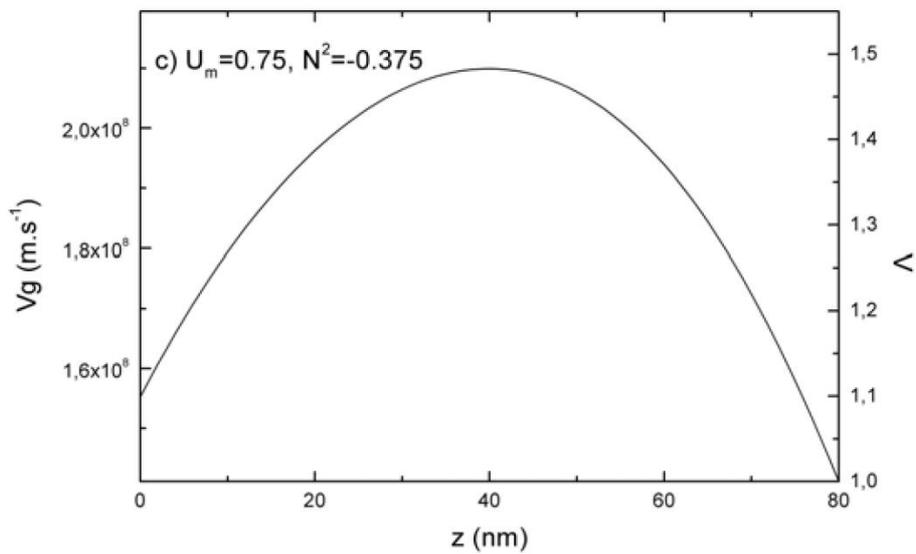

Shvartsburg et al. Figure 5

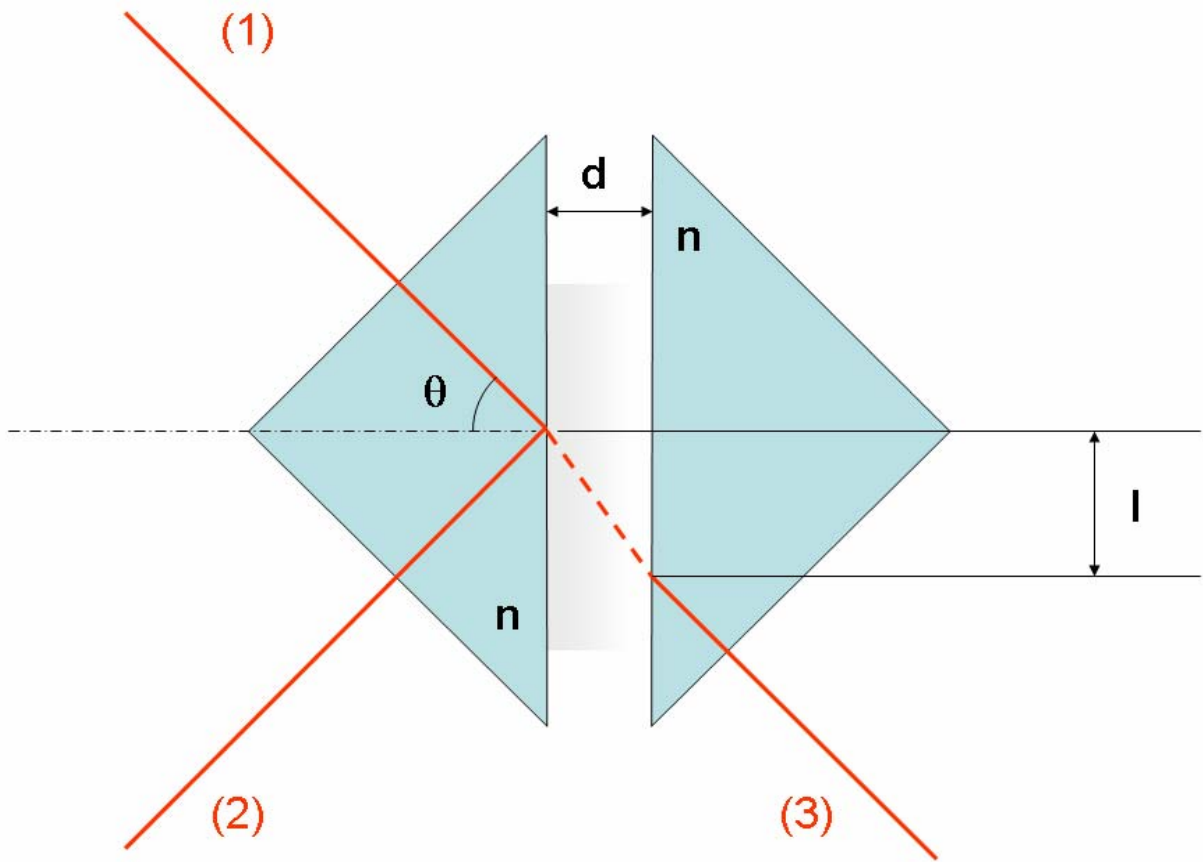

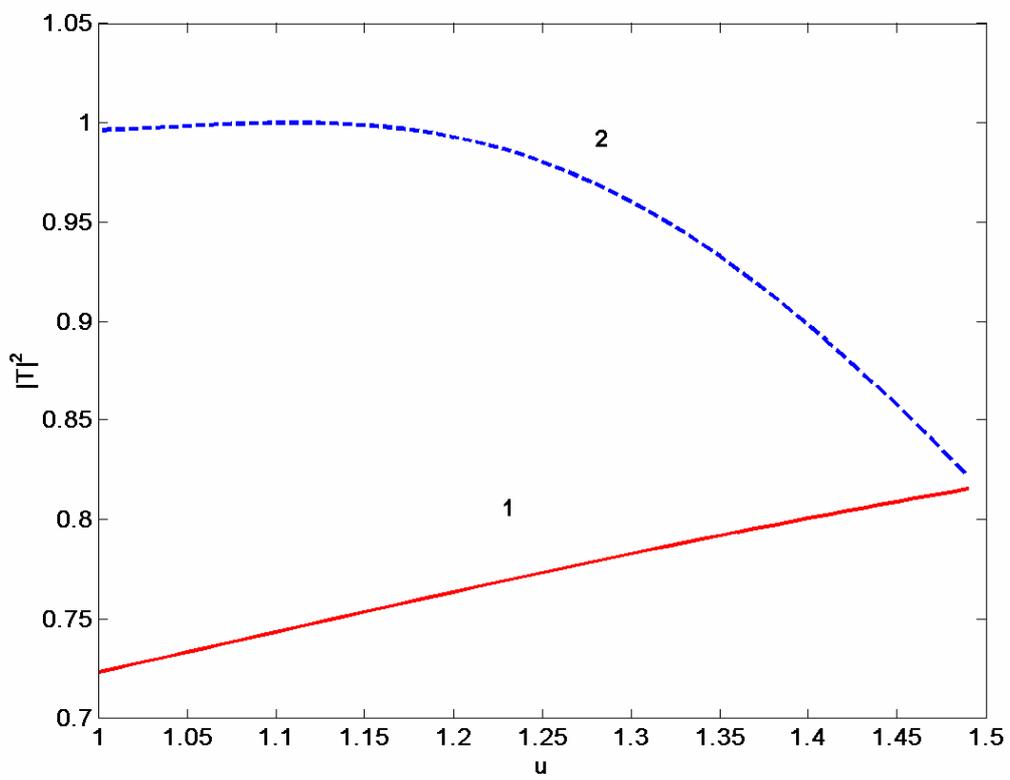

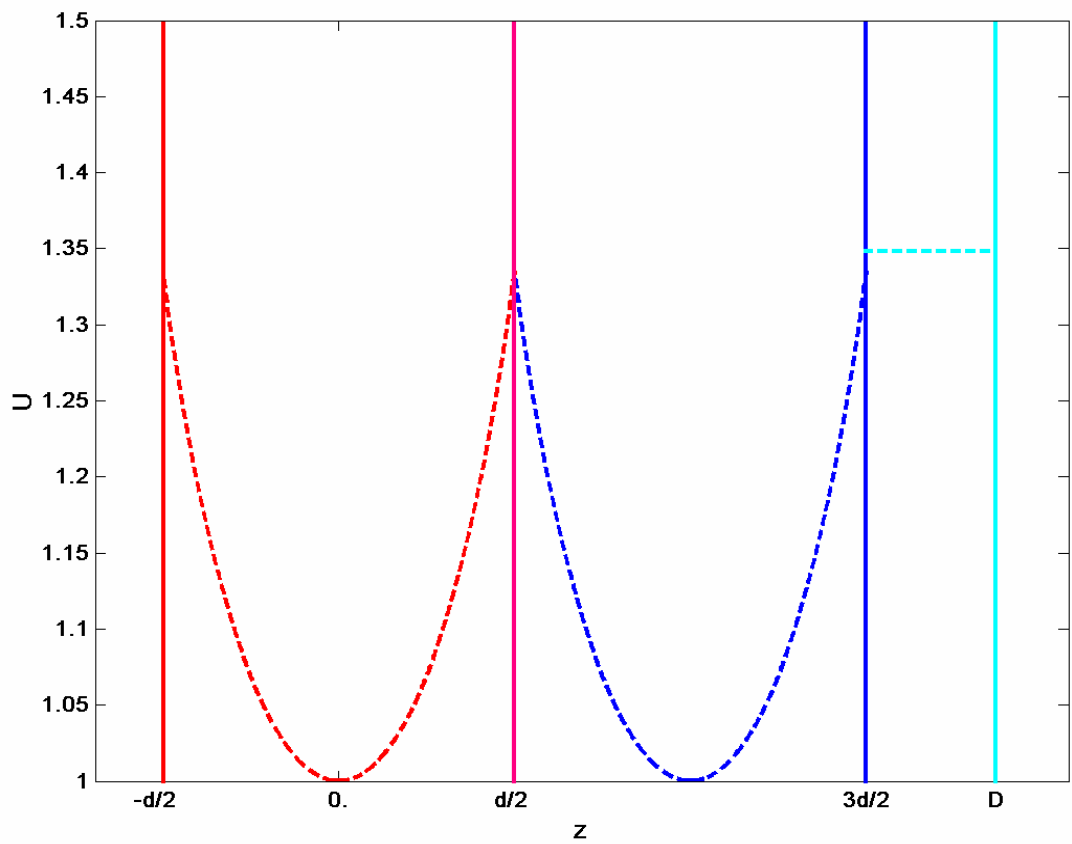

Shvartsburg et al. : Figure 8

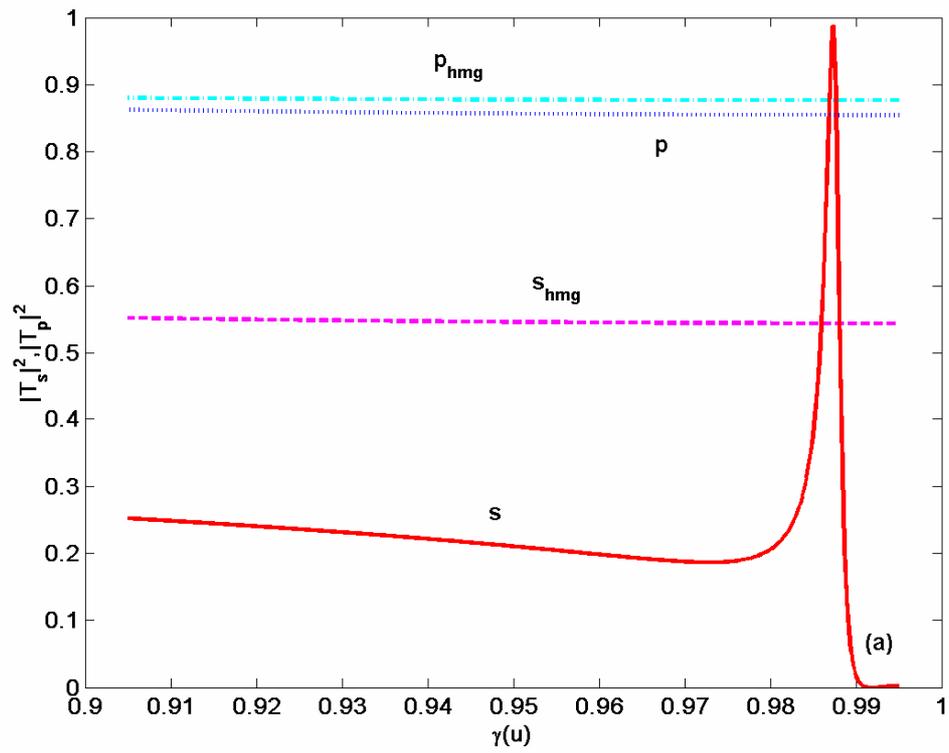

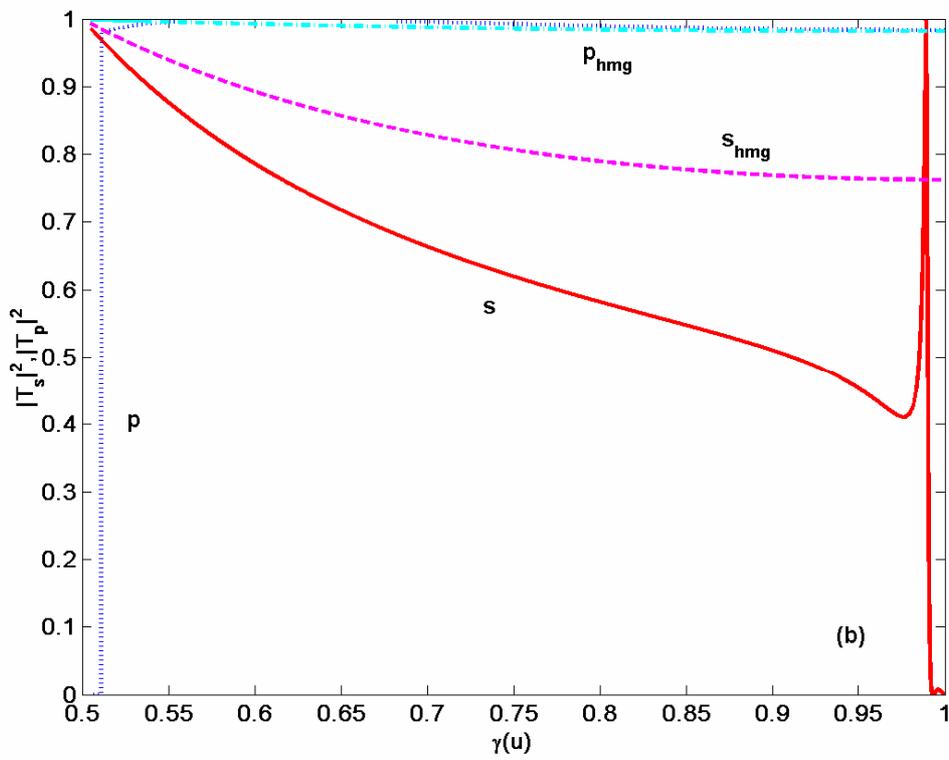

Shvartsburg et al. : Figure 9

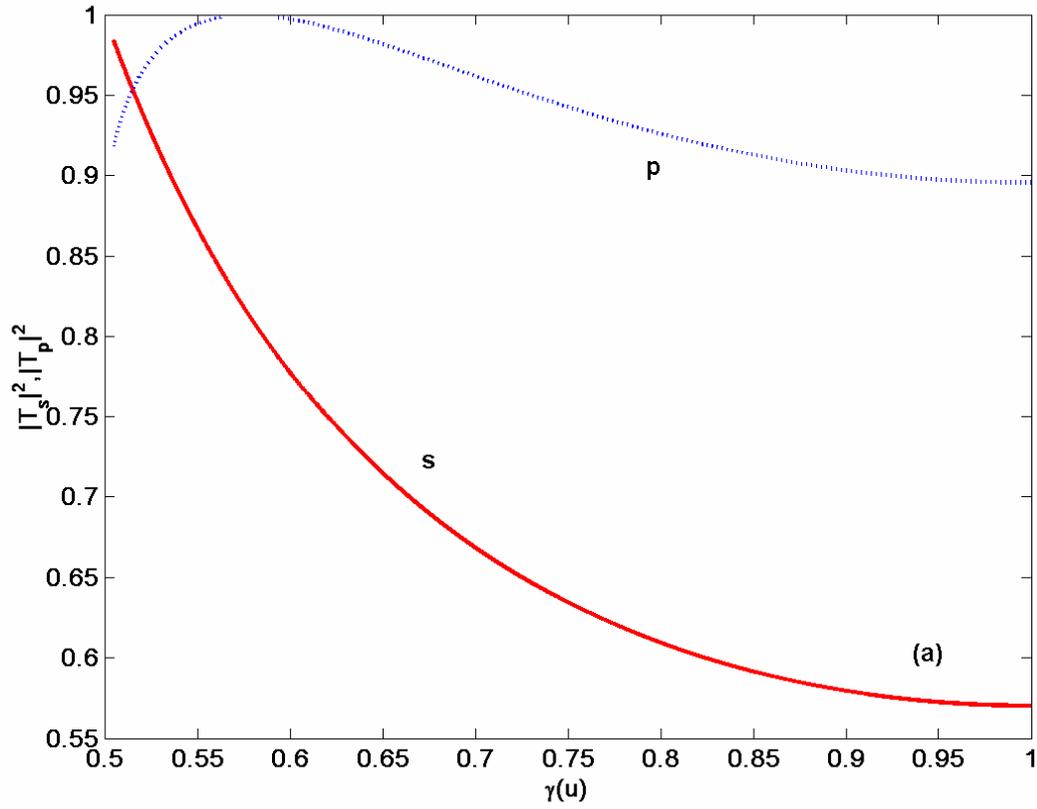

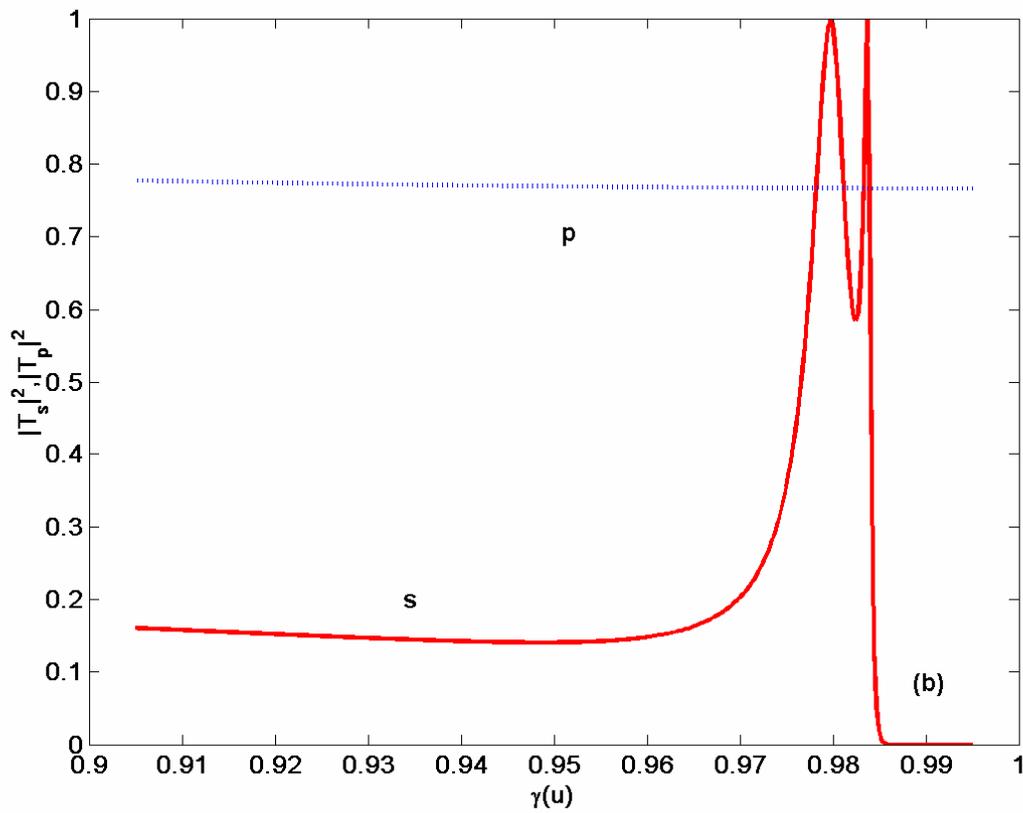

Shvartsburg et al. : Figure 10

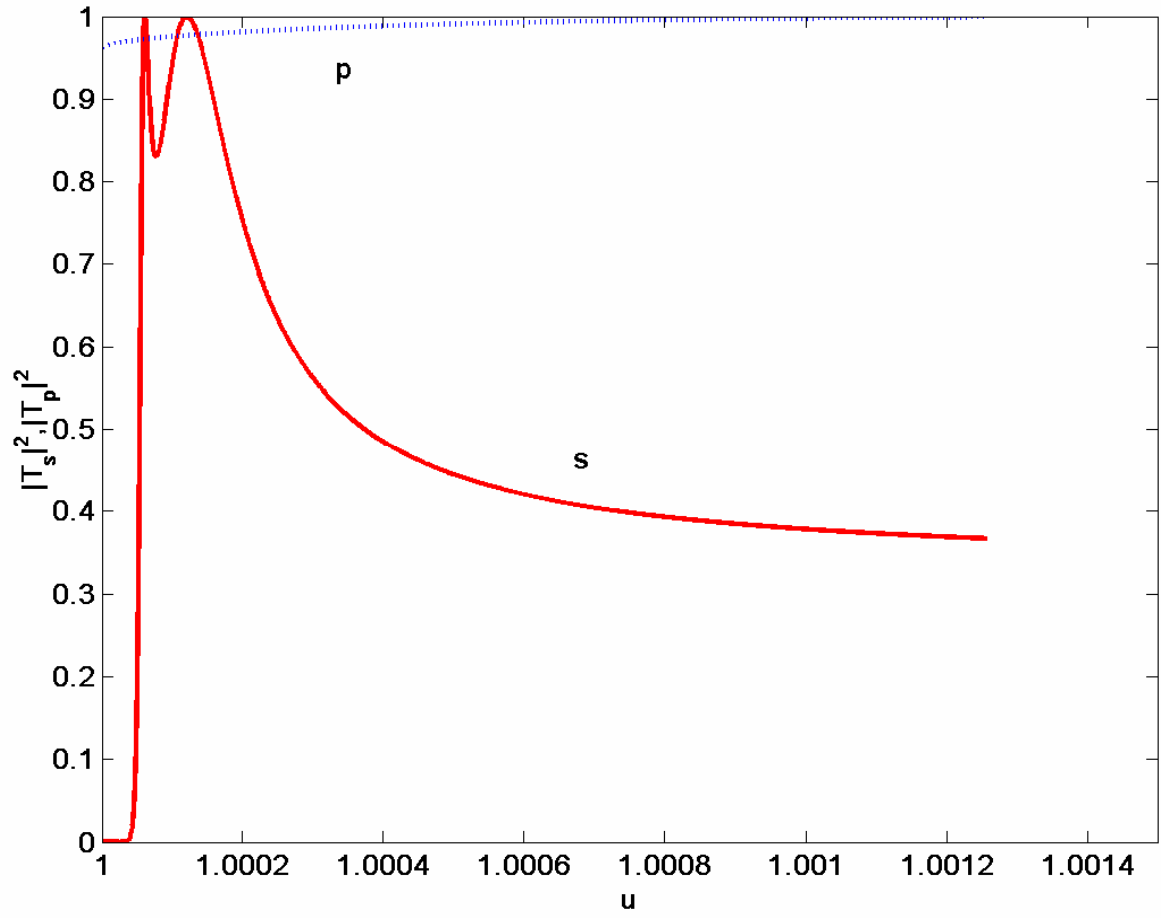

Shvartsburg et al. : Figure 11

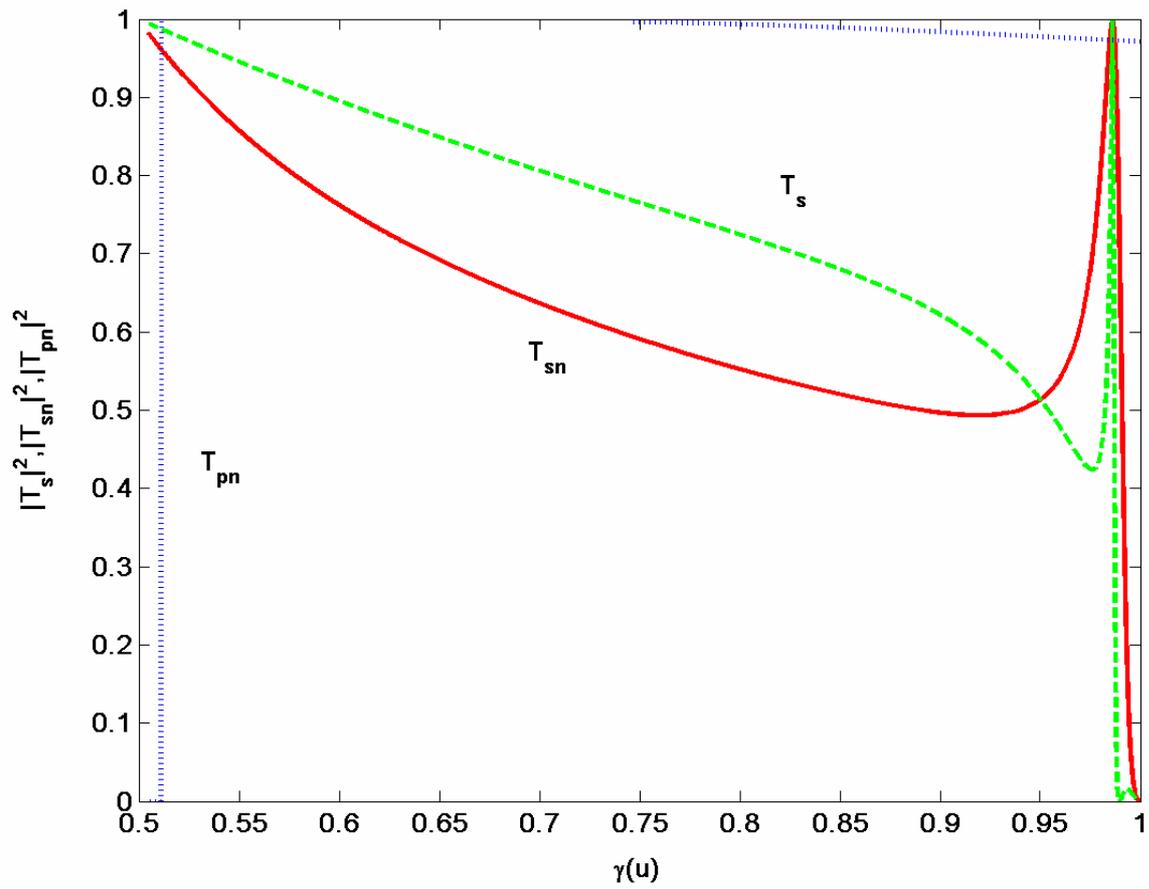

Shvartsburg et al. : Figure 12

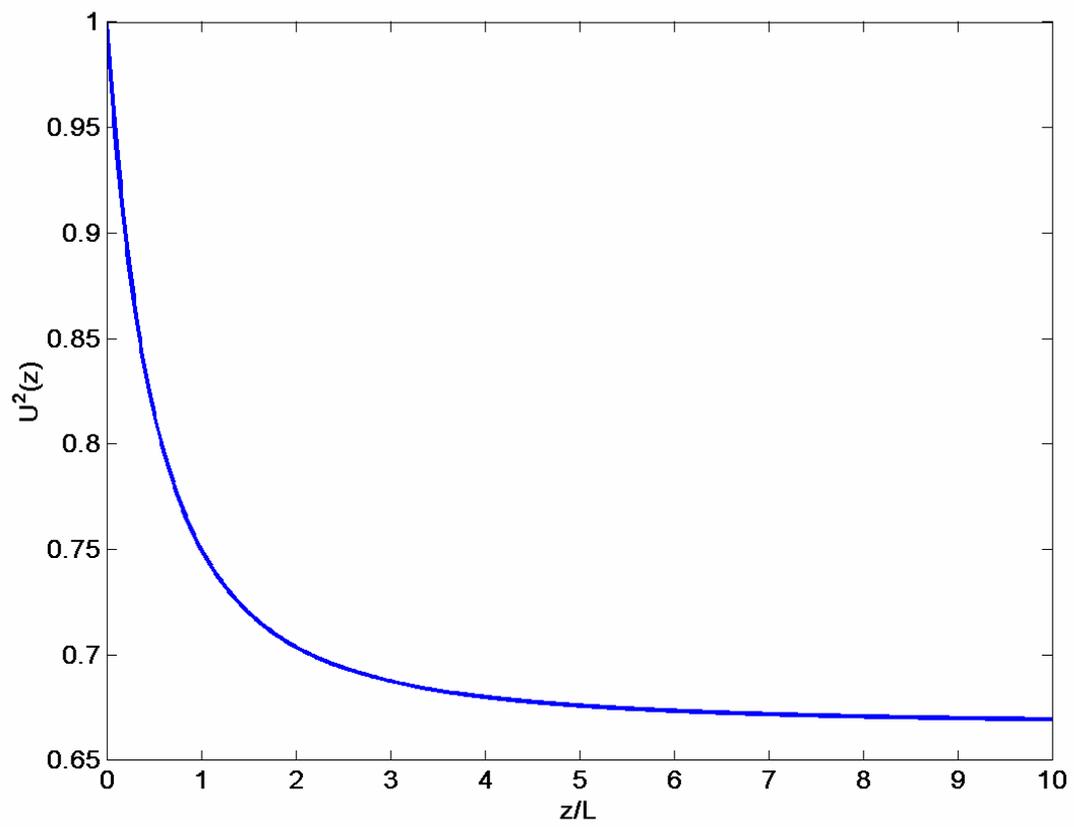

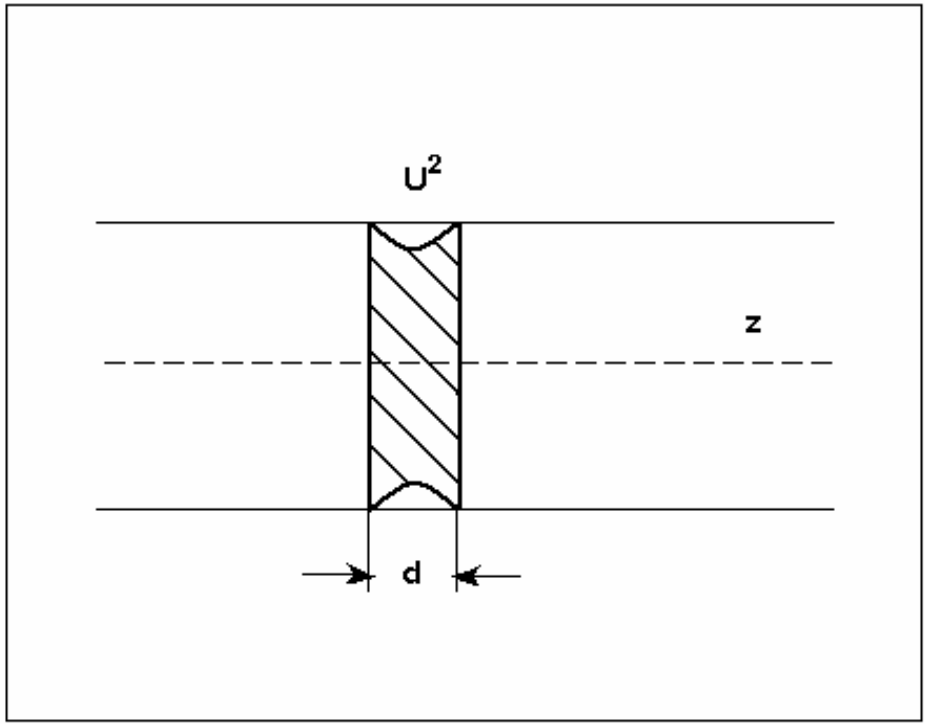

Shvartsburg et al. : Figure 14